\pdfoutput=1
\documentclass[acmsmall]{acmart}

\AtBeginDocument{%
  }

\setcopyright{acmlicensed}
\acmJournal{PACMHCI}
\acmYear{2024} \acmVolume{8} \acmNumber{CSCW1} \acmArticle{44} \acmMonth{4} \acmPrice{15.00}\acmDOI{10.1145/3637321}

\newcommand{\pzh}[1]{{\color{black} #1}}
\newcommand{\zh}[1]{{\color{black} #1}}
\newcommand{\peng}[1]{{\color{black} #1}}

\newcommand{\cqy}[1]{{\color{black} #1}}

\newcommand{\zhenhui}[1]{{\color{black} #1}}

\usepackage{multirow}
\usepackage{array}
\usepackage{tikz}
\usepackage{xcolor}
\definecolor{ui_element}{HTML}{FF7F00}

\usepackage{listings}
\usepackage{xcolor}
\usepackage{footmisc}

\definecolor{eclipseStrings}{RGB}{42,0.0,255}
\definecolor{eclipseKeywords}{RGB}{127,0,85}
\colorlet{numb}{magenta!60!black}

\lstdefinelanguage{json}{
    basicstyle=\normalfont\ttfamily,
    commentstyle=\color{eclipseStrings}, 
    stringstyle=\color{eclipseKeywords}, 
    numbers=left,
    numberstyle=\scriptsize,
    stepnumber=1,
    numbersep=8pt,
    showstringspaces=false,
    breaklines=true,
    frame=lines,
    backgroundcolor=\color{gray}, 
    string=[s]{"}{"},
    comment=[l]{:\ "},
    morecomment=[l]{:"},
    literate=
        *{0}{{{\color{numb}0}}}{1}
         {1}{{{\color{numb}1}}}{1}
         {2}{{{\color{numb}2}}}{1}
         {3}{{{\color{numb}3}}}{1}
         {4}{{{\color{numb}4}}}{1}
         {5}{{{\color{numb}5}}}{1}
         {6}{{{\color{numb}6}}}{1}
         {7}{{{\color{numb}7}}}{1}
         {8}{{{\color{numb}8}}}{1}
         {9}{{{\color{numb}9}}}{1}
}

\begin{document}

\title{DesignQuizzer: A Community-Powered Conversational Agent for Learning Visual Design
}

\author{Zhenhui Peng}
\authornote{Corresponding author.}
\email{pengzhh29@mail.sysu.edu.cn}
\affiliation{%
  \institution{Sun Yat-sen University}
  \city{Zhuhai}
  \country{China}
}

\author{Qiaoyi Chen}
\email{chenqy99@mail2.sysu.edu.cn}
\affiliation{%
  \institution{Sun Yat-sen University}
  \city{Zhuhai}
  \country{China}
}

\author{Zhiyu Shen}
\email{shenzhy23@mail2.sysu.edu.cn}
\affiliation{%
  \institution{Sun Yat-sen University}
  \city{Zhuhai}
  \country{China}
}

\author{Xiaojuan Ma}
\email{mxj@cse.ust.hk}
\affiliation{%
  \institution{The Hong Kong University of Science and Technology}
  \city{Hong Kong}
  \country{China}
}

\author{Antti Oulasvirta}
\email{antti.oulasvirta@aalto.fi}
\affiliation{%
  \institution{Aalto University}
  \city{Helsinki}
  \country{Finland}
}



\renewcommand{\shortauthors}{Zhenhui Peng et al.}

\begin{abstract}
Online design communities, where members exchange free-form views on others’ designs, offer a space for beginners to learn visual design. However, the content of these communities is often unorganized for learners, containing many redundancies and irrelevant comments. In this paper, we propose a computational approach for leveraging online design communities to run a conversational agent that assists informal learning of visual elements (e.g., color and space). Our method extracts critiques, suggestions, and rationales on visual elements from comments. We present DesignQuizzer, which asks questions about visual design in UI examples and provides structured comment summaries. 
\peng{Two user studies demonstrate the engagement and usefulness of DesignQuizzer compared with the baseline (reading reddit.com/r/UI\_design). We also \cqy{showcase} how effectively novices can apply what they learn with DesignQuizzer in a design critique task and a visual design task.} 
We discuss how to use our approach with other communities and offer design considerations for community-powered learning support tools.

\end{abstract}



\received{January 2023}
\received[revised]{October 2023}
\received[accepted]{December 2023}
    \maketitle






\begin{figure}
  \centering
  \includegraphics[width=\linewidth]{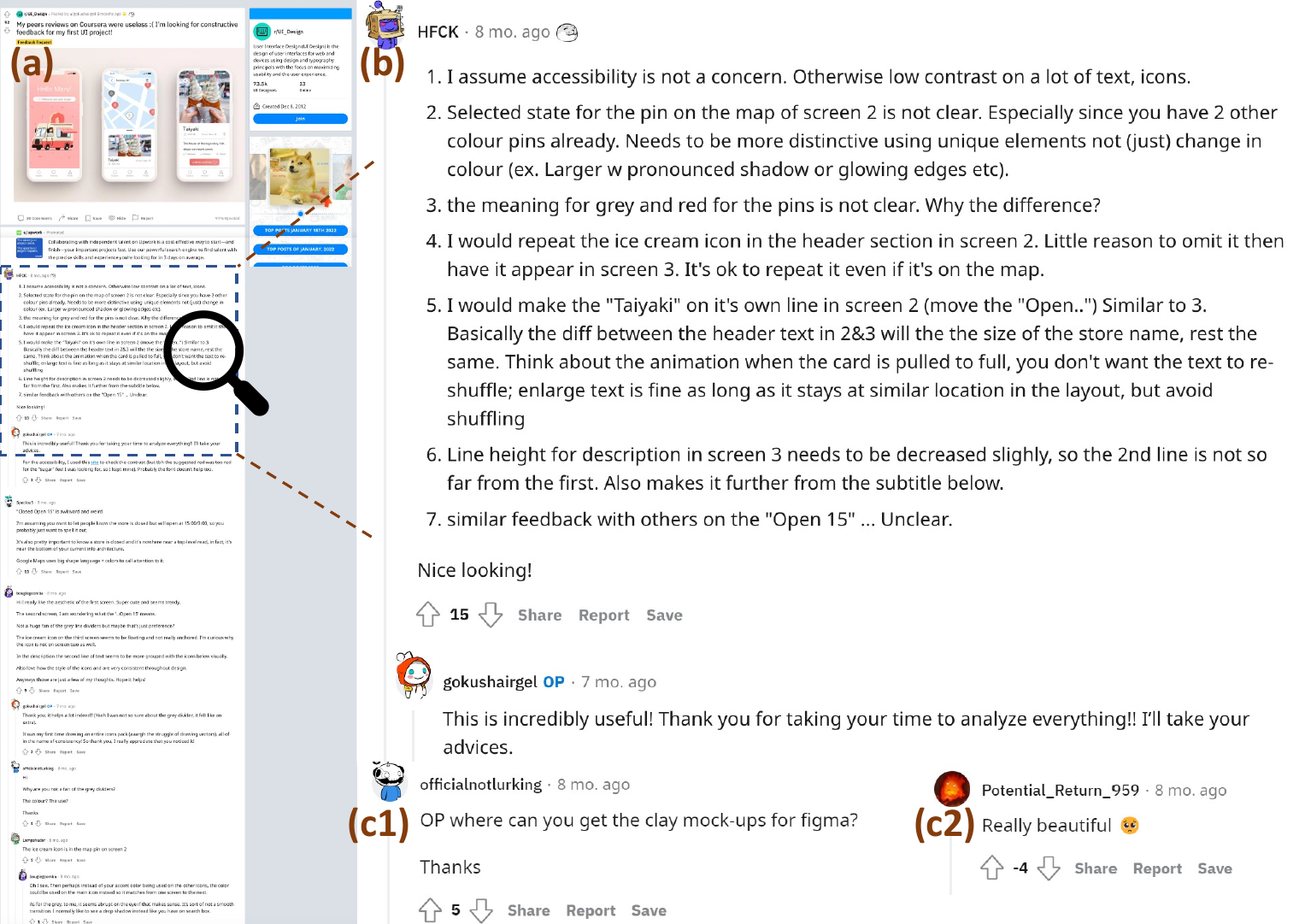}
  \caption{Online design communities like Reddit r/UI\_Design offer design examples and comments as learning resources for novices. For instance, under a UI feedback-request post (part \textit{a}), there are comments (e.g., \textit{b}) providing meaningful feedback on the example's visual design. However, many comments are general (e.g., \textit{c1, c2}) and often unstructured from learners' aspects regarding the types of feedback, UI components, and visual elements. In this paper, we develop computational approaches for extracting and structuring meaningful feedback on UI design examples, based on which we present DesignQuizzer to assist beginners in learning visual design.}
  \label{fig:reddit}
\end{figure}

\section{Introduction}


Online design communities, e.g., Reddit r/UI\_design (\autoref{fig:reddit}), offer a public space for individuals to learn about design \cite{online_feedback_exchange,critique_me,example_for_learning,learnfromexamples,designing_with_examples,Barrett1988ACO, Lundstrom2009ToGI}. 
For instance, the shared design examples and constructive feedback can help novices learn how a key design principle (e.g., the color theme of a webpage) is enacted or violated in practice \cite{learning_and_transfer,case_comparison,feldman1994practical}. 
Professional designers also participate in design communities to seek inspiring samples and share opinions \cite{koch2018,getting_inspire}. 
This makes these communities more appealing for novices, because it offers a way to learn through peripheral participation.
\peng{This paper focuses on facilitating novices to learn the elementary design principles governing visual elements, such as shape, color, space, form, line, value, and texture \cite{visual_elements}, from design examples and comments in online communities.}
Our focus is motivated, on the one hand, by the fact that these elements are a good starting point for novices to learn about visual design, because they are important, visible, and concrete \cite{best_practice}. 
On the other hand, the commentary in design communities commonly talks about the visual elements of the designs \cite{increase_quality_cscw2017}, making it possible to build computational models to support learning at scale in informal settings outside classrooms.   
\peng{
Our primary targeted user groups are novice designers who only take the user-generated content in the communities as learning materials (merely readers) but do not participate in the content creation. 
This target is different from previous work on design support tools \cite{cocolor_iui2023,MetaMap} or on learning in communities by directly interacting with others via posting and commenting
\cite{kou2017supporting,informal_learning_chi2022,ruijia_cscw2022}. 
}
By exploring the materials in design communities, \peng{novices} are expected to develop their domain skills in visual design,
\peng{which could be reflected by their critiques of others' designs using learned knowledge and \zhenhui{applications of the knowledge to their visual design} \cite{ruijiacscw22,dasgupta2016remixing,dasgupta2018wide,scaffidi2012skill,learning_and_transfer,yang2015uncovering}. 
}

However, it is challenging for beginners to effectively and engagingly leverage the examples and comments in online communities \zhenhui{to learn visual design}. 
For one thing, \cqy{peer comments} are of varying quality \cite{high-quality-cscw2012, high-quality-cscw2014,content_quality_1,structured_feedback} and \cqy{are} typically unstructured for learning purposes (\autoref{fig:reddit}), making it time-consuming for users to find helpful content.  
Existing works have predicted the helpfulness of a comment for the poster of designs based on its language features \cite{critique_style_guide} and qualitatively categorized the content of the feedback \cite{increase_quality_cscw2017}.
Nevertheless, few of them have computationally structured the comments regarding the types of meaningful feedback (e.g., critique, suggestion \cite{increase_quality_cscw2017}) and visual elements that learners are interested in. 

For another, people tend to passively read the posts and comments, which could be less engaging and effective than interactively conversing with a conversational agent (CA) for learning \cite{ICAP_framework}. 
\zh{
According to the ICAP framework \cite{ICAP_framework} that details \cqy{the} potential benefits of different learning activities (Interactive, Constructive, Active, and Passive), CAs can act as human partners to help users construct their own understandings (constructive engagement) or co-develop explanations with others (interactive engagement).
These forms of engagement are more conducive to learning than just reading materials (passive engagement) or manipulating them (e.g., underlining, copying) without new ideas that go beyond the information given (active engagement) \cite{ICAP_framework}.
}
CAs have been demonstrated to be engaging and effective for learning tasks like factual knowledge \cite{quizbot10.1145/3290605.3300587}, programming concepts \cite{sara10.1145/3313831.3376781}, and argumentation skills \cite{arguetutor}. 
\zh{Many of them adopt the quiz-based interaction design that prompts users \cqy{with} a question or a task, requires them to answer or finish it, and provides feedback on their performance \cite{quizbot10.1145/3290605.3300587, sara10.1145/3313831.3376781, arguetutor,peng2022crebot}.}
Nevertheless, little work has looked into the design, effectiveness, and user experience of a community-powered conversational agent for learning visual design. 
Unlike previous knowledge acquisition tasks in which there is usually a correct answer for each question, the creative nature of the visual design makes it challenging to prepare quizzes and corrective feedback. 
\zh{
Moreover, previous educational CAs usually require experts to curate and label learning materials from official sources, which would be less scalable compared to computationally leveraging rich resources from online communities. 
}
Our vision is a computer-supported mode of informal learning where novices can pick up knowledge points by efficiently exploring their interesting content and get engaged in this process by interacting with a conversational agent that generates meaningful learning materials from online communities. 

To this end, we develop a computational workflow to structure comments from online design communities into categories of meaningful feedback, based on which we present DesignQuizzer, which facilitates users to learn visual elements by prompting relevant quizzes around given design examples. 
We present methods that summarize the meaningful feedback from the original comments, classify the summarized feedback sentences into  ``critique'', ``suggestion'', and ``rationale'', recognize the keywords about visual elements and UI components (e.g., button, card) in the feedback sentence, and cluster these keywords into \peng{more abstract} design concepts.  
We apply this workflow to the comments of UI feedback-request posts in Reddit r/UI\_Design, \peng{creating} a quiz pool that supports DesignQuizzer to ask questions on a given design example's visual elements, highlight the classified sentences and keywords in the comment summary, and retrieve questions relevant to users' interesting visual elements or UI components. 

\peng{
We first conduct a within-subjects experiment \uppercase\expandafter{\romannumeral1} with 24 participants to evaluate DesignQuizzer's user experience and usefulness compared to the r/UI\_Design web page baseline. 
The results suggest that DesignQuizzer improved participants' efficiency in exploring helpful UI examples and comments and engagement in the visual design learning process. 
Participants produced more points about the targeted visual elements (e.g., color and typography) with Designquizzer than with the baseline tool when criticizing others' designs.
Nevertheless, due to the potential learning effect on novices' performance in the design tasks, the experiment \uppercase\expandafter{\romannumeral1} did not evaluate DesignQuizzer's impact on the application of the learned knowledge \zhenhui{to} their designs, which is also an important visual design skill. 
We then conduct a between-subjects experiment \uppercase\expandafter{\romannumeral2} with another 28 participants. 
The results suggest that participants in both groups of DesignQuizzer and the r/UI\_Design baseline could \zhenhui{use} the learned knowledge about color and typography to design \cqy{more consistent, distinct, and intentional} UIs in the post-test visual design task compared to the pre-test \zhenhui{task}.
In both experiments, participants felt that DesignQuizzer was significantly more useful and easier to use. 
They favored its structured comments that ease the reading workload and its quiz-like interaction that encourages active thinking. 
Yet, they suggested that DesignQuizzer should further incorporate professional knowledge from external resources and enhance users' sense of community.
We further discuss ways to apply our approach to other communities for computer-supported learning activities.
}


This work contributes to CSCW communities from three aspects. 
First, we develop a computational workflow \zhenhui{that makes} use of the large-scale comment data in online communities for learning purposes. 
This workflow can recognize the critique, suggestion, and rationale of the visual elements and UI components in the comments of UI design feedback-request posts. 
Second, we present an educational application DesignQuizzer for facilitating informal learning activities in online communities. 
\peng{Third, we extend empirical understandings of how people can learn with online user-generated content via two user studies and provide insights into future community-powered learning support tools.} 

\section{Related Work}
Our work is built on previous studies on learning in online communities, online design communities, computational methods for modeling comments, and conversational agents in educational domains.
 \subsection{Learning in Online Communities} 
 \label{section2.1}
\zh{
In the last two decades, we have witnessed the proliferation of online communities that seek to promote informal learning through unstructured activities and social interactions with others. 
Cheng \cqy{et al.} described three distinct categories of the learning outcomes common to informal learning communities: development of domain skills, development of community identity, and development of community practices \cite{ruijia_cscw2022}. 
Developing domain skills refers to the acquisition of knowledge necessary for a person to carry out the core tasks, such as computer programming \cite{scratch,ruijia_cscw2022}, fan fiction writing \cite{positive_reviews_cscw2016}, and encyclopedia article editing \cite{Peripheral_cscw2013}. 
The second type of outcome involves the development of identity as a member of the community like developing relationships, affinities, and a sense of belonging \cite{ruijia_cscw2022}, while the third type means assimilating ``cultural artifacts, norms, and values'' developed in the community over time \cite{barab2012practice}. 
Our work aims to facilitate novices to a discipline in developing their domain skills by exploring the content in online communities. 
As suggested by previous quantitative studies on learning in communities, the learning of domain knowledge can be captured by the size of the learners’ repertoire in terms of the number of types of concepts users can enact after the learning sessions, e.g., computational thinking concepts demonstrated in their projects \cite{ruijiacscw22,dasgupta2016remixing,dasgupta2018wide,scaffidi2012skill,yang2015uncovering}. 
Following these works, we measure learners' developed \cqy{skills} by how many enacted or violated principles about the target visual elements they can recognize on an unseen UI example after the learning session. 
\peng{Apart from this measure, we also examine learners' design skills and compare their performance in a visual design task before and after the learning session in experiment \uppercase\expandafter{\romannumeral2}.}

Previous work mostly seeks to understand and support community members' learning that happens through sharing their creative artifacts \cite{dasgupta2016remixing,gender_learner_community,high-quality-cscw2014,toshare_cscw2017} like design mock-ups \cite{critique_me,ruijia_cscw2020} and/or through social interactions around these artifacts like commenting and critiquing \cite{tausczik2014collaborative,shorey2021hanging,ruijia_cscw2020}.
For example, social computing scholars have documented the way users work together to learn writing and web development skills in fan communities such as FanFiction.net and Archive of Our Own (AO3) \cite{campbell2016thousands,QA_chi2018} as well as programming skills in creative coding communities such as Scratch \cite{ruijia_cscw2022,scratch,informal_learning_chi2022}.
In \cite{ruijia_cscw2022}, Cheng \cqy{et al.} further provided a quantitative analysis of legitimate peripheral participation (e.g., engagement with practice proxies and feedback exchange) and learning outcomes in a programming community, suggesting that users’ early participation in an online community is associated with long-term learning outcomes \cite{ruijia_cscw2022}. 
However, little work has explored how to facilitate those who explore others' artifacts and interactions in the communities without direct participation in the posting and commenting activities. 
Our study fills this gap by supporting these ``mere readers'' in leveraging community-generated resources for learning purposes.

}

\subsection{Online Design Communities}
Online design communities have been found to be beneficial for learners in design. 
For example, people can post their creative works in these communities to get more timely and ``more equal, collaborative, and interactive'' \cite{kou2017supporting} critiques for improvement compared to traditional classrooms and workplaces. 
In fact, many instructors have adopted online peer feedback to support students' learning as their professional and personalized feedback are not always available \cite{peer_assessment_classes,peerfeedback_chi2021}. 
Similar to many other communities in domains like creative writing \cite{campbell2016thousands} and programming \cite{informal_learning_chi2022}, 
online design communities also support interest-driven learners with shared design examples and associated feedback to gain knowledge \cite{kang2018paragon}. 
For instance, the members' critiques on the design examples and suggestions can help learners discern what concepts have been executed effectively in the examples \cite{InteractiveGuidance,increase_quality_cscw2017}.

However, it is non-trivial for interest-driven learners to \cqy{find helpful and needed content in online design communities efficiently} and get engaged in this informal learning activity. 
On the one hand, the peer feedback online is often not as meaningful as that from experts \cite{hui2014crowd,yuan2016almost, frens2018supporting}. 
Although the communities have enabled sorting comments and posts based on keywords and ranking mechanisms (e.g., ``Best'', ``Top'', ``New'', ``Controversial'', and ``Q\&A''), 
novices who lack domain knowledge still find it hard to specify relevant keywords to search \cite{example_for_learning}. 
Besides, the comments after sorting may still \cqy{be unstructured} and contain redundant information \cite{structured_feedback,luther2015structuring}, which hinders learners from locating the content they are interested in. 
On the other hand, the organization of \cqy{posts} and comments in online design communities tends to offer people a passive-receiving learning experience. 
According to the ICAP framework \cite{ICAP_framework}, the learners' engagement with learning materials can range ``from passive to active to constructive to interactive'' \cite{ICAP_framework} and would result in an improved learning outcome. 
Reading entire text passages (e.g., the comments of the design examples) silently without active highlighting and \cqy{note-taking} is a typical passive-receiving activity \cite{ICAP_framework}, while asking and answering comprehension questions with a partner is an interactive learning activity \cite{ICAP_framework}. 
\zh{Conversational agents can play the roles of such partners to promote interactive engagement in learning activities and have been demonstrated engaging and effective in learning tasks like argumentative writing \cite{arguetutor} and factual knowledge \cite{quizbot10.1145/3290605.3300587}.}

In all, our research is motivated by the \zhenhui{benefit of online design communities for novices' learning tasks}, and we explore an effective and engaging way for them to leverage the shared design examples and comments \zhenhui{to learn visual design}.

\subsection{Computational Methods for Modeling Comments} 
To support efficient comment exploration, existing HCI work has explored a variety of computational techniques to help users filter, structure, and digest the comments in online communities. 
For example, researchers on community-based question-answering platforms exploit text summarization techniques to extract concise takeaways from threads \cite{keikha2014evaluating,song2017summarizing}. 
Literature about online health communities widely applies document/sentence classification methods, e.g., random forests, linear regressors, and neural networks, to predict the satisfaction level and the amount of sought/received support expressed in members' text \cite{zhenhui2020,zhenhui2021chi,yangseeker,yang2019channel,sharma2018mental,wambsganss2020conversational}. 
As for the design domain, Krause et al. rated the helpfulness of the feedback from 176 online providers on students' design solutions and extract a set of natural  language features (e.g., specificity, sentiment\cqy{, etc.}) that correlated with the ratings \cite{critique_style_guide}. 
Yen et al. developed an interactive visualization tool named Decipher that helps designers group the received feedback based on sentiment, keywords (e.g., typography, color) and their interpretation (e.g., fix, keep in mind, need clarification) \cite{yen2020decipher}. 
However, little work has tried to computationally structure the comments in online design communities from learners' perspectives, e.g., about the mentioned UI components and visual elements of their interests. 
While there are qualitative findings about what types of design feedback (e.g., the specific, actionable, and justified ones) are perceived \cqy{as helpful} and of a high quality \cite{InteractiveGuidance,structured_feedback}, it is under-investigated how to model these types. 
In this paper, we complement these previous works with a computational approach to model comments regarding the critiques, suggestions, and rationales on visual elements and UI components in design examples shared in online design communities.


\subsection{Conversational Agents in Educational Domains} 
To engage users in their learning tasks, HCI communities have designed and developed conversational agents that converse with users in diverse knowledge domains \cite{weber_pca,auto_tutor, divekar2021conversational}.
For example, AutoTutor has been used to teach college students in computer literacy and critical thinking \cite{auto_tutor}. 
AutoTutor provides explanations, feedback,  scaffolding, deep reasoning questions, and subject content in online courses, and multiple studies have demonstrated its effectiveness in improving learning gains \cite{auto_tutor}. 
Similarly, Wambsganss et al. designed ArgueTutor that judges \cqy{the argumentative} writing performance of users' \cqy{essays} and suggests how to improve \cite{arguetutor}.
As for the interactive strategies, these agents commonly adopt the quiz-like design by asking questions and giving feedback. 
For example, Ruan et al. created QuizBot, an interactive agent that asks questions and provides corrective feedback to users' answers in learning factual knowledge about science, safety, and English vocabulary \cite{quizbot10.1145/3290605.3300587}. 
They showed that QuizBot engaged users better in the learning process than the traditional flashcard tool, and users preferred the bot strongly for casual learning \cite{quizbot10.1145/3290605.3300587}. 
Winkler et al. developed Sara which acts like a teacher to ask students questions during an online video lecture about programming \cite{sara10.1145/3313831.3376781}. 
They demonstrated in a lab experiment that Sara could significantly improve learning gains compared to the without-Sara condition \cite{sara10.1145/3313831.3376781}. 
Peng et al. proposed a CReBot that prompts critical thinking questions and showed its engagement and usefulness for routine paper readers in their critical paper reading process when compared to a static question list \cite{peng2022crebot}. 
\zh{
To power these agents, previous work normally collects quiz questions, prepares and labels answers, and then builds computational models (e.g., classifiers) to provide adaptive feedback to learners \cite{arguetutor,quizbot10.1145/3290605.3300587,sara10.1145/3313831.3376781}. 
They would require additional human effort to extend the quiz pool when incorporating new learning materials, e.g., factual knowledge about medicine (QuizBot \cite{quizbot10.1145/3290605.3300587}), argumentative writings about scientific papers (Arguetutor \cite{arguetutor}), or another online video lecture (Sara \cite{sara10.1145/3313831.3376781}). 
}

In this paper, we explore the possibility of a conversational agent helping novices learn visual design. 
Similar to previous educational agents \cite{quizbot10.1145/3290605.3300587,sara10.1145/3313831.3376781,peng2022crebot}, our proposed DesignQuizzer uses questions to drive the learning process because they are generally effective in encouraging thinking \cite{bloom_taxonomy,questioningAI,question_generation,lubold2018automated}. 
Unlike previous agents that could be limited to a small number of labeled learning materials, we seek to power DesignQuizzer with computer-generated meaningful materials from online communities. 
In all, to the best of our knowledge, our work is the first to probe the design, effectiveness, and user experience of a community-powered conversational agent for learning visual design. 

\section{A Computational Workflow for Structuring Comments in Design Communities}
\label{section:3}

\begin{figure}
  \centering
  \includegraphics[width=\linewidth]{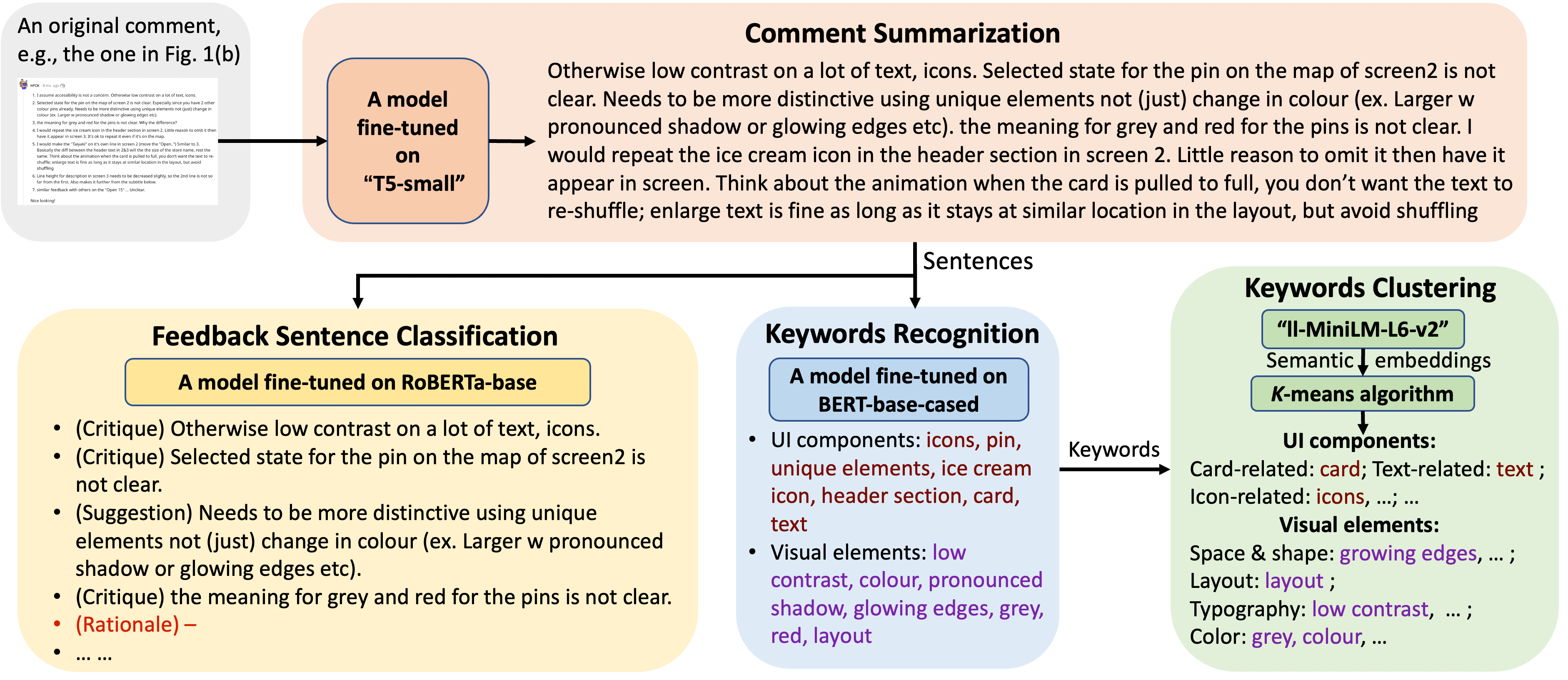}
  \caption{Computational workflow that structures the information of \textcolor[HTML]{000000}{critique}, \textcolor[HTML]{2F3756}{suggestion}, \textcolor[HTML]{CC0000}{rationale}, \textcolor[HTML]{660000}{UI component} and \textcolor[HTML]{660099}{visual element} keywords, and the clusters of these keywords from the comments under feedback-request posts in online design communities. 
  }
  \label{fig:pipeline}
\end{figure}

\pzh{
In this section, we present our computational workload (\autoref{fig:pipeline}) to extract useful content from comments as ``design quizzes'' to facilitate novices in visual design learning. 
We outline four desirable properties of such quizzers. 
First, they should include meaningful feedback (specific \cite{critique_style_guide,yuan2016almost} - \zh{critique}, actionable \cite{dow2012shepherding,kulkarni2015peerstudio} - \zh{suggestion}, and justified \cite{InteractiveGuidance} - \zh{rationale}) on the designs.  
This yields the need for a text summarization model. 
Second, the quizzes should contain classified feedback sentences that help learners understand how well the design example performs (\zh{critique}), how to improve it (\zh{suggestion}), and why giving this critique or suggestion (\zh{rationale}) \cite{bloom_taxonomy}. 
Third, they should be customized based on learners' interests, e.g., on visual elements or UI components, which requires a token classification method to identify the related keywords. 
Fourth, they should support the learning session with a learning focus, e.g., on the color-related visual elements. 
Therefore, we need to cluster the identified keywords about visual elements into higher-level concept groups. 
In all, our computational approach can extract the following terms from comments of UI feedback-request posts: 
}

\begin{itemize}
    \item \textbf{UI component}: keywords that describe the building blocks for creating the UI examples, e.g., bar, backdrop, button, card, menus, sliders, login page, the title of the post, etc \cite{materialdesign}. 
    \item \textbf{Visual element}: keywords that are the basic units of any visual design which form its structure and convey visual messages, e.g., color, space, shape, font, and size, as well as keywords that describe such units, e.g., aligned, alignment, white, black, hierarchy, readability \cite{visual_elements}. 
    \item \textbf{Critique}: \pzh{sentences that tell the advantages or drawbacks of  specific UI components or visual  elements in the design examples \cite{critique_style_guide,yuan2016almost}. For example, ``the locking system seems great and very protective of the space'' is a critique of our interests, while ``good work'' is not.}
    \item \textbf{Suggestion}: sentences that provide actionable recommendations about specific UI or visual elements \cite{dow2012shepherding,kulkarni2015peerstudio}, e.g., ``You may consider having the shape resemble that of a cup''.
    \item \textbf{Rationale}: \pzh{sentences that justify the critique or suggestion \cite{InteractiveGuidance}, e.g., ``I'd think the title of the post is the most important part of it''.}
\end{itemize}


\subsection{Data Collection and labeling}
\label{data_collection}
\subsubsection{Data collection.}
\label{collect_data}
\pzh{To support the supervised learning tasks in the workflow,} we used the PushShift api \cite{pushshift} to collect all posts with a ``Feedback Request'' flair and associated comments created between 2019.5.30 and 2022.5.30 in Reddit r/UI\_Design (\autoref{fig:reddit}).
\peng{
We chose the r/UI\_Design for three reasons. 
First, it is a typical online design community in Reddit, offering a free place for designers,
especially those with little chance to receive feedback from private feedback exchange groups or professional critique services~\cite{crain2017share}, to discuss artworks and exchange critiques~\cite{cheng2020critique,crain2017share}. 
Second, the shared artworks and exchanged critiques in r/UI\_Design focus on the design of graphic user interfaces for websites and mobile devices, which matches our focus on facilitating the learning of visual design. 
Third, founded in 2012, r/UI\_Design has over 119K members until May 2023, ranking top 1\% by size in Reddit. 
In other words, r/UI\_Design is a representative online design community with valuable resources for learning visual design and developing our computational workflow. 
}

To mitigate the data labeling workload brought by the low-quality comments, we took one more step to increase the likelihood that meaningful feedback exists in our intended labeled comment dataset. 
Specifically, we used the words ``thank'', ``great'', ``good'', `agree'', ``advice'', and ``suggest'' (adapted from \cite{zhenhui2021chi}) to filter the original posters' replies to the comments, as previous work suggests that higher-quality comments tend to evoke the original posters to reply and express satisfaction \cite{zhenhui2021chi}. 
This step resulted in $2250$ comment-reply pairs. 
\peng{To validate if the filtered replies express posters' satisfaction and if the comments contain meaningful feedback, 
three authors of this paper independently assigned binary labels to 100 comment-reply pairs and \cqy{resolved} the disagreement via majority voting. 
The result indicates that 96 out of the 100 replies do express gratefulness (ICC = 0.851) and 72 comments contain meaningful feedback (ICC = 0.751). 
}
Thus, we use these $2250$ comments under the feedback-seeking posts as the source for the following fine-grained data labeling step and qualitative analyses of the trained computational models. 


\subsubsection{Data labeling}
We randomly sampled 200 of the 2250 comments for the data labeling task. 
We recruited \zh{two annotators (females, ages: 20, 20)} from a local university to work on a sequence labeling project in doccanno \cite{annotation_tool}. 
For each comment, the annotators needed to 1) assign ``critique'', ``suggestion'', or ``rationale'' to the appropriate sentence and 2) tag ``UI component'' or ``visual element'' to the related keywords \footnote{The sentence can be the parts separated by ``,'', ``and'', ``but'', etc. The keywords can consist of one or multiple successive words.}. 
\zh{We invited a master student (male, age: 24) who majors in industrial design engineering and has completed over ten UI design projects to train our annotators for this labeling task. 
The training sessions started by asking the annotators to familiarize themselves with the comments in r/UI\_Design for one hour and learn visual elements and UI components in Material Design \cite{materialdesign} for three days. 
Next, the master student showcased how he labeled 10 comments, following which our two annotators independently labeled another 20 comments. 
They then met with the master student to discuss and refine the code book, after which they applied it to another 30 comments. 
}
Next, they discussed with the master student again and slightly revised the code book. 
Finally, they applied the code book for each label (introduced above) to all 200 comments. 
\zh{To evaluate the level of agreement between two annotators, we adopted the ROUGE criteria for the labeled sentences \cite{lin2004rouge} \footnote{A common NLP practice to measure the text difference. A higher ROUGE score \cqy{ranging} from 0 to 1 indicates that the texts are more similar.}. 
Specifically, we concatenated the labeled critique sentences separately for each comment to form a paragraph and calculated the ROUGE scores between the paragraphs from two annotators. 
We did the same for the labeled suggestion and rationale sentences. 
The results (\autoref{table:labeling}) showed that the ROUGE scores reached around 0.8 in all common metrics, indicating a good level of agreement. 
As for the labeled keywords, there are 80 (8.8\%) and 59 (10.6\%) times of \cqy{disagreement} on the annotated UI components and visual elements. 
}
\zh{The disagreements were resolved by discussions among the two annotators, the master student, and the first author who coordinated and participated in the training process.} 
In the end, we have 307 critique sentences, 366 suggestions, 155 rationales, 873 keywords about UI components, and 583 visual elements from 173 comments, while the rest 27 do not have any aforementioned labels. 

\begin{table}
\centering
\caption{The average ROUGE scores of labeled comment sentences (in percentage).} 
\label{table:labeling}
\scalebox{0.9}{
\begin{tabular}{llll}
\hline
Sentence Type & ROUGE-1 & ROUGE-2 & ROUGE-L  
\\ \hline
Critique & 80.42\% & 79.84\% & 80.38\%
\\
Suggestion & 78.22\% & 77.47\% & 78.22\%  
\\
Rationale & 90.24\% & 90.06\% & 90.24\%  
\\ \hline
\end{tabular}
}
\end{table}

\subsection{Comment Summarization}

To extract meaningful feedback from the comments, we developed a text summarization model that takes a comment as input and \cqy{outputs} sentences about critiques, suggestions, and rationales. 
\zh{
Following \cite{planhelper}, we first took an abstractive summarization approach that could better capture the overall meanings of the comment and then fuzzy-matched the output sentences to those in the original comment to help users locate the \peng{information they need}.
}
For each labeled comment, the sentences with labels ``critique'', ``suggestion'' and ``rationale'' are concatenated as the ground truth of the output. 
For the comments (N = 23) without meaningful feedback, the ground truth is an empty string ``''. 
We followed the tutorial from HuggingFace on how to fine-tune transformers for downstream summarization task \cite{huggingface_summarization}, with high regularization and length penalty to ensure that the output sentences used a similar vocabulary to the original text. 
We experimented with the pre-trained ``sshleifer/distilbart-cnn-12-6'' model \cite{wolf2020transformers} and ``T5-small'' model \cite{2020t5} using a 7-3 training-testing split method and found that the later one achieves better performance \peng{on} the test set (e.g., 46.23\% vs. 59.90\% regarding ROUGE-1 F1 score \cite{lin2004rouge}). 
Given the small size of our labeled dataset, we used \cqy{all 200 comments} to fine-tune our final ``T5-small''  summarization model.
These summarized texts achieved a high ROUGE-1 / ROUGE-2 / ROUGE-L / ROUGE-L-sum \cite{lin2004rouge} similarity score of 80.98\% / 81.53\% / 80.98\% / 82.29\% compared to the ground truth. 
To provide a concrete context for the summarized texts, we fuzzy-matched each output sentence to the best correspondent sentence and concatenated them as the final output of the comment summarization model. 


\begin{table}
\centering
\caption{Examples of the unlabeled comment under feedback-seeking posts and corresponding summary from our model.} 
\label{table:summarization_examples}
\scalebox{0.88}{
\begin{tabular}{ll}
\hline
No. & Original comment and its \textbf{summary} (marked in \textbf{bold}) from our model                                                                                                                                                                                                                                                                                 \\ \hline
1   & \textbf{My one suggestion would be to divide the card number, expiry and cvc into visually}\\&\textbf{separate input boxes. Or even just add a dividing line. }                                                                                                                                                                                                                                                           \\ \hline
2   & Not much to complain about here, looking good. \textbf{I might suggest that the tab bar icons have a}\\&\textbf{lot of varying complexity, maybe they could be more unified or simplified.} Also really really \\&nit picky, but you have a pattern of capitalizing the first letter of a statement, except for “Apply \\&Filters”. \textbf{Maybe just “Apply” would work if “Apply filters” looked weird in a CTA.} Thanks for \\&sharing! \\ \hline

3   & This is nice, like it a lot actually. \textbf{Only thing I'd say is maybe the icons could be more unified }\\& \textbf{\&amp; there's possibly not enough difference between the search field and the tiles below.} \\& Other than that it's great. Good job mate!                                                                                                                                                     \\ \hline
4   & \textbf{I'd bump the line height of the paragraph a bit.}                                                                                                                                                                                                                                                                                                                                             \\ \hline
5   & They sorta feel odd.                                                                                                                                                                                                                                                                                                                                                                                  \\ \hline
6   & \textbf{Maybe take the highlight of the background down, make it darker. Thus your product will}\\& \textbf{be the main focus of the page.} At the moment it kinda blends with the background.                                                                                                                                                                                                                    \\ \hline
7   & I’m curious as to why the train trip isn’t listed at the top with the hotel and flight?                                                                                                                                                                                                                                                                                                               \\ \hline
8   & Really nice take on the stepper component.                                                                                                                                                                                                                                                                                                                                                            \\ \hline
\end{tabular}

}
\end{table}

\autoref{table:summarization_examples} shows the comment summarization results of eight randomly sampled unlabeled comments.
The summaries of comments No.1-4 contain meaningful feedback about critique, suggestion, and rationale on specific UI components (e.g., ``Apply Filters'' in No.2) or visual elements (e.g., line height in No.4). 
Our model did not capture the general feedback like ``That is nice, like it a lot actually''. 
However, it missed some sentences that could be meaningful (e.g., ``At the moment it kinda blends with the background'' in No.6). 
Therefore, we consider that our model's summary may not cover all the helpful information in the comment, but all the output sentences are meaningful feedback on the UI designs. 

\subsection{Feedback Sentence Classification} 

Our feedback sentence classifier inputs a sentence from the comment summary  \cqy{and outputs} a label ``critique'', ``suggestion'', or ``rationale''. 
Our dataset for this supervised multi-class classification task consists of 307 critique, 366 suggestion, and 155 rationale sentences. 
Using the 8-2 hold-out method, we had 661 sentences for training and the rest 167 for testing. 
Given the small sample size, we decided to exploit the strength of pre-trained language models to boost the classifier's performance. 
\peng{We experimented with various BERT-style models including BERT \cite{bert_devlin-etal-2019-bert}, DistilBERT \cite{sanh2019distilbert}, ALBERT \cite{lan2019albert}, and RoBERTa-base \cite{reberta} for this classification task.
We found that RoBERTa-base outperformed the others \peng{on} the test set regarding accuracy, precision, recall, and F1 score (\autoref{model performance}).}
RoBERTa-base is a robustly optimized model using BERT-style pretraining methods and achieved the start-of-the-art results on downstream sequence classification tasks (e.g., 94.8\% accuracy in the SST-2 sentiment analysis dataset) in 2019 \cite{reberta}. 
\peng{We finetuned this model in our dataset using a batch size of 32 and 20 epochs, with an early stopping mechanism that selects the model with the highest accuracy on the validation set during the training process. 
The warm-up steps were 500, and the learning rate was 5e-5. 
We exploited a weight decay coefficient of 0.01 for regularization and the \cqy{cross-entropy} loss function.}
After fine-tuning, our sentence classifier achieves a 0.83 accuracy, a 0.83 weighted F1, and a 0.81 macro F1 score \peng{on} the test set, indicating its appropriate validity. 
\autoref{table:sentence_classification} shows the classification results on eight randomly sampled sentences. 
Our master student, who is experienced in UI design, suggested that sentences No.1-4 and No.6 were accurately classified, No.5 should be a rationale, No.7 should be a suggestion, and No.8 had better be treated as a critique.  
Our feedback sentence classifier has a chance to make an incorrect or incomplete decision. 
Nevertheless, it is overall acceptable in our usage scenario with the goal of helping users locate and digest meaningful feedback. 
%


\begin{table}[]
\caption{\peng{The performance of experimented models for feedback sentence classification on our test set.}}
\label{model performance}
\scalebox{0.78}{
\begin{tabular}{llllllll}
\hline
Model      & Accuracy & macro Precision & weighted Precision & macro Recall & weighted Recall & macro F1 & weighted F1 \\ \hline
\textbf{RoBERTa}    & \textbf{82.54\%}  & \textbf{83.32\%}         & \textbf{83.23\%}            & \textbf{80.03\%}      & \textbf{82.93\%}         & \textbf{81.02\%}  & \textbf{82.60\%}    \\ \hline
BERT       & 73.87\%  & 73.14\%         & 74.85\%            & 70.49\%      & 75.61\%         & 70.72\%  & 74.43\%     \\ \hline
ALBERT     & 76.90\%  & 77.79\%         & 78.01\%            & 71.91\%      & 77.24\%         & 72.51\%  & 75.90\%     \\ \hline
DistilBERT & 71.11\%  & 72.36\%         & 73.17\%            & 65.81\%      & 72.36\%         & 64.98\%  & 69.75\%     \\ \hline
\end{tabular}
}
\end{table}

\begin{table}
\caption{Examples of the unlabeled sentences from the comment summary and corresponding classification results from our model.}
\label{table:sentence_classification}
\scalebox{0.88}{
\begin{tabular}{lll}
\hline
No. & Sentence                                                                                                                                                                    & Classification \\ \hline
1   & \begin{tabular}[c]{@{}l@{}}With traditional news apps, users can scroll through content and find things that are \\ interesting to them (or relevant to them).\end{tabular} & rationale      \\ \hline
2   & \begin{tabular}[c]{@{}l@{}}One thing would be maybe adding a + icon to the 'add' button and shortening \\the width a bit?.\end{tabular}                                     & suggestion      \\ \hline
3   & \begin{tabular}[c]{@{}l@{}}The illustration in the last mock seems a little out of place considering the other mocks. \end{tabular}                                       & critique       \\ \hline
4   & \begin{tabular}[c]{@{}l@{}}My one suggestion would be to divide the card number, expiry and cvc into \\visually separate input boxes.\end{tabular}                         & suggestion      \\ \hline
5   & This would give that big chunk of real-estate a purpose.                                                                                                                    & critique       \\ \hline
6   & The space can be used more effectively.                                                                                                                                     & rationale      \\ \hline
7   & Title of the left section should match the tab/step you're in ("Payment Details")                                                                                           & critique       \\ \hline
8   & \begin{tabular}[c]{@{}l@{}}I don't see any point in including 4K labels on the movie unless you're also an app\\ for booking virtual viewing.\end{tabular}                  & suggestion      \\ \hline
\end{tabular}

}
\end{table}

\subsection{Keywords Recognition and Clustering}
To \peng{help users explore comments} based on their interests, our computational workflow further processes the sentences in the comment summary and outputs 1) the recognized keywords about UI or visual elements and 2) associated clusters that can reflect a higher-level design concept like color and space \cite{best_practice}. 
\peng{After this step, each classified feedback sentence is tagged with the types of its UI components (\autoref{table:ui_element}) and visual elements mentioned (\autoref{table:visual_element}).}
\subsubsection{Keywords recognition}
We approximate it as a token classification task 
following the related tutorial in Huggingface \cite{huggingface_token_classification}  which details how to fine-tune the pre-trained ``bert-base-cased'' language model \cite{bert_devlin-etal-2019-bert} for name entity recognition. 
Specifically, we attributed labels ``B-ui'' / ``B-visual'' to the tokens at the beginning of relevant keywords, ``I-ui'' / ``I-visual'' to those inside the keywords, and ``O'' to the others. 
For this supervised learning task, the labeled dataset contains 828 sentences (i.e., those labeled critiques, suggestions, or rationales), which are tokenized and tagged with the BIO labels based on the 837 UI components and 583 visual element keywords. 
We used 680 sentences (80\% of the data) for training and the rest sentences (20\%) for testing. 
We evaluated the resulting model by considering both classified ``B-ui'' and ``I-ui'' tokens as UI components and both ``B-visual'' and ``I-visual'' tokens as visual elements, since these tokens, regardless of their positions in the keywords, can help users locate the content of their interests in the comments. 
It achieved a 0.88 accuracy, a 0.71 F1, a 0.80 precision, and a 0.64 recall score \peng{on} the test set, suggesting an acceptable performance for a classification problem. 
\autoref{table:example_recognition} shows the recognition results of eight randomly selected sentences. 
Our master student, who is experienced in UI design, suggested that the UI components and visual elements in sentences No.1-5 and No.8 were correctly detected. 
However, the model made mistakes of classifying ``the'' as a UI component (which could be avoided with a rule to ignore words like ``the'' and ``a'') in sentence No.6 and forgetting the UI component ``labels'' in sentence No.7. 
In general, our element keywords recognition model has a proper performance that can power our DesignQuizzer to personalize the learning materials. 

\begin{table}
\caption{\zhenhui{Examples of the recognized UI components and visual elements from sentences.}} 
\label{table:example_recognition}
\scalebox{0.88}{
\begin{tabular}{llll}
\hline
No. & Sentence                                                                                                                                                                                                                        & Output UI component                                                                           & Output visual element                                                    \\ \hline
1   & \begin{tabular}[c]{@{}l@{}}At first glance, i think you should make the cal\\to action buy  button a brighter color to catch\\peoples eyes.\end{tabular}                                                                       & call to action, buy button                                                                    & brighter color                                                           \\ \hline
2   & \begin{tabular}[c]{@{}l@{}}It doesn't have to be real website to be properly \\designed...\end{tabular}                                                                                                                                                                   & -                                                                                          & -                                                                     \\ \hline
3   & The heading again needs more contrast                                                                                                                                                                                           & heading                                                                                       & contrast                                                                 \\ \hline
4   & \begin{tabular}[c]{@{}l@{}}Gradient/contrast on the progress track (top of\\screen)looks  weird, this should be more clear, lines\\are too thickContent columns/containers/boundaries\\ unclear and unbalanced.\end{tabular} & \begin{tabular}[c]{@{}l@{}}the progress track,\\ lines\end{tabular}                           & \begin{tabular}[c]{@{}l@{}}Gradient, contrast,\\ boundaries\end{tabular} \\ \hline
5   & \begin{tabular}[c]{@{}l@{}}Things I'd change are, font weight in the CTA's\\plus the Add button in the investment block make\\it aligned with the rest of the text fields.\end{tabular}                                       & \begin{tabular}[c]{@{}l@{}}CTA's,  Add button,\\ investment block,\\ text fields\end{tabular} & -                                                                     \\ \hline
6   & \begin{tabular}[c]{@{}l@{}}It looks like the left column is slightly more narrow\\than the right.\end{tabular}                                                                                                                 & \begin{tabular}[c]{@{}l@{}}the, left column, right\end{tabular}                             & -                                                                     \\ \hline
7   & \begin{tabular}[c]{@{}l@{}}I remember seeing spelled-out "close" buttons on\\the left, but not as an icon.\end{tabular}                                                                                                     & "close" buttons, icon                                                                         & -                                                                     \\ \hline
8   & I like the overall structure and layout.                                                                                                                                                                                        & -                                                                                          & structure, layout                                                        \\ \hline
\end{tabular}
}
\end{table}

\begin{table}
\caption{UI component clusters with representative keywords.}
\label{table:ui_element}
\scalebox{0.88}{
\begin{tabular}{lll}
\hline
No & Clusters        & Keywords                                                                                                                                            \\ \hline
1  & button-related  & \begin{tabular}[c]{@{}l@{}}button, sign up button, login button, button background, close buttons, buy button,\\ white botton\end{tabular}           \\ \hline
2  & text-related    & lines, right-hand content, text, the second line of text, fonts, body text, content                                                                 \\ \hline
3  & card-related    & badges, "Your Cart", card number, item card, price, shopping cart, product card                                                                      \\ \hline
4  & divider-related & labels, slider, dividers, the back arrow, the left of the image, dividing line, filter                                                               \\ \hline
5  & color-related   & \begin{tabular}[c]{@{}l@{}}colors, color palette, "add colour", "create colour", the color settings,\\ background elements, dark theme\end{tabular} \\ \hline
6  & menus-related   & \begin{tabular}[c]{@{}l@{}}title of the left section, menu options, clickable links, this page, drop-down menu,\\ tab bar, header\end{tabular}      \\ \hline
7  & icon-related    & \begin{tabular}[c]{@{}l@{}}"trash bin icon", "lock icon", the ice cream icon, tab bar icons, iconography,\\ profile icon, logos\end{tabular}        \\ \hline
8  & general element & elements, container element, components, UI component, Control elements                                                                             \\ \hline
9  & others          & the title of the post, username, planes, stats, ads, the progress track, delivery                                                                       \\ \hline
\end{tabular}
}
\end{table}

\begin{table}
\caption{Visual element clusters with representative keywords.}
\label{table:visual_element}
\scalebox{0.88}{
\begin{tabular}{lll}
\hline
No. & Cluster & Example keywords                                                                                                                                                       \\ \hline
1   & space, shape       & \begin{tabular}[c]{@{}l@{}}padding, space, whitespace, shape, rounded edges, align, margins, consistency,\\line height,  width, floating\end{tabular}         \\ \hline
2   & layout             & \begin{tabular}[c]{@{}l@{}}layout, responsive layout, second screen, accessibility, information architecture,\\user flows, readability\end{tabular} \\ \hline
3   & typography         & \begin{tabular}[c]{@{}l@{}}typography, fonts, the visual hierarchy of text, contrast, styling, the font size, visible,\\saturation, sans-serif\end{tabular}   \\ \hline
4   & color              & \begin{tabular}[c]{@{}l@{}}color, calm colors, lighter, black, color usage, medium gray, red, background color,\\yellow, dark, pink\end{tabular}             \\ \hline
\end{tabular}
}
\end{table}
	
\subsubsection{Element keywords Clustering}
We further attach the detected keywords with tags that can reflect a higher-level design concept, such as ``color'' and ``space'', to support learning with a specific focus. 
We first adopted the pre-trained language model with the most downloads in the Huggingface's Sentence Transformer tutorial, ``all-MiniLM-L6-v2'', to map the keywords into a 384-dimensional dense vector space that can be used for tasks like clustering  \cite{huggingface_sentence_transformers}. 
Then, we applied the well-developed $K$-means clustering algorithm to the semantic embeddings of UI components and visual elements, respectively. 
We manually went through the output results with the $K \in [3, 10]$ and found that most of the clusters make sense with $K=4$ for grouping visual elements and with $K=9$ for grouping UI components. 
Again, we invited our master student with UI design experience to select one or two representative words that reflect the main higher-level design concepts of each cluster. 
\autoref{table:ui_element} and \autoref{table:visual_element} present the resulting clusters.

\section{DesignQuizzer System}
\label{section:designquizzer}
\begin{figure}
  \centering
  \includegraphics[width=0.82\linewidth]{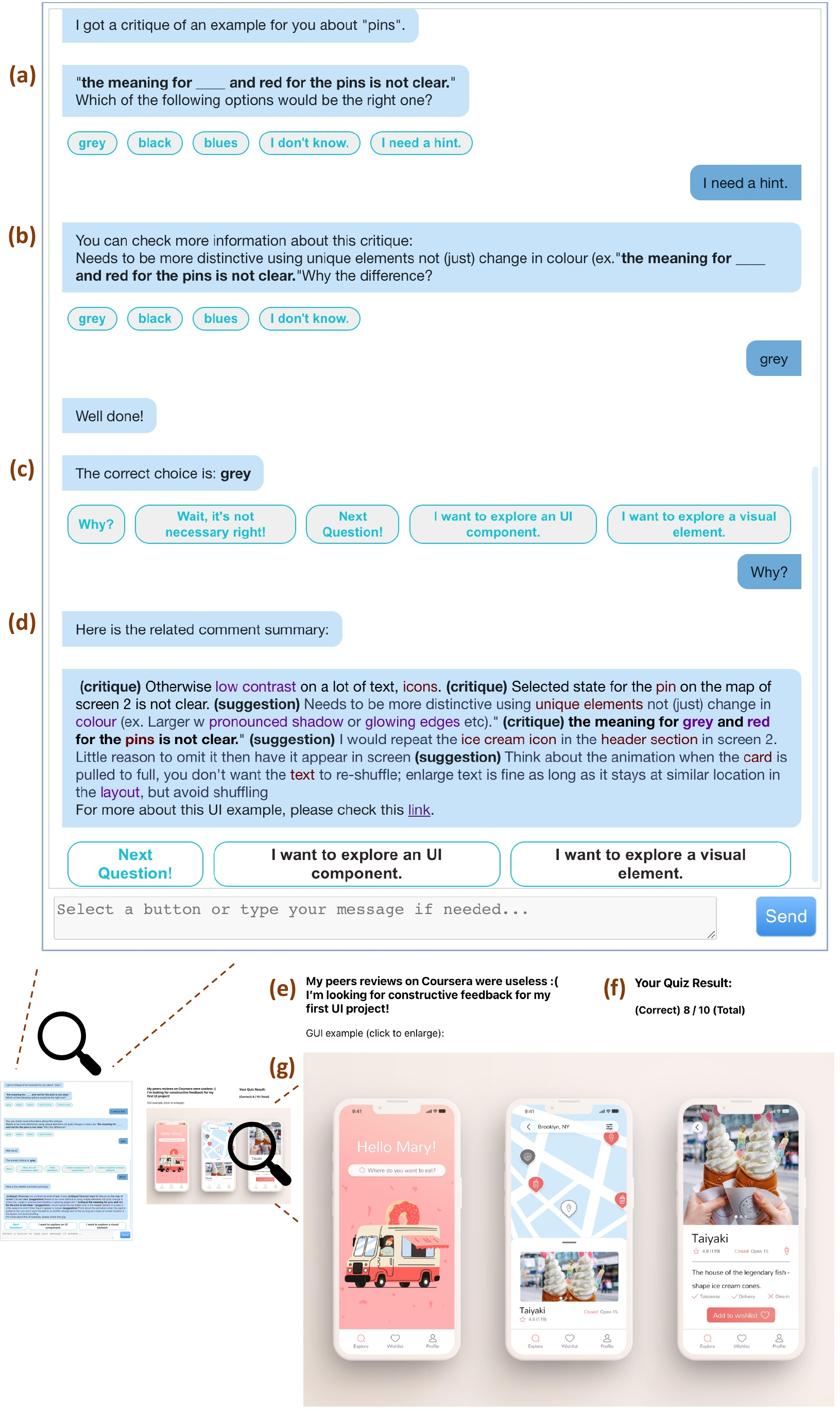}
  \caption{Interface of DesignQuizzer with a sample UI design and a conversation log. (a) Question. (b) Hint. (c) Answer. (d) Structured comment summary. (e) Post title. (f) Performance track. (g) UI example.}
  \label{fig:design_quizzer}
\end{figure}

To facilitate users in visual design learning activities, we develop DesignQuizzer which prompts questions on the UI designs in online communities and presents structured feedback for explanations. 
\zh{
We choose a quiz-based interaction design as previous learning support tools (e.g., QuizBot \cite{quizbot10.1145/3290605.3300587}, Sara \cite{sara10.1145/3313831.3376781}, CReBot \cite{peng2022crebot}) have demonstrated its engagement and effectiveness. 
}
As our objective is to allow learning anywhere at any time, we build DesignQuizzer as a responsive web-based application (\autoref{fig:design_quizzer}).
Further, to support learning at scale based on users' interests, DesignQuizzer should computationally generate a quiz pool that covers different types of visual elements. 
Below, we demonstrate the DesignQuizzer's design and evaluation with the processed data from Reddit r/UI\_Design and the learners' tasks as getting familiar with ``space'', ``shape'' (cluster No.1 in \autoref{table:visual_element}), ``typography'' (No.3),  ``color'' (No.4). 
We discuss the extended applications and limitations of the DesignQuizzer in \autoref{section:discussion}.




\begin{figure}
  \centering
  \includegraphics[width=\linewidth]{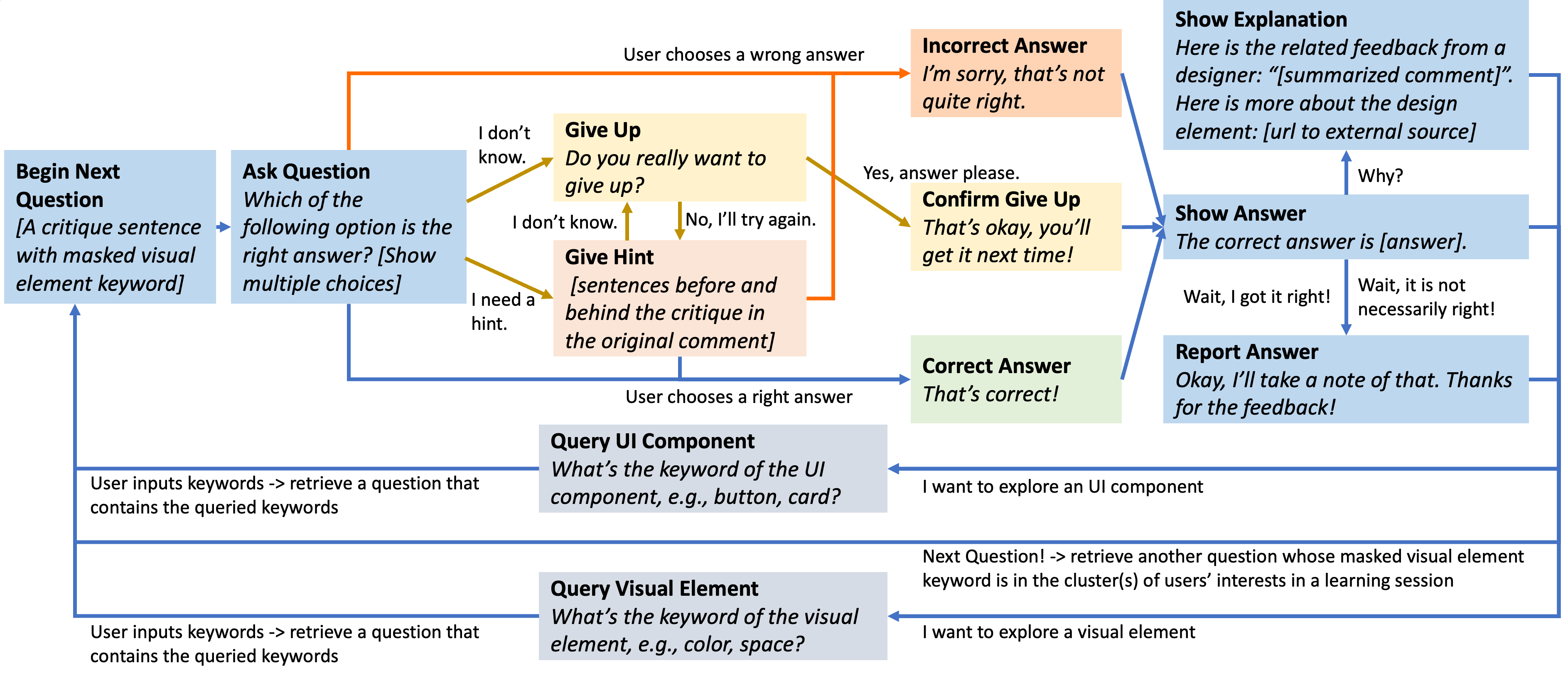}
  \caption{The dialogue flow of DesignQuizzer including typical sample responses.}
  \label{fig:dialog_flow}
\end{figure}

\subsection{DesignQuizzer Interface and Interaction Design}
As shown in \autoref{fig:design_quizzer}, 
the left-hand side is a chat window for users to interact with DesignQuizzer. 
To interact with the DesignQuizzer, users can click one of the buttons below its message or type down keywords if they want to explore a visual element or a UI component. 
The reasoning behind this mixed modality is to ensure both flexibility and efficiency regarding user interactions with the DesignQuizzer. 
Besides, these buttons help to maintain the conversational flow  \cite{chatbot_social_intelligence}. 
The right-hand side of the interface presents the title (part \textit{e}) and the attached UI example (\textit{g}) of the original feedback-request post to provide the main learning context. 
Users can click the UI example to view it in full screen and enlarge it to check its details. 
Part \textit{f} offers a simple way for users to keep track of their quiz progress and performance \cite{quizbot10.1145/3290605.3300587,arguetutor}. 

Following \cite{quizbot10.1145/3290605.3300587}, \autoref{fig:dialog_flow} illustrates one round of the interaction between a user and DesignQuizzer. 
At the beginning of each round, DesignQuizzer asks a user a single-choice question (part \textit{a} in \autoref{fig:design_quizzer}) on the UI example selected according to our quiz strategy (\autoref{section:quiz_logic}). 
The question masks the keywords about visual elements in a sentence that critiques the UI example (e.g., ``the meaning for \_\_\_\_ and red for the pins is not clear''), and the options consist of the masked keywords (e.g., ``grey'') and another two words (e.g., ``black'' and ``blues'') as distractors. 
As our focus is to help users learn about visual elements, the critique with such a masked keyword used as a quiz prompt could direct their attention to a specific visual aspect of the UI example, and the options would encourage them to think of and differentiate related visual elements.
We choose single-choice questions instead of open-ended questions to allow an easy way for user's input and ensure the system's robustness. 
A user can click one of the three options, click the ``I don't know'' button and confirm whether ``give up'' or not, or click the ``I need a hint'' button. 
If users ask for a hint, DesignQuizzer will give more contexts about the critique sentence by presenting the sentences before and behind it in the original comment (part \textit{b}). 
The DesignQuizzer will provide positive reinforcement feedback that is typical of a study partner if a user chooses the correct answer, and it will give encouragement if a user chooses a \peng{incorrect} option or gives up. 
After the user sees the correct answer (part \textit{c}), they can click the ``Why'' button to check the structured summary of the comment that may provide suggestions and rationales for the critiques (\textit{d}). 
To facilitate the visual search for users' interested content, we encode the \textcolor[HTML]{000000}{critique}, \textcolor[HTML]{2F3756}{suggestion}, \textcolor[HTML]{CC0000}{rationale}, \textcolor[HTML]{660000}{UI component}, and \textcolor[HTML]{660099}{visual element} in the comment summary with different colors. 
At the end of each quiz, the user can click the ``I want to explore a UI component'' or ``I want to explore a visual element'' button to type down and send related keywords. 
The DesignQuizzer will then retrieve a question about the queried keywords and the associated UI example as the next quiz. 
To simulate the experience of human-human conversations, we provide a variety of different responses of DesignQuizzer at each state in the dialogue flow. 

\subsection{Quiz Pool}
\label{section:quiz_pool}
Our quiz pool contains single-choice questions about visual elements in UI design examples (\autoref{table:visual_element}). 
Every question has five parts: a feedback-request post with a title and a UI example, a critique that has a keyword of visual elements masked, options that contain the masked keyword and two distractors, a hint that includes the sentences around the critique, and a structured comment summary for providing explanations. 
Different from previous pedagogical conversational agents (e.g., \cite{quizbot10.1145/3290605.3300587,sara10.1145/3313831.3376781}) that need manual collections of the questions and their answers, we prepare the quiz pool by applying our trained computational models to the comments in online design communities. 
The procedure of the quiz preparation is detailed below: 

\textbf{1)} \textbf{Get feedback-request posts with downloadable UI examples}.  We first used the PushShift api to collect all the 1521 posts with the ``Feedback Request'' flair and with at least one comment (excluding the one from the bot AutoModerator and those from original posters) within 2019.5.30 and 2022.5.30 in Reddit r/UI\_Design. 
Then we kept those posts attached with downloadable UI design images by checking if the post body contains an image link and if the image is publicly available. 
This step results in 502 posts, and each is associated with a UI design example. 
To reduce the human workload for the semi-automatic question and option preparation in step 3 below, we randomly select half of them, i.e., 251 posts, to continue with. 

\textbf{2)} \textbf{Process the comments with our computational models.} 
Next, we applied our computational workflow (\autoref{fig:pipeline}) to all the comments of the sampled 251 posts. 
Note that the input to the feedback sentence classifier is a full sentence separated by ``.'', ``?'' or ``!'' in this case rather than the sub-sentences (e.g., can be part of a full sentence) that we labeled and were used to train the classifier. 
This is because we want to provide learners with a concrete context via full sentences in the questions and explanations. 
In total, there are 4437 comments on these 251 posts, which include 867 critiques, 2103 suggestions, 189 rationales sentences, 8171 occurrences of UI component keywords, and 5115 occurrences of visual element keywords. 
The unbalanced numbers between the suggestion and rationale sentences could be due to the fact that some full sentences may contain both suggestion and rationale sub-sentences, while our model can not give multiple labels for one input sentence and favor labeling them as suggestions.

\textbf{3)} \textbf{Prepare questions and options.} We generated a multiple-choice question for each of the 867 critiques by masking the first keyword about visual elements if it exists, resulting in 509 questions. We adopted a semi-automatic method to prepare the distractors for these 509 questions. 
First, from the 5115 visual element keywords, we randomly selected two different words that fall into the same cluster (\autoref{table:visual_element}) and have the same POS tag (assigned by the ntlk package) with the right answer for each question. 
Then, the first author manually went through the 509 questions and their options to validate the quality and make revisions. 
He identified 28 generated questions that do not make sense, e.g., the sentence is not a critique or the masked word is not about visual elements. 
Among the 962 distractors of the rest 481 questions, he revised 373 items that could be easily excluded by checking the \peng{grammar of the question.} 

\textbf{4)} \textbf{Prepare hints and explanations.} To prepare hints and explanations for users, we further processed the 481 questions and kept 337 of them that had at least one suggestion or one rationale sentence in the sourced comment of the question. 
\textbf{After the full procedure, we have 337 questions from the critiques of 152 posts. }
Among them, the masked keywords of 133 questions are about visual elements ``space'' and ``shape'' (cluster No.1 in \autoref{table:visual_element}), 23 are about ``layout'', 92 are about ``typography'', and 89 items are about ``color''. 
The numbers of unique keywords under each visual element cluster are 84, 13, 47, and 39, respectively. 

\subsection{Quiz Strategy and System Implementation}
\label{section:quiz_logic}
The DesignQuizzer's strategy for selecting the next quiz depends on three conditions. 
First, \peng{
if users click the ``Next Question'' button, it will randomly retrieve another question that aligns with users' interests in a learning session. 
To be more specific, this question contains (masked) visual element keywords related to the target concept to learn but can be about any type (\autoref{table:ui_element}) of UI component in our dataset.}
For example, if a user wants to learn visual elements about ``space'' and ``shape'' in the learning session (one task in the later user study), DesignQuizzer's next quiz will come from the 133 questions whose masked keywords are in cluster No.1 (\autoref{table:visual_element}). 
If a user wants to get familiar with ``typography'' and ``color'' (the other task), DesignQuizzer will retrieve a quiz from the 92 questions with answers in cluster No.3 or the 89 questions with answers in cluster No.4. 
Second, if users click the ``I want to explore a visual element'' button, DesignQuizzer will have a 50\% chance to present a question with the queried keywords as the right answer and the other 50\% chance to get a question whose answer is in the same visual element cluster of the queried word. 
This design choice could satisfy users' interests while not making the quiz too easy for them. 
Third, if users click the ``I want to explore a UI component'' button, DesignQuizzer will retrieve a question that contains the queried keywords. 
\peng{For example, if a user queries ``icons'', the Quizzer will randomly prompt a question that originates from a critique sentence containing icon-related UI component keywords (\autoref{table:ui_element}). 
In our quiz pool for user study, there are 30, 0, 19, and 16 icon-related questions that mention visual elements about space/shape, layout, typography, and color, respectively (\autoref{table:visual_element}). }
If there are no satisfactory questions in the second and third conditions, DesignQuizzer will apologize for not having related quizzes and switch to the first condition. 
To mitigate this issue, DesignQuizzer will randomly suggest two candidate keywords for references. 

We implement DesignQuizzer's frontend based on React.js and the react-viewer and chat-ui-kit-react components \footnote{Links of the main components: \url{https://github.com/infeng/react-viewer}, and \url{https://github.com/chatscope/chat-ui-kit-react}}. 
Its backend server is based on a python flask framework to maintain the dialogue flow. 
The UI design images are locally stored using python http.server package. 
We use the ElasticSearch engine to store the quiz pool and retrieve quizzes. 



\peng{\section{Experiment \uppercase\expandafter{\romannumeral1}}}
\peng{
We conducted two experiments with novice designers to study DesignQuizzer's benefits and drawbacks when compared with the community-like baseline interface for learning visual design.
The experiment \uppercase\expandafter{\romannumeral1} with 24 participants adopts a within-subjects study design to reduce the possible effect of individual differences in the learning process and \cqy{user perceptions of the tools. }
We counterbalance the order of used interfaces (DesignQuizzer vs. baseline) to mitigate the learning effects and confirm that the order does not significantly impact the reported results, as detailed in \autoref{subsubsection:order_effect}. 
Based on the insights from experiment \uppercase\expandafter{\romannumeral1}, experiment \uppercase\expandafter{\romannumeral2} with 28 participants further evaluates what users can learn with DesignQuizzer and how well they can apply the learned knowledge in a subsequent design activity. 
This experiment adopts a between-subjects design as the knowledge transition from one learning task to the other could largely affect the measured outcome of knowledge application. 
In this section, we present the design and results of experiment \uppercase\expandafter{\romannumeral1}. 
}
Our research questions are:


\textbf{RQ1}: How would DesignQuizzer affect the amount and the novices' perceptions of their explored design examples and comments in the learning session? 

\textbf{RQ2}: How would DesignQuizzer affect novices' engagement and cognitive load in the learning session? 

\peng{\textbf{RQ3}: How would DesignQuizzer affect the novices' outcome on recognizing enacted or violated principles about the learned visual elements in others' design examples?}


\textbf{RQ4}: How would novices interact and perceive with DesignQuizzer in their learning sessions? 

\pzh{
\subsection{Experimental Design}
Our experiment is a within-subject design. 
Each participant has one learning task with DesignQuizzer and the other with a baseline tool. 
}
\subsubsection{Baseline: Reddit r/UI\_Design Interface}
\label{baseline_interface}
To simulate how users normally explore examples and comments in online design communities, we use the Reddit r/UI\_Design webpage as the baseline (\autoref{fig:reddit}). 
\peng{We use the Reddit r/UI\_Design webpage as the baseline for two reasons. 
First, exploring examples and comments in online design communities is a common learning practice \cite{online_feedback_exchange,critique_me,example_for_learning,learning_and_transfer,case_comparison}, and as stated in \autoref{collect_data}, r/UI\_Design is a representative one for learning visual design. 
Second, since the current quiz pool of DesignQuizzer is sourced from the r/UI\_Design community, using it as the baseline can ensure the coherence of learning materials. 
}
Participants can use its embedded functions to search the posts and comments inside the communities based on keywords and sort them based on ``Hot'', ``New'', ``Top'', and ``Rising''.
We use the ``Filter by Flair'' feature to list all ``Feedback Request'' posts on the webpage in advance to avoid distraction from other types of posts, e.g., ``Portfolio Review Requests''. 
Participants in the DesignQuizzer condition can also access this community webpage, e.g., via a link for more \cqy{explanations} (\autoref{fig:design_quizzer} part \textit{d}). 

\subsubsection{Learning tasks}
\label{task}
\peng{Each \cqy{participant} has two learning tasks}. 
The task prompt is: 
``You will learn design knowledge about (\peng{task A}) `space' and `shape' / (\peng{task B}) `color' and `typography' in this session. You will explore the online UI examples and comments to learn how these concepts are enacted or violated in practice. After the learning session, you need to critique a new UI example from the aspects of (\peng{task A}) `space' and `shape' / (\peng{task B}) `color' and `typography', give suggestions for improvement, and provide corresponding rationales if any''. 
Our quiz pool contains 84 unique keywords about ``space'' and ``shape'' and 86 unique keywords about ``color'' and ``typography'', which helps to balance the difficulty of the two tasks. 
DesignQuizzer will adjust the quiz strategy for ``Next Question'' (detailed in \autoref{section:quiz_logic}) to support users' learning goal in their task, while in the baseline condition users are presented with all feedback-request posts and comments in the Reddit interface. 
\peng{After being counterbalanced with Latin Square, there are four task-interface combinations, each with six participants: (a) task A (DesignQuizzer) - task B (Baseline), (b) task B (DesignQuizzer) - task A (Baseline), (c) task A (Baseline) - task B (DesignQuizzer) and (d) task B (Baseline) - task A (DesignQuizzer).
}

\subsection{Participants}
We recruited 24 students (10 \textbf{F}emales,  11 \textbf{M}ales, 3 \textbf{N}ot \textbf{A}vailable; age range 18-21, $Mean = 19.25, SD = 0.74$; noted as P1-24) from a local university via a post in a group chat and word of mouth. 
The inclusion criteria are that participants are novices in visual design \zh{but have interests to learn more about it}. 
\zh{We do not require the participants to be design students, because our DesignTutor would support any students who are interested in learning visual design online.}
All of them are undergraduates and have passed the national College English Test for general requirements. 
Twenty major in Artificial Intelligence, two in Software Engineering, one in Micro-electronics, and one in Ocean Engineering. 
In general, our participants have little or no experience in learning UI visual design ($M=1.42, SD=0.88$) and exploring online design communities ($M=2.83, SD=1.90$; 1 - No experience at all, 7 - A lot of experience) but they are interested in learning it in our study ($M=5.42, SD=1.18$; 1 - No interest at all, 7 - A great deal of interest). 
\zh{The usage frequency of online communities where people create posts and comments is: 17 daily, 5 2-6 days a week, 1 once a week, and 1 less than once a week. }


\subsection{Measures}
\label{measures}


\textbf{RQ1. Explored design examples and comments.} 
We log the numbers of explored design examples and comments by analyzing the interaction log of DesignQuizzer and the screen video record of the Reddit condition, e.g., plus one to the numbers \zh{if users' web page stays in the unique example or comment for more than four seconds as indicated by the authors' trials.}
Besides, after each task, we measure participants' satisfaction and perceived helpfulness of the explored examples and comments using two 7-point Likert scale items (1 - strongly disagree; 7 - strongly agree) adapted from \cite{Zhenhui2019,zhenhui2020}. 

\textbf{RQ2. Engagement in the learning process.} 
We derive six 7-point Likert scale items (Cronbach's $\alpha = 0.793$) to measure user engagement based on Brien's theoretical model \cite{OBrien2016} regarding the flow theory for a positive experience \cite{csikszentmihalyi1990flow}. 
Specifically, they are concentration (``completely involved, focused, and concentrating''), a sense of ecstasy (``feel doing something is special''), doability (``skills are adequate, neither anxious nor bored''), sense of serenity (``forgot about myself doing something''), timeless feeling (``time passed quickly''), and intrinsic motivation (``feel self-rewarded'') \footnote{The full items are provided in the supplementary materials.}. 
As for the cognitive load, we use one 7-point item ``I think the cognitive load of exploring examples and comments to learn the targeted visual design elements with this tool is very high'' (1 - strongly disagree, 7 - strongly agree) adapted from \cite{nasatlx}. 

\peng{\textbf{RQ3. Outcome of the learning session on recognizing enacted or violated design principles.}}
\zh{
Because online design communities like Reddit do not make any systematic attempt to measure learning, a user study like ours must rely heavily on proxy measures of the learning outcomes. 
Our measure for the development of visual design skills is inspired by previous quantitative studies on computational learning in the Scratch community, which capture learning gain via the size of the learners’ repertoire in terms of the number of different types of computational thinking concepts a user has demonstrated in their projects \cite{ruijiacscw22,dasgupta2016remixing,dasgupta2018wide,scaffidi2012skill,yang2015uncovering}. 
Specifically, we construct an outcome measure on the numbers of enacted or violated principles about the visual elements (i.e., space and layout in one condition, color and typography in the other) mentioned in the participants' feedback comments to the unseen UI examples in each post-test. 
For instance, ``LOGO and navigation bar are located on the same line and aligned'' in the comment counts as one point as it recognizes the enacted layout alignment principles of the design example. 
Two annotators first separately labeled 12 randomly sampled comments and then discussed and reached a consensus on the rating scheme, e.g., the sentence should include a specific UI component or visual element. 
Next, they independently labeled all 48 comments in the post-test; ICC(intraclass correlation coefficient) $ = 0.856, p < 0.001$. 
They resolved the disagreement by discussions. 
}

\textbf{RQ4. Interaction and perception with DesignQuizzer.}
We log the number of participants' clicks on each button to understand how participants use DesignQuizzer. 
We adapt the technology acceptance model \cite{Technology_acceptance_model:doi:10.1111/j.1540-5915.2008.00192.x} that has been used for evaluating educational chatbots \cite{peng2022crebot,arguetutor} to measure the following in each condition: 
usefulness (four items, Cronbach's $\alpha = 0.897$); easy to use (four items, Cronbach's $\alpha = 0.737$); and intention to use (two items, Cronbach's $\alpha = 0.782$). 
We average the ratings of multiple questions as the final score for each factor in the acceptance model. 

\subsection{Procedure}
\begin{figure}
  \centering
\includegraphics[width=\linewidth]{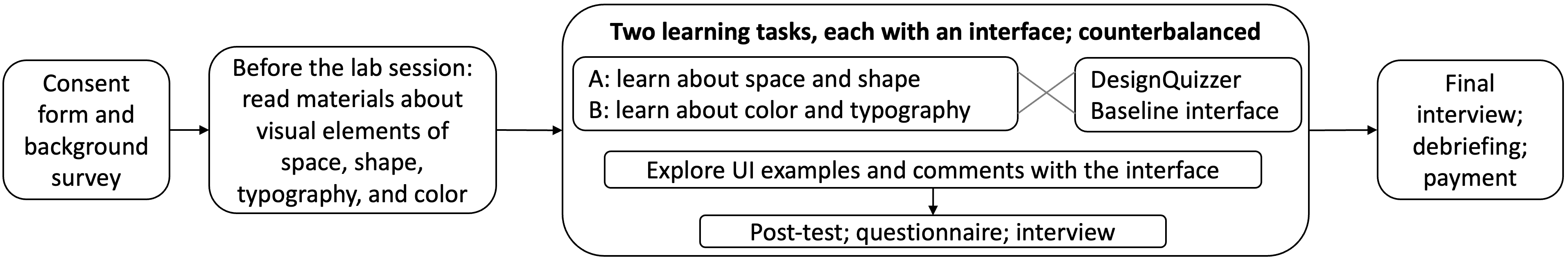}
  \caption{\peng{Procedure of the within-subjects (tool: DesignQuizzer vs. Baseline Reddit interface) experiment \uppercase\expandafter{\romannumeral1}. In each learning task, participants explore design examples and comments with the assigned interface to learn the required visual elements.}}
  \label{fig:procedure}
\end{figure}
\autoref{fig:procedure} illustrates the procedure of \peng{experiment \uppercase\expandafter{\romannumeral1}}. 
After obtaining participants' consent, we sent them the links to an online background survey and a document that lists the basic definitions and concepts of visual elements about space, shape, typography, and color. 
The document contains 13 screenshots from related pages of \cite{materialdesign} and \cite{best_practice} that introduce the basic design terminology. 
At the beginning of each task, we first introduced the task and the tool interface. 
Then participants explored online UI examples and comments with the assigned tool. We allocated 25 minutes for each learning session based on a pilot study with two users. 
After each task, participants wrote a comment to criticize a new UI example in the post-test and filled in the questionnaire to rate their perceptions of the learning process and the tool. 
We further conducted a semi-structured interview to make sense of the ratings and suggestions to improve the tool. 
Upon completion of two tasks, we asked which tool they preferred and why. 
The whole procedure lasted for 100-120 minutes. 
After debriefing, each participant received a \$8.5 compensation following the local payment policy.

\subsection{Analysis and Results}
For the self-report items and \zh{numbers of enacted or violated principles about the visual elements mentioned in the post-tests' comments}, we performed Wilcoxon signed-rank tests \cite{wolcoxon_test}, as used in previous HCI studies \cite{MetaMap,visualize_example,forkit}, to assess the difference between the DesignQuizzer and baseline conditions (\autoref{table:results}). 
\peng{We also conducted a set of statistical tests (detailed in \autoref{test_order_effect}) to affirm that neither the factor of tool/task orders nor their interactions with the tool factor impact the significances of the reported results. 
This \cqy{indicates} that the learning bias did not significantly impact our results.
}
For the interview recordings, two of the authors transcribed them into text. 
They first familiarized themselves by reviewing all the text scripts independently. After several rounds of coding with comparison and discussion, they finalized the codes of all the interview data regarding each RQ aspect. 
We counted the occurrences of codes \peng{(\autoref{table:experiment_1_pros_cons})} and incorporated these qualitative findings in the following presentation of our results.

\begin{table}[]
\caption{
\peng{
The experiment \uppercase\expandafter{\romannumeral1}'s RQ1-4 statistical results about DesignQuizzer and the community-like baseline interface. All items except the one for RQ3 are measured using a standard 7-point Likert scale (1 - strongly disagree; 7 - strongly agree). Note: -: $p > .1, +: .05 < p < .10, *:p < .05, **:p < .01, ***:p < .001$; Wilcoxon signed-rank test; within-subjects; $N = 24$.
}}
\label{table:results}
\scalebox{0.9}{
\begin{tabular}{lllllll}
\hline
\multirow{2}{*}{Research Question}                                                                                 & \multirow{2}{*}{Item}                                                                                                       & \multirow{2}{*}{\begin{tabular}[c]{@{}l@{}}DesignQuizzer\\ Mean (SD)\end{tabular}} & \multirow{2}{*}{\begin{tabular}[c]{@{}l@{}}Baseline\\ Mean (SD)\end{tabular}} & \multicolumn{3}{c}{Statistics}                                                                      \\ \cline{5-7} 
                                                                                                                   &                                                                                                                             &                                                                                    &                                                                               & \multicolumn{1}{c}{\textit{Z}} & \multicolumn{1}{c}{\textit{p}} & \multicolumn{1}{c}{\textit{Sig.}} \\ \hline
\multirow{2}{*}{\begin{tabular}[c]{@{}l@{}}(RQ1) Explored\\ examples and comments\end{tabular}}                   & Satisfaction                                                                                                                & 5.29 (0.86)                                                                        & 4.67 (1.61)                                                                   & -1.774                         & 0.076                          & +                                 \\
                                                                                                                   & Helpfulness                                                                                                                 & \textbf{5.83 (1.20)}                                                               & 3.83 (1.61)                                                                   & -3.453                         & 0.001                          & **                                \\ \hline
\multirow{8}{*}{\begin{tabular}[c]{@{}l@{}}(RQ2)\\ Engagement and\\ cognitive load in\\ the process\end{tabular}} & Mean engagement                                                                                                             & \textbf{5.35 (0.82)}                                                               & 4.39 (1.14)                                                                   & -3.137                         & 0.002                          & **                                \\
                                                                                                                   & - Concentration                                                                                                             & \textbf{5.67 (1.24)}                                                               & 4.24 (1.51)                                                                   & -2.890                         & 0.004                          & **                                \\
                                                                                                                   & - Sense of Ecstasy                                                                                                          & \textbf{6.25 (0.85)}                                                               & 4.33 (1.52)                                                                   & -4.073                         & 0.000                          & ***                               \\
                                                                                                                   & - Doability                                                                                                                 & 5.04 (1.37)                                                                        & 4.75 (1.42)                                                                   & -0.821                         & 0.412                          & -                                 \\
                                                                                                                   & - Sense of Serenity                                                                                                         & 4.38 (1.61)                                                                        & 4.25 (1.94)                                                                   & -0.292                         & 0.771                          & -                                 \\
                                                                                                                   & - Timelessness Feeling                                                                                                      & \textbf{5.50 (1.29)}                                                               & 4.50 (1.79)                                                                   & -2.027                         & 0.043                          & *                                 \\
                                                                                                                   & - Intrinsic Motivation                                                                                                      & \textbf{5.29 (1.30)}                                                               & 4.25 (1.54)                                                                   & -2.427                         & 0.015                          & *                                 \\ \cline{2-7} 
                                                                                                                   & Cognitive load                                                                                                              & 5.04 (1.04)                                                                        & 4.42 (1.25)                                                                   & -1.557                         & 0.119                          & -                                 \\ \hline
\begin{tabular}[c]{@{}l@{}}\peng{(RQ3) Outcome on} \\ \peng{recognizing}\\ \peng{enacted or violated} \\ \peng{design principles}\end{tabular}               & \begin{tabular}[c]{@{}l@{}}Numbers of \\ design principles about visual \\ elements mentioned in\\ the post-test's comment\end{tabular} & \textbf{3.38 (1.50)}                                                                        & 2.29 (1.23)                                                                   & -2.994                         & 0.003                          & **                                \\ \hline
\multirow{3}{*}{\begin{tabular}[c]{@{}l@{}}(RQ4)\\ Perception\\ towards the tool\end{tabular}}                     & Usefulness                                                                                                                  & \textbf{5.65 (0.80)}                                                               & 4.38 (1.08)                                                                   & -3.754                         & 0.000                          & ***                               \\
                                                                                                                   & Easy to use                                                                                                                 & \textbf{5.29 (0.97)}                                                               & 4.13 (1.10)                                                                   & -3.411                         & 0.001                          & **                                \\
                                                                                                                   & Intention to use                                                                                                            & \textbf{5.46 (1.01)}                                                               & 4.54 (1.06)                                                                   & -2.725                         & 0.006                          & **                                \\ \hline
\end{tabular}
}
\end{table}

\begin{table}[]
\caption{\peng{Summarized pros and cons of DesignQuizzer and the community-like baseline interface in experiment \uppercase\expandafter{\romannumeral1}. These findings are incorporated into subsubsections 5.5.2 - 5.5.5 to make sense of the \cqy{statistical} results. The number next to each point is the number of participants who mention it; within-subjects; $N = 24$.}}
\label{table:experiment_1_pros_cons}
\scalebox{0.9}{
\begin{tabular}{lll}
\hline
     & DesignQuizzer                                                                                                                                                                                                       & Baseline                                                                                                                                                                                                                                                             \\ \hline
Pros & \begin{tabular}[c]{@{}l@{}}Related comments (6); "Query" function (11); \\ Quiz-like interaction (16); Facilitate deep \\ understanding (4); Structured comment \\ summary (11); Clear interaction (7)\end{tabular} & \begin{tabular}[c]{@{}l@{}}Sense of a community (9);\\ Entertaining content (5);\\ Broad knowledge view (3)\end{tabular}                                                                                                                                             \\ \hline
Cons & \begin{tabular}[c]{@{}l@{}}Boring interaction (3);\\ Lack external knowledge (4);\\ Insufficient explanation in the summary (7);\\ Overwhelming visual encodings (5)\end{tabular}                                   & \begin{tabular}[c]{@{}l@{}}Unrelated comments (15); Low-quality comments (5); \\ Passive learning experience (7); Unstructured \\ comments (5); Lack external knowledge (4); \\ Unaligned comments to the UI image (4);\\ Messing comment structure (4)\end{tabular} \\ \hline
\end{tabular}
}
\end{table}

\peng{
\subsubsection{The Impact of Tool/Task Orders on the Measures}
\label{test_order_effect}
\label{subsubsection:order_effect}
We first conducted a set of mixed ANOVA analyses using SPSS software to examine the impact of tool/task order on our measures. 
The within-subjects factor is the used tool (DesignQuizzer vs. baseline interface), and the between-subjects factors are the order of the learning task and the order of \cqy{the experienced} tool (\autoref{fig:procedure}). 
As \cqy{shown} in \autoref{table:experiment_2_order_effect}, most of the main effects of tool/task orders and their interaction effects with the tool condition are not significant. 
One exception is the effect ($F = 6.835, p = 0.017$) of the task order factor \cqy{on perceived satisfaction.}
Specifically, the twelve participants with the task order ``B (color and typography) -> A (space and shape)'' ($M = 5.46$) were generally more satisfied with the explored design examples and comments compared to the other twelve participants with ``A -> B'' task order ($M = 4.50$). 
Nevertheless, the ordering effects do not affect our key findings on helpfulness, engagement, number of design \cqy{principles}, and perceived usefulness. 
Therefore, in the following reported results, we focus on the effects of the within-subjects tool factor. 
While ANOVA helps mixed analyses for testing the ordering effect, it is generally used for the comparisons of more than two means \cite{statistic_test_guide}. 
In our case, we performed Wilcoxon signed-rank tests \cite{wolcoxon_test} to assess the difference between the DesignQuizzer and the community-like baseline conditions. 
}

\begin{table}[]
\caption{\peng{The \zhenhui{p-values} of the mixed-ANOVA tests to examine the impact of tool/task orders on the measures in experiment \uppercase\expandafter{\romannumeral1}. Within-subjects: \textbf{tool} (DesignQuizzer vs. baseline); between-subjects: the \textbf{order} of the used tool, the order of learning \textbf{task}; $N = 24$. The tests affirm that the orders do not significantly impact the significance of the reported results.}}
\label{table:experiment_2_order_effect}
\scalebox{0.86}{
\begin{tabular}{rccccccc}
\cline{2-8}
\multicolumn{1}{l}{}                                                   & \multicolumn{4}{c}{Within-subjects effects}                                                                                             & \multicolumn{3}{c}{Between-subjects effects}                                            \\ \cline{2-8} 
\multicolumn{1}{l}{}                                                   & \multicolumn{1}{l}{Tool} & \multicolumn{1}{l}{Tool * Task} & \multicolumn{1}{l}{Tool * Order} & \multicolumn{1}{l}{Tool * Task * Order} & \multicolumn{1}{l}{Task} & \multicolumn{1}{l}{Order} & \multicolumn{1}{l}{Task * Order} \\ \hline
Satisfaction                                                           & 0.075                    & 0.538                           & 0.538                            & 0.711                                   & 0.017                    & 0.319                     & 0.155                            \\
Helpfulness                                                            & 0.000                    & 0.844                           & 0.844                            & 1.000                                   & 0.194                    & 0.850                     & 0.451                            \\ \hline
Mean engagement                                                        & 0.001                    & 0.597                           & 0.889                            & 0.090                                   & 0.304                    & 0.248                     & 0.137                            \\
Cognitive load                                                         & 0.107                    & 0.739                           & 0.739                            & 0.580                                   & 0.897                    & 0.165                     & 0.698                            \\ \hline
\begin{tabular}[c]{@{}r@{}}Numbers of\\ design principles\end{tabular} & 0.002                    & 0.793                           & 0.600                            & 1.000                                   & 0.325                    & 0.619                     & 0.868                            \\ \hline
Usefulness                                                             & 0.000                    & 0.538                           & 0.194                            & 0.538                                   & 0.119                    & 0.193                     & 0.301                            \\
Easy to use                                                            & 0.000                    & 0.473                           & 0.139                            & 0.631                                   & 0.173                    & 0.586                     & 0.903                            \\
Intention to use                                                       & 0.003                    & 0.112                           & 0.085                            & 0.665                                   & 0.519                    & 0.605                     & 0.519                            \\ \hline
\end{tabular}
}
\end{table}

\subsubsection{Explored Design Examples and Comments (RQ1)}
In general, participants explored more UI examples ($Mean = 13.21, SD = 6.62$ vs. $M = 6.71, SD = 1.55$) and comments ($M = 20.67, SD = 10.07$ vs. $M = 13.71, SD = 5.53$) with DesignQuizzer than with the baseline tool during the 25-minutes learning session. 
Overall, there is a tendency that participants are more satisfied with the explored examples and comments for learning required visual elements ($Median = 5.00$) than in the baseline condition ($Mdn = 5.00$); $Z = -1.774, p = 0.076$. 
Compared with the baseline condition ($Mdn = 4.00$), they felt that most of their explored examples and comments were significantly more helpful for their learning goals in the DesignQuizzer condition ($Mdn = 6.00$); $Z = -3.453, p = 0.001$. 
Fifteen users mentioned that most of the Reddit comments were unrelated to their interested visual elements. 
Five of them reported that they encountered low-quality comments in Reddit. 
``\textit{It is hard to find the needed information, especially from long comments, in the Reddit community}'' (P19, Female, age: 19). 
Six people reported that DesignQuizzer's prompted examples and comments matched their learning interests, and eleven people appreciated its ``query'' function. 
``\textit{I like the DesignQuizzer’s `I want to explore a UI component / visual element' buttons that allow me quickly explore examples and comments of interests}'' (P7, Not Available, 20). 
These results suggest that DesignQuizzer helped users explore helpful UI design examples and comments more efficiently in comparison to the community-like interface.


\subsubsection{Engagement in the Process (RQ2)} 
In general, participants felt that they were significantly more engaged in the learning session with the DesignQuizzer condition ($Mdn = 5.50$) than with the baseline interface ($Mdn = 4.58$); $Z = -3.137, p = 0.002$ (\autoref{table:results}). 
Specifically, DesignQuizzer improves users’ concentration ($Z = -2.890, p = 0.004$), sense of ecstasy ($Z = -4.073, p < 0.001$),  timelessness feeling ($Z = -2.027, p = 0.043$), and intrinsic motivation ($Z = -2.427, p = 0.015$) during the visual design learning process. 
Sixteen participants highlighted that the quiz-like interaction of DesignQuizzer encouraged active thinking, which helped them focus on the learning materials. 
``\textit{These questions made me more focused. I want to answer them correctly}'' (P16, F, 19).  
``\textit{DesignQuizzer’s hint guided me to think actively by examining the image with the critique}'' (P17, Male, 19).  
As for the Reddit interface, seven users reported having a passive learning experience, and five people encountered unstructured comments with scattered knowledge points. 
``\textit{I gradually lost interest in learning because the resources in the comments are not organized and I was passively receiving them}'' (P2, M, 21). 
\peng{Nevertheless,} there is no significant difference between the tool conditions regarding the doability and sense of serenity. 
\peng{Participants’ comments on the cons of DesignQuizzer and pros of Reddit baseline could explain this result. For example, } 
three participants commented that they got bored with DesignQuizzer after having several rounds of interactions. 
``\textit{I was tired of many single-choice questions, which are monotonous}'' (P9, M, 19). 
In contrast, nine people mentioned that the Reddit design community offered them a ``\textit{relaxing learning environment with the sense of a community}''. 
Besides, five users liked the  ``\textit{entertaining content}'' (P15, M, 19) in Reddit, while DesignQuizzer filters out this content in the comment summarization step. 

\subsubsection{Outcome on recognizing enacted or violated design principles (RQ3)}
Compared to the conditions with the community-like interface ($Mdn = 3.00$), when using DesignQuizzer, participants mentioned significantly more points of design principles about the learned visual elements in their written comments to new examples in the post-test; $Z = -2.994, p = 0.003$ (\autoref{table:results}).
\peng{This suggests that compared to the baseline interface, DesignQuizzer could better help users develop their design skills in recognizing enacted or violated design principles about learned visual design elements in the UI examples. 
}
In the interview after each task, we asked participants if they could give examples of what they had learned in the learning session. Seventeen / twelve people provided such examples when they \cqy{learned} with the DesignQuizzer / Reddit interface. 
``(With DesignQuizzer) \textit{I learned how to choose the shape of icons. Beginners of design may prefer to use pointed edges for the icons, but customers would prefer the rounded edges}'' (P7, NA, 20). 
``(With Reddit) \textit{I learned that we could use the hero title with a large font size to draw people’s attention}'' (P13, NA, 19). 
Four users mentioned that DesignQuizzer helped them gain a deep understanding of these elements, while three people indicated that the comment list of a UI design in the community provides a broad knowledge view. 
``\textit{The DesignQuizzer helped me understand the design principles about color more deeply, especially when I answered the question incorrectly}'' (P8, M, 18). 
``\textit{The Reddit interface gave me a sense of learning in a community where the members' opinions are diverse and comprehensive}'' (P11, M, 18).  
Nevertheless, the learning activity in both conditions could suffer from the lack of knowledge input from external resources, as commented by four participants. 
``\textit{Some comments did not give reasonable rationales for the critiques. It would be better to provide the definition and a knowledge graph for the mentioned visual elements}'' (P11, M, 18). 


\subsubsection{Interaction and Perception with DesignQuizzer (RQ4)}
Overall, participants actively responded to DesignQuizzer's questions ($M = 20.67, SD = 10.07$) and answer correctly for $13.54 (SD = 7.37)$ times. 
They seldom clicked the ``I need a hint'' button ($M = 2.13, SD = 2.38$) but frequently hit the ``Why'' button to assess the structured comment summary after seeing the correct answer ($M = 9.42, SD = 5.19$). 
Besides, they sometimes input their interested keywords via the ``I want to explore a UI component / visual element'' buttons ($M = 3.00, SD = 1.31$). 
In total, they reported the correctness of the answers eight times. 

In terms of the perceptions of the tools, participants rate DesignQuizzer  ($Mdn = 5.75$) to be significantly more useful for visual design learning than the community-like baseline ($Mdn = 4.50$); $Z =-3.754, p < .001$ (\autoref{table:results}). 
Eleven participants valued DesignQuizzer’s structured comment summary. 
``\textit{My favorite feature of DesignQuizzer is the `Why' button, which provided me meaningful feedback sentences marked in different colors. I can also go to the original post thread with the provided link}'' (P24, F, 19). 
However, seven users mentioned that the comment summary sometimes could not address their confusions, as P24 (F, 19) further said: ``\textit{But sometimes the comment summary did not provide sufficient information for me to understand why it is this answer. It would be better to have more explanations}''. 
Moreover, participants felt that DesignQuizzer ($Mdn = 5.50$) is significantly easier to use than the community-like interface ($Mdn = 4.13$) for visual design learning; $Z = -3.411, p = .001$. 
Seven people commented that the interaction with DesignQuizzer was ``\textit{clear, funny, and flexible}'' (P1, F, 19). 
While the structured comment summary was valuable for eleven users, five suggested that its visual design could be further improved. 
``\textit{The visual codes of the `why' message were too much. Sometimes it was hard to find the key points as I did not know the meanings of these text colors}'' (P18, M, 21).  
As for the Reddit community interface, four participants indicated that its layout design was not convenient for aligning the comments to the UI image, and four users felt that the nesting structure of comments was a mess for their learning purpose. 
``\textit{The design example and comments are far from each other. I can not check them simultaneously but need to scroll the page up and down frequently}'' (P7, NA, 20).  
All in all, participants have a significantly stronger intention to use DesignQuizzer ($Mdn = 5.50$) than the baseline tool ($Mdn = 4.50$) in their future informal learning practices of visual design; $Z = -2.725, p = .006$. 
In the interview after two tasks, all participants indicated their preferences for DesignQuizzer over the community-like interface for learning visual design. 
In addition to the benefits mentioned above, six people favored DesignQuizzer’s single-choice questioning design, which ``\textit{is easy to get started and reduces the learning cost of the tool}'' (P9, M, 19). 
However, twelve participants were reminded that the preferences of DesignQuizzer depend on their roles as novices and their goals to learn specific visual elements. 
``\textit{The Reddit webpage may be better if I want to broadly explore the examples and comments without a specific learning focus}'' (P13, NA, 19). 

\subsubsection{Summary of the Findings in \zhenhui{Experiment} \uppercase\expandafter{\romannumeral1}}
\peng{Compared with the community-like baseline interface, DesignQuizzer improves novices’ efficiency and overall engagement in exploring helpful UI examples and comments for learning specific visual elements.
They favor its question-answering interaction that stimulates active thinking and its structured comment summary that reduces the reading workload. 
Besides, we find a significant improvement with DesignQuizzer in participants' ability to critique others' designs using the learned knowledge about targeted visual elements. 
However, our experiment \uppercase\expandafter{\romannumeral1} does not evaluate the impact of DesignQuizzer on participants' ability to apply what they learn in future design activities, which is also an important aspect of learning design knowledge. 
Studying this aspect would require a between-subjects experiment design that is different from experiment \uppercase\expandafter{\romannumeral1}, because the knowledge transition from one learning task to the other could largely affect the participants' performance in the subsequent design activities. 
Furthermore, we incorporate feedback from design experts on the proposed DesignQuizzer in the following \zhenhui{experiment} \uppercase\expandafter{\romannumeral2} to provide a more comprehensive assessment of the tool. 
}




\peng{
\section{Experiment \uppercase\expandafter{\romannumeral2}}
The primary goal of our between-subjects Experiment \uppercase\expandafter{\romannumeral2} with 28 participants is to explore an \cqy{additional} research question: 

\textbf{RQ5:} How would DesignQuizzer affect the novices' learning outcome on \textbf{a)} design knowledge and \textbf{b)} application of the knowledge in the visual design activity?

Specifically, we measure learning outcome on RQ5b via participants' performance in the visual design tasks. 
This measure can reveal people's skill in creating new or original work using the learned knowledge, which is the most complex stage of the learning process compared to the ``remember'', ``understand'', ``apply, ``analyse'', and ``evaluate'' stages in Bloom Taxonomy \cite{bloom_taxonomy}. 

To complement the evaluation of DesignQuizzer, \zhenhui{experiment}  \uppercase\expandafter{\romannumeral2} also serves the goals of inputting more evidence on the RQ1-4 results and collecting qualitative feedback from visual design experts. 
}

\peng{
The \textbf{baseline} Reddit r/UI\_Design interface is identical to the one used in experiment \uppercase\expandafter{\romannumeral1}, as described in \autoref{baseline_interface}. 

The \textbf{learning task} is similar to the one in experiment \uppercase\expandafter{\romannumeral1}, as described in \autoref{task}. 
However, this time, each participant only needs to conduct one learning task: learn design knowledge about 'color' and 'typography'. 

We recruited 28 student \textbf{participants} (14 \textbf{F}emales,  14 \textbf{M}ales; age range 19-23, $Mean = 21.25, SD = 1.08$) via a recruitment post in a local university. 
We conducted a power analysis using the G*power software to determine the number of participants. 
Specifically, in the 2 x 2 mixed-design ANOVA test for the RQ5b's measure described in \autoref{section:experiment_2_result}, we input to G*power \cqy{an effect size} of 0.5 (expected means difference of measures: 1, SD: 1), a power of 0.8, and a p-value of 0.05. 
This outputs that the recommended smallest sample size is 26 (13 in each group). 
The inclusion criteria are that participants \cqy{are novices in visual design but are interested in learning} more about it. 
All of them have passed the \cqy{National} College English Test for general requirements. 
In general, our participants have little or no experience in learning UI visual design ($M=2.00, SD=1.09$) and exploring online design communities ($M=3.18, SD=1.85$; 1 - No experience at all, 7 - A lot of experience) but they are interested in learning it in our study ($M=5.36, SD=0.83$; 1 - No interest at all, 7 - A great deal of interest). 
The usage frequency of online communities where people create posts and comments is: 22 daily, 1 2-6 days a week, 3 once a week, and 2 less than once a week. 
We randomly assign our participants into the DesignQuizzer (noted as PD1-14) and Baseline groups (PB1-14), each with 14 participants. 
}

\peng{
\subsection{Measures}
We input more evidences on the RQ1-4 results regarding the following measures (detailed in \autoref{measures}): 
\begin{itemize}
    \item \textbf{RQ1:} Satisfaction and perceived helpfulness of the explored examples and comments for learning design knowledge about color and typography. 
    \item \textbf{RQ2:} Perceived engagement in the learning process. 
    \item \textbf{RQ3:} Numbers of enacted or violated principles about color and typography mentioned in the participants' feedback comments to others' UI examples after learning session. 
    \item \textbf{RQ4:} Perceived usefulness, easy to use, and intention to use regarding the DesignQuizzer and baseline interface. 
\end{itemize}

We also offer more qualitative findings regarding \textbf{RQ4} about perception with the used tool.
In the questionnaire after the learning task, we ask participants to write down their perceived pros/cons of the tool, the possible scenarios in their long-term usage of the tool, and suggestions for improvement. 

}

\peng{
To address the added \textbf{RQ5a} about the learning outcome on design knowledge, we ask participants in the questionnaire  to list the knowledge related to color and typography learned from the explored UI examples and comments. 

As for the \textbf{RQ5b} about applying learned knowledge in the design activity, we capture the change in participants' performance on a visual design activity before and after the learning task. 
Specifically, in the \textbf{pre-test} before the learning task, participants are required to color the UI components and adjust the typography of two given uncolored mockups (\autoref{fig:web_page_examples}a and \autoref{fig:mobile_page_examples}a). 
This is a common graphic or visual design activity in design firms, in which designers usually deal with a high-fidelity representation of \cqy{a UI} that shows exactly what the UI is supposed to look like (i.e., mockup) \cite{graphic_design_common}. 
The goal of visual design is to literally enable proper visual communication of the UI's information using elements such as color, images, typography, and layout \cite{graphic_design_common}. 
Therefore, this design activity enables us to capture the participants' performance \cqy{in} applying the knowledge about color and typography that they need to learn. 
We collect two mockups that have different types of UI components from the Figma design community \footnote{\url{https://www.figma.com/community/file/1074977530633823056} and \url{https://www.figma.com/community/file/1079825061633332655}}. 
One is a web page of a travel agency and has UI components like menu, button, and hero title.
The other is a mobile payment page of an educational app and contains UI components like card, button, and form. 
We adjust the mockups as \cqy{follows}. 
First, we only keep one page of each original mockup to ease the design activity for our novice participants. 
Second, we uncolor the kept pages by assigning black, white, and grey colors to all UI components except the images (e.g., the background of the web page mockup). 
Third, we normalize the font style, size (=20), weight (regular) and color (\# 000000) of the texts in each mockup. 

In the \textbf{post-test} after the learning task, participants are required to conduct the same visual design activity on the same two uncolored mockups as those in the pre-test. 
We choose the same two mockups in pre- and post-tests to avoid noises from the design materials on the measured performance. 
We invite two visual design experts (E1-2, both males, ages: 28 and 27) to score the participants' outcome UIs from the visual design activity. 
E1 has experience in UI research projects and has been in the UI design teams of a travel agency company and a learning app startup. 
E2 majors in engineering design and has contributed to more than three UI design projects, including websites and corresponding mobile apps. 
They score each outcome UI from three aspects based on the suggested design principles from MaterialDesign \cite{materialdesign}: 
\textbf{Consistent}: color and typography should be applied throughout a UI consistently and \cqy{be compatible with the brand. }
\textbf{Distinct}: color and typography should create a distinction between elements, with sufficient contrast between them. 
\textbf{Intentional}: color and typography should be applied purposefully as it can convey meaning in multiple ways. 
We randomize the order of outcome UIs for the scoring session.
The experts do not know whether the scoring UI is from the pre- or post-test and whether it is from the DesignQuizzer or Baseline group. 
For each aspect of an outcome UI, we average the two experts' scores (range: 0-10 points) as the final score. 
}

\begin{figure}
  \centering
\includegraphics[width=\linewidth]{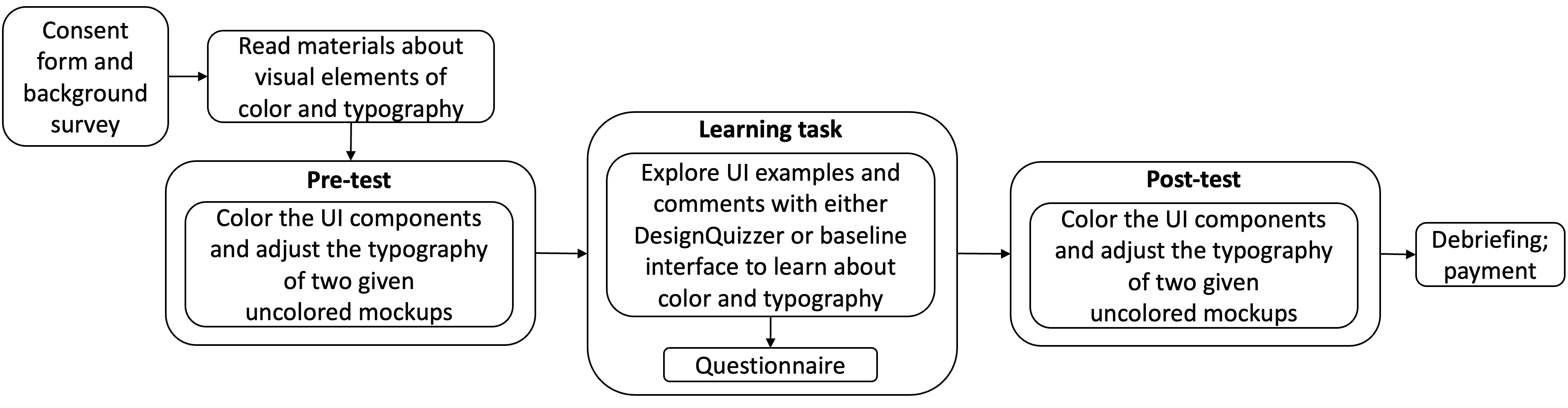}
  \caption{\peng{Procedure of the between-subjects (tool: DesignQuizzer vs. Baseline Reddit interface) experiment \uppercase\expandafter{\romannumeral2}.}}
  \label{fig:procedure_2}
\end{figure}

\peng{
\subsection{Procedure}

\autoref{fig:procedure_2} illustrates the procedure of \peng{experiment \uppercase\expandafter{\romannumeral2}}. 
Participants used the desktop computer with a 23.8-inch monitor in our lab to conduct the study. 
After filling in the consent form and background survey, participants walked through the basic definitions and concepts of color and typography in a document compiled from \cite{materialdesign} and \cite{best_practice}. 
We then demonstrated how to use the Figma desktop application to color the UI components and adjust the typography of \cqy{a UI} example. 
Next, participants conducted the pre-test in the visual design activity on two mockups. 
We allocated twelve minutes for the pre-test based on a pilot study with two users. 
After the pre-test, we introduced the used tool, i.e., either DesignQuizzer or Baseline, for the learning task. 
Participants started to explore online UI examples and comments with the assigned tool for 25 minutes. 
Upon completion of the learning task, participants filled in a questionnaire that \cqy{asked} them to 1) rate their perceptions of the learning process and the tool, 2) \cqy{comment on a} given new UI from the aspects of color and typography, and 3) write down what knowledge they learned, perceived pros/cons of the tool, how would they use it in the long term, and their suggestions for improvement. 
Finally, participants conducted the post-test in the visual design activity on the same two mockups as in pre-test within twelve minutes. 
The whole procedure lasted for 90-120 minutes. 
After debriefing, each participant received a \$12.5 compensation.
}

\peng{
\subsection{Analysis and Results}
\label{section:experiment_2_result}
For the self-report items and numbers of enacted or violated principles about the visual elements mentioned in the post-tests' comments, we used the Mann-Whitney U test \cite{mann1947test} to compare the ratings between two user groups (\autoref{table:experiment_2_RQ1-4}). 
The Mann-Whitney U is a non-parametric test commonly used to compare differences between independent conditions (e.g., in HCI studies \cite{chandrasekharan2017you,umar2019detection,kin2022stylette}) especially when the data normality is violated, as confirmed in our cases. 
For the written responses in the questionnaire about RQ4 and RQ5a, two of the authors conducted several rounds of open coding with comparison and discussion. 
We counted the occurrences of codes and reported them in \autoref{table:experiment_2_pros_cons} and \autoref{table:experiment_2_learned_knowledge}. 
To evaluate the changes in the participants' performance in the visual design activity (\textbf{RQ5b}), we conducted a two-way mixed ANOVA to compare the performance of participants in each group (as the between-subjects factor) in the pre-test and post-test (as the within-subjects factor). 
}

\begin{table}[]
\caption{
\peng{
The experiment \uppercase\expandafter{\romannumeral2}'s RQ1-4 statistical results about DesignQuizzer and the community-like baseline interface. All items except the one for RQ3 are measured using a standard 7-point Likert scale (1 - strongly disagree; 7 - strongly agree). Note: -: $p > .1, +: .05 < p < .10, *:p < .05, **:p < .01$; Mann-Whitney U test; between-subjects; $N = 28$.
}}
\label{table:experiment_2_RQ1-4}
\scalebox{0.9}{
\begin{tabular}{lllllll}
\hline
\multirow{2}{*}{Research Question}                                                                                & \multirow{2}{*}{Item}                                                                                                                 & \multirow{2}{*}{\begin{tabular}[c]{@{}l@{}}DesignQuizzer\\ Mean (SD)\end{tabular}} & \multirow{2}{*}{\begin{tabular}[c]{@{}l@{}}Baseline\\ Mean (SD)\end{tabular}} & \multicolumn{3}{c}{Statistics}                                           \\ \cline{5-7} 
                                                                                                                  &                                                                                                                                       &                                                                                    &                                                                               & \multicolumn{1}{c}{U} & \multicolumn{1}{c}{p} & \multicolumn{1}{c}{Sig.} \\ \hline
\multirow{2}{*}{\begin{tabular}[c]{@{}l@{}}(RQ1) Explored\\ examples and comments\end{tabular}}                   & Satisfaction                                                                                                                          & 5.07 (1.07)                                                                        & 5.00 (0.78)                                                                   & 87.00                 & 0.592                 & -                        \\
                                                                                                                  & Helpfulness                                                                                                                           & 5.50 (1.02)                                                                        & 4.79 (1.42)                                                                   & 75.00                 & 0.272                 & -                        \\ \hline
\multirow{8}{*}{\begin{tabular}[c]{@{}l@{}}(RQ2)\\ Engagement and\\ cognitive load in\\ the process\end{tabular}} & Mean engagement                                                                                                                       & \textbf{5.40 (1.08)}                                                               & 4.67 (0.80)                                                                   & 51.00                 & 0.030                 & *                        \\
                                                                                                                  & - Concentration                                                                                                                       & 5.71 (1.33)                                                                        & 5.43 (1.22)                                                                   & 83.00                 & 0.469                 & -                        \\
                                                                                                                  & - Sense of Ecstasy                                                                                                                    & \textbf{6.29 (0.61)}                                                               & 4.86 (1.41)                                                                   & 34.00                 & 0.002                 & **                       \\
                                                                                                                  & - Doability                                                                                                                           & 4.79 (1.25)                                                                        & 5.14 (0.86)                                                                   & 80.50                 & 0.395                 & -                        \\
                                                                                                                  & - Sense of Serenity                                                                                                                   & \textbf{4.93 (1.44)}                                                               & 3.47 (1.28)                                                                   & 47.00                 & 0.017                 & *                        \\
                                                                                                                  & - Timelessness Feeling                                                                                                                & 5.21 (1.53)                                                                        & 4.57 (1.50)                                                                   & 62.50                 & 0.097                 & +                        \\
                                                                                                                  & - Intrinsic Motivation                                                                                                                & 5.40 (1.08)                                                                        & 4.68 (0.80)                                                                   & 72.50                  & 0.232                 & -                        \\ \cline{2-7} 
                                                                                                                  & Cognitive load                                                                                                                        & 4.64 (1.22)                                                                        & 4.00 (1.62)                                                                   & 71.50                 & 0.212                 & -                        \\ \hline
\begin{tabular}[c]{@{}l@{}}(RQ3) Outcome on \\recognizing\\ enacted or violated\\ design principles\end{tabular}     & \begin{tabular}[c]{@{}l@{}}Numbers of design\\ principles about visual\\ elements mentioned in\\ the post-test’s comment\end{tabular} & 4.14 (2.32)                                                                        & 3.29 (1.44)                                                                   & 91.50                 & 0.756                 & -                        \\ \hline
\multirow{3}{*}{\begin{tabular}[c]{@{}l@{}}(RQ4)\\ Perception\\ towards the tool\end{tabular}}                    & Usefulness                                                                                                                            & \textbf{5.61 (0.94)}                                                               & 4.38 (0.96)                                                                   & 34.00                 & 0.003                 & **                       \\
                                                                                                                  & Easy to use                                                                                                                           & \textbf{5.13 (1.40)}                                                               & 3.95 (0.92)                                                                   & 40.50                 & 0.008                 & **                       \\
                                                                                                                  & Intention to use                                                                                                                      & 5.18 (1.27)                                                                        & 4.54 (1.63)                                                                   & 75.50                 & 0.297                 & -                        \\ \hline
\end{tabular}
}
\end{table}

\peng{
\subsubsection{Statistical Results of RQ1-4}
\autoref{table:experiment_2_RQ1-4} summarizes the statistical results of RQ1-4 in experiment \uppercase\expandafter{\romannumeral2}. 
Overall, there is no significant difference between the DesignQuizzer ($Mean = 5.07, SD = 1.07$) and baseline conditions ($M = 5.00, SD = 0.78$) regarding perceived satisfaction with explored examples and comments for learning color and typography; $U = 87.00, p = 0.592$. 
The perceived helpfulness of the explored examples and comments tends to be higher in DesignQuizzer ($M = 5.50, SD = 1.02$) than that in the baseline condition ($M = 4.79, SD = 1.42$); $U = 75.00, p = 0.272$. 
These results moderately support RQ1 findings in experiment \uppercase\expandafter{\romannumeral1} that DesignQuizzer could help users explore helpful UI design examples and comments for their learning tasks in comparison to the community-like interface. 

In general, participants with DesignQuizzer ($M = 5.40, SD = 1.08$) in experiment \uppercase\expandafter{\romannumeral2} felt that they were significantly more engaged in the learning session that those with the baseline interface ($M = 4.67, SD = 0.80$); $U = 51.00, p = 0.030$. 
Specifically, DesignQuizzer significantly improved users' sense of ecstasy ($U = 34.00, p = 0.002$) and serenity ($U = 47.00, p = 0.017$) during the visual design learning process. 
There are no significant differences regarding other aspects of engagement and cognitive load between the two user groups. 
These results also moderately support RQ2 results in experiment \uppercase\expandafter{\romannumeral1} that compared to the community-like interface, DesignQuizzer can provide more engaging learning experience to novices. 

On average, compared to the user group with the baseline interface ($M = 4.14, SD = 2.32$), the group with DesignQuizzer ($M = 3.29, SD = 1.44$) mentions more points of color and typography design principles in their comments to the new UI example.
However, there is no significant difference between the two conditions regarding the learning outcome on recognizing enacted or violated design principles; $U = 91.50, p = 0.756$. 
It moderately supports the RQ3 result in experiment \uppercase\expandafter{\romannumeral1} that DesignQuizzer could better help users develop their visual design skills in recognizing enacted or violated design principles about targeted visual elements. 

Participants with DesignQuizzer ($M = 5.61, SD = 0.94$)  rated it to be significantly more useful for visual design learning than those with the community-line baseline interface ($M = 4.38, SD = 0.96$); $U = 34.00, p = 0.003$. 
Besides, the DesignQuizzer users ($M = 5.13, SD = 1.40$)  found it to be significantly easier to use for visual design learning than the baseline users ($M = 3.95, SD = 0.92$); $U = 40.50, p = 0.008$.
On average, participants with DesignQuizzer ($M = 5.18, SD = 1.27$) had a higher intention to use it for future visual design learning than those with the baseline interface ($M = 4.54, SD = 1.63$); $U = 75.50, p = 0.297$. 
These results offer strong evidences to RQ4 results in experiment \uppercase\expandafter{\romannumeral1} that DesignQuizzer could be more useful and easier to \cqy{use} than the community-like interface for learning visual design knowledge. 
}

\begin{table}[]
\caption{\peng{Summarized pros and cons of DesignQuizzer and the community-like baseline interface, possible long-term usage scenarios, and suggestions for improvement in experiment \uppercase\expandafter{\romannumeral2}. The number next to each point is the number of participants who mention it; between-subjects; $N = 28$.}}
\label{table:experiment_2_pros_cons}
\scalebox{0.9}{
\begin{tabular}{lll}
\hline
                                                          & DesignQuizzer                                                                                                                                                                                                              & Baseline                                                                                                                                                                                        \\ \hline
Pros                                                      & \begin{tabular}[c]{@{}l@{}}Related comments (5); \\ Quiz-like interaction (5); Facilitate deep \\ understanding (4); Structured comment \\ summary (3); Focus on critiques (4)\end{tabular} & \begin{tabular}[c]{@{}l@{}}Sense of a community (6);\\ Entertaining content (1);\\ Broad knowledge view (6);\\ Diverse examples (2)\end{tabular}                                                \\ \hline
Cons                                                      & \begin{tabular}[c]{@{}l@{}}Boring interaction (3);\\ Lack external knowledge (3);\\ Insufficient explanation in the summary (6);\\ Unable to select interested examples (1)\end{tabular}                                   & \begin{tabular}[c]{@{}l@{}}Unrelated comments (6); Low-quality \\ comments (1); Conflicting opinions (1)\\ Unaligned comments to the UI image (5);\\ Messing comment structure (1)\end{tabular} \\ \hline
\begin{tabular}[c]{@{}l@{}}Long-term\\ usage\end{tabular} & \begin{tabular}[c]{@{}l@{}}Poster, ppt, painting, photoshop (1); \\ UI design project (7); Seek inspiration (2);\\ Study at leisure time (3)\end{tabular}                                                                  & \begin{tabular}[c]{@{}l@{}}UI design project (3); Seek inspiration (3);\\ Study at leisure time (3);\\ Seek feedback (4)\end{tabular}                                                           \\ \hline
Suggestion                                                & \begin{tabular}[c]{@{}l@{}}Give feedback to users' designs (2); Involve\\ structured professional knowledge (6); \\ Bookmarking (1); Can answer questions (1)\end{tabular}                                                  & \begin{tabular}[c]{@{}l@{}}Provide design exercises (3); \\ Content classification (5); \\ Align comments to the UI image (3)\end{tabular}                                                      \\ \hline
\end{tabular}
}
\end{table}

\peng{
\subsubsection{Qualitative Results of RQ4}
\autoref{table:experiment_2_pros_cons} \cqy{offers} more qualitative findings about users' perceived pros/cons of DesignQuizzer and baseline interface, the possible long-term usage scenarios, and suggestions for improvements. 

\textbf{Pros and cons.} Similar to the qualitative feedback in experiment \uppercase\expandafter{\romannumeral1} (\autoref{table:experiment_1_pros_cons}), participants with DesignQuizzer reported that the encountered comments are related to knowledge about color and typography (Number of participants who mention it $ = 5$). 
They favored its quiz-like interaction ($N = 5$) and structured comment summary ($N = 3$) and felt that it facilitated \cqy{a deep understanding} of the knowledge points ($N = 4$). 
Four participants further commented that the Quizzer can help them focus the critical opinions on the UI design example. 
\textit{``With DesignQuizzer, I can pay attention to the good and bad points of the design example and the specific reasons. It deepened my understanding and memory of the design details''} (PD8, Female, age: 20). 
However, similar to the findings in experiment \uppercase\expandafter{\romannumeral1}, some participants felt bored after several rounds of interaction with DesignQuizzer ($N = 3$), felt that it lacked knowledge from external sources ($N = 3$), or felt that the comment summary did not provide sufficient explanation for the quiz ($N = 6$). 
One participant (PD11, F, 20) also pointed out that she was unable to select her interested UI design examples in the Quizzer. 

We get similar comments on the baseline interface to those in experiment \uppercase\expandafter{\romannumeral1}. 
Participants favored the sense of a community ($N = 6$), the entertaining content in the community  ($N = 1$), the broad knowledge view in the comments ($N = 6$), and the diverse UI design examples ($N = 2$). 
However, they often encountered unrelated comments ($N = 6$) to the knowledge about color and typography, low-quality comments ($N = 2$), or conflicting opinions on the design examples ($N = 1$). 
Five participants also found that the baseline Reddit interface did not align the comments well to the UI design image, and one participant was unhappy with its messing comment structure. 
\textit{``I had to scroll back and forth to check the comment and UI design, which is troublesome''} (PB3, F, 22). 

\textbf{Possible long-term usage scenarios.} 
Participants in both groups anticipated that they would use DesignQuizzer ($N = 7$) or the baseline interface ($N = 3$) when they \cqy{had} UI design projects. 
They would seek design inspiration from the Quizzer ($N = 2$) or the community baseline ($N = 3$). 
They prefer to study the UI design knowledge with Quizzer ($N = 3$) or baseline ($N = 3$) \cqy{in} their leisure time. 
One participant (PD1, M, 23) also mentioned that he would use DesignQuizzer when he \cqy{needed} to design a poster, present a \cqy{PowerPoint}, or have painting or \cqy{Photoshop} tasks. 
Four participants commented that they would like to seek feedback from the design community. 

\textbf{Suggestions for improvement.} 
We asked participants to write down suggestions for improving DesignQuizzer or the community baseline interface especially in their long-term usage. 
Participants suggested that DesignQuizzer should give feedback to users' designs ($N = 2$), involve more structured knowledge from professional sources ($N = 7$), provide a bookmarking feature ($N = 1$), and be able to answer users' questions on design knowledge ($N = 1$). 
Users of the baseline interface recommended that it should provide in-situ design exercises that can get feedback from the community ($N = 3$), classify the UI examples and comments ($N = 5$), and adjust the interface to better align comments to the UI images ($N = 3$). 
These suggestions point out the directions to improve learning support interfaces that leverage the design examples and comments in online communities. 
}

\begin{table}[]
\caption{\peng{(RQ5a) Learned knowledge points about color and typography with DesignQuizzer and the community-like baseline interface in \zhenhui{experiment} \uppercase\expandafter{\romannumeral2}. The number next to each point is the number of participants who mention it; between-subjects; $N = 28$.}}
\label{table:experiment_2_learned_knowledge}
\scalebox{0.9}{
\begin{tabular}{lll}
\hline
           & DesignQuizzer                                                                                                                                                                                    & Baseline                                                                                                                                                                                         \\ \hline
Color      & \begin{tabular}[c]{@{}l@{}}Number (5); Contrast and attention (12);\\ Saturation (1); Palette (7); Specific\\ components (1); Uncommon colors (3) \\Assignment in the space (2) \end{tabular} & \begin{tabular}[c]{@{}l@{}}Number (1); Contrast and attention (11);\\ Saturation (1); Palette (1); Specific\\ components (4); Uncommon colors (2) \\Assignment in the space (4) \end{tabular} \\ \hline
Typography & \begin{tabular}[c]{@{}l@{}}Hierarchy (1); Style (5); \\ Size and attention (6)\end{tabular}                                                                                                      & \begin{tabular}[c]{@{}l@{}}Style (4); \\ Size and attention (4)\end{tabular}                                                                                                                     \\ \hline
\end{tabular}
}
\end{table}

\peng{
\subsubsection{Learned Knowledge and Application of it in the Visual Design Activity (RQ5)}
(\textbf{RQ5a}) \autoref{table:experiment_2_learned_knowledge} summarizes the learned knowledge points about color and typography that participants list in the questionnaire. 
Both user groups reported that they  learned visual design knowledge via the explored design examples and comments. 
For example, they learned that \textit{``an UI should not use a large number of colors; otherwise, it would be a chaos''} (PD9, F, 21). 
Most participants learned that there should be enough color contrast among the UI components to direct viewers' attention.
\textit{``The UI with a dark background should be careful about the color choices to ensure high readability. The usage of contrasting colors \cqy{is} attractive and informative''} (PD11, F, 20). 
\textit{``The color contrast should be obvious and allow users to capture the UI's key information at a glance''} (PB6, M, 20). 
Besides, the palette can be \textit{``monochromatic''} (PD7, F, 22) with \textit{``gradually varied colors''} (PD10, M, 21). 
There are also common practices of color usage in specific components, and we'd better not use uncommon colors. 
\textit{``We should pay attention to \cqy{the manner} reflected by the text color, e.g., do not use red for information such as names of people. Sometimes red can make the UI components stand out, and using contrasting colors can also achieve this effect''} (PD8, F, 20). 
Last but not least, the color assignment in the UI space should follow principles like \textit{``components of similar types should have the same color''} (PB12, M, 22) and \textit{``there should not be too many colors in the left part and one color in the right part of the UI''} (PD4, F, 21). 

The learned knowledge points about typography are mainly about the style and size of fonts. 
\textit{``The choice of font depends on the UI style. For example, a vivid UI page should not use the boldface that looks serious''} (PD7, F, 22). 
The size of fonts should direct viewers' attention to the target area and reflect the hierarchy of the UI page. 
For instance, \textit{``the text that conveys more important should have larger font size, but it should not be too large; otherwise, it will be stressful''} (PB3, F, 22).

\begin{figure}
  \centering
  \includegraphics[width=\linewidth]{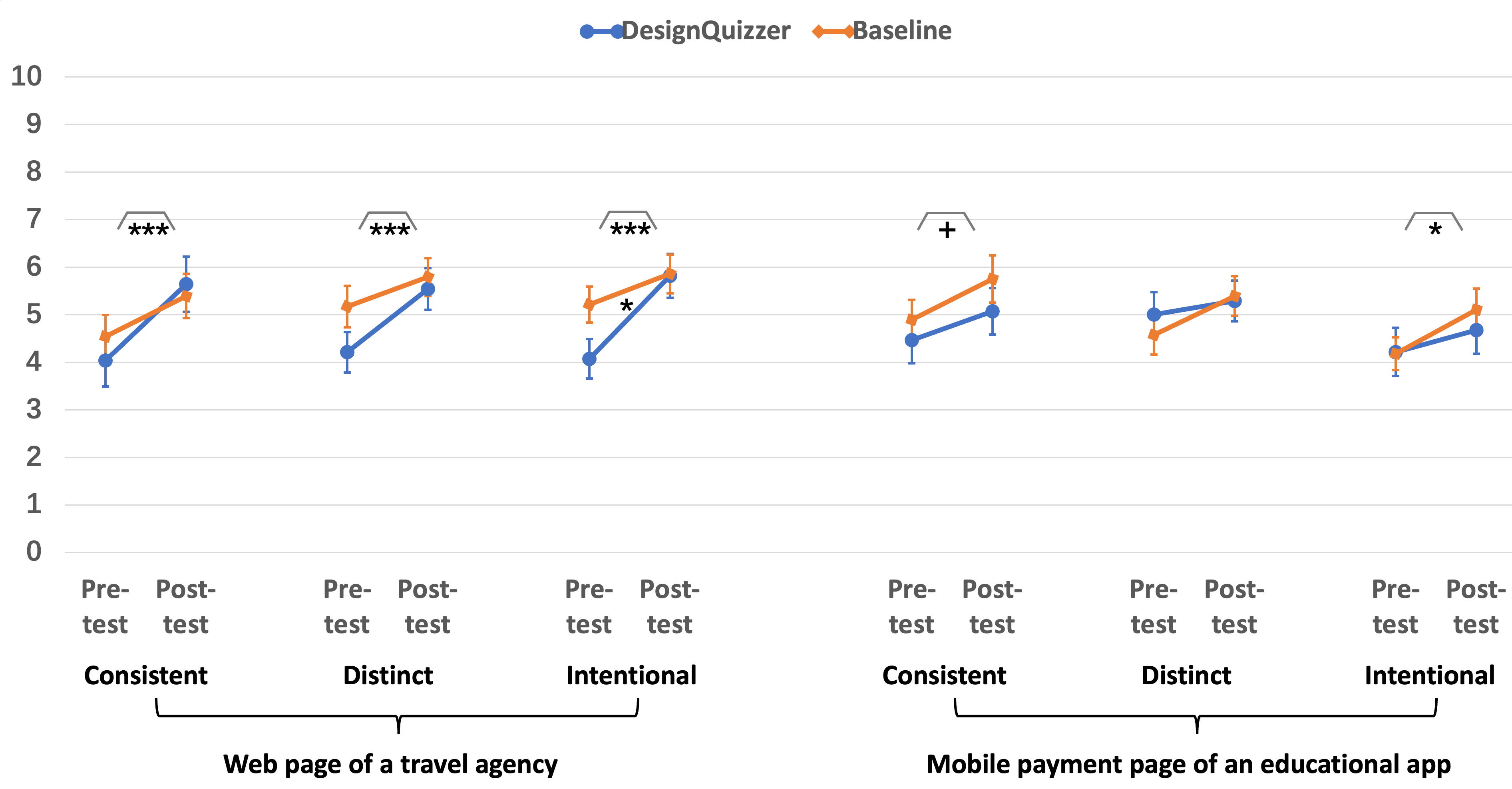}
  \caption{\peng{RQ5b results regarding the changes in participants' performance on the visual design activity before and after the learning session with either DesignQuizzer or baseline interface. In both pre-test and post-test, participants need to color the UI components and adjust the typography of two given uncolored mockups. Two visual design experts rate the designs from aspects of consistency, distinction, and intention, each from 0 to 10 points. $***: p < 0.001, *: p < 0.05, +: p < 0.1.$}}
  \label{fig:application_design}
\end{figure}

(\textbf{RQ5b}) \autoref{fig:application_design} shows the results regarding changes in participants' performance on the visual design activity before and after the learning task. 
For the visual design of a travel agency web page, our results indicate that participants significantly improve their performances in matching the consistent ($F = 21.54, p < 0.001$), distinct ($F = 16.97, p < 0.001$), and intentional ($F = 24.18, p < 0.001$) design principles of color and typography after the learning session. 
The used learning interface does not significantly impact their performance on this web page design task. 
However, we observe a significant interaction effect ($F = 5.18, p = 0.031$) between the used interface and time factors on participants' performance on matching the intentional principle. 
This indicates that the matched degree with the intentional principle of DesignQuizzer users improved significantly more than the baseline users after the learning session. 

For the visual design of a mobile payment page, our results indicate a significant improvement regarding the measure of intention ($F = 5.14, p = 0.032$) in post-test than that in pre-test. 
There is also a tendency that the measure of consistency improves ($F = 3.89, p = 0.059$) after the learning session.
Neither the used interface nor its interaction with the time factor significantly affects the \cqy{participants' performance} in the visual design activity. 

\begin{figure}
  \centering
  \includegraphics[width=\linewidth]{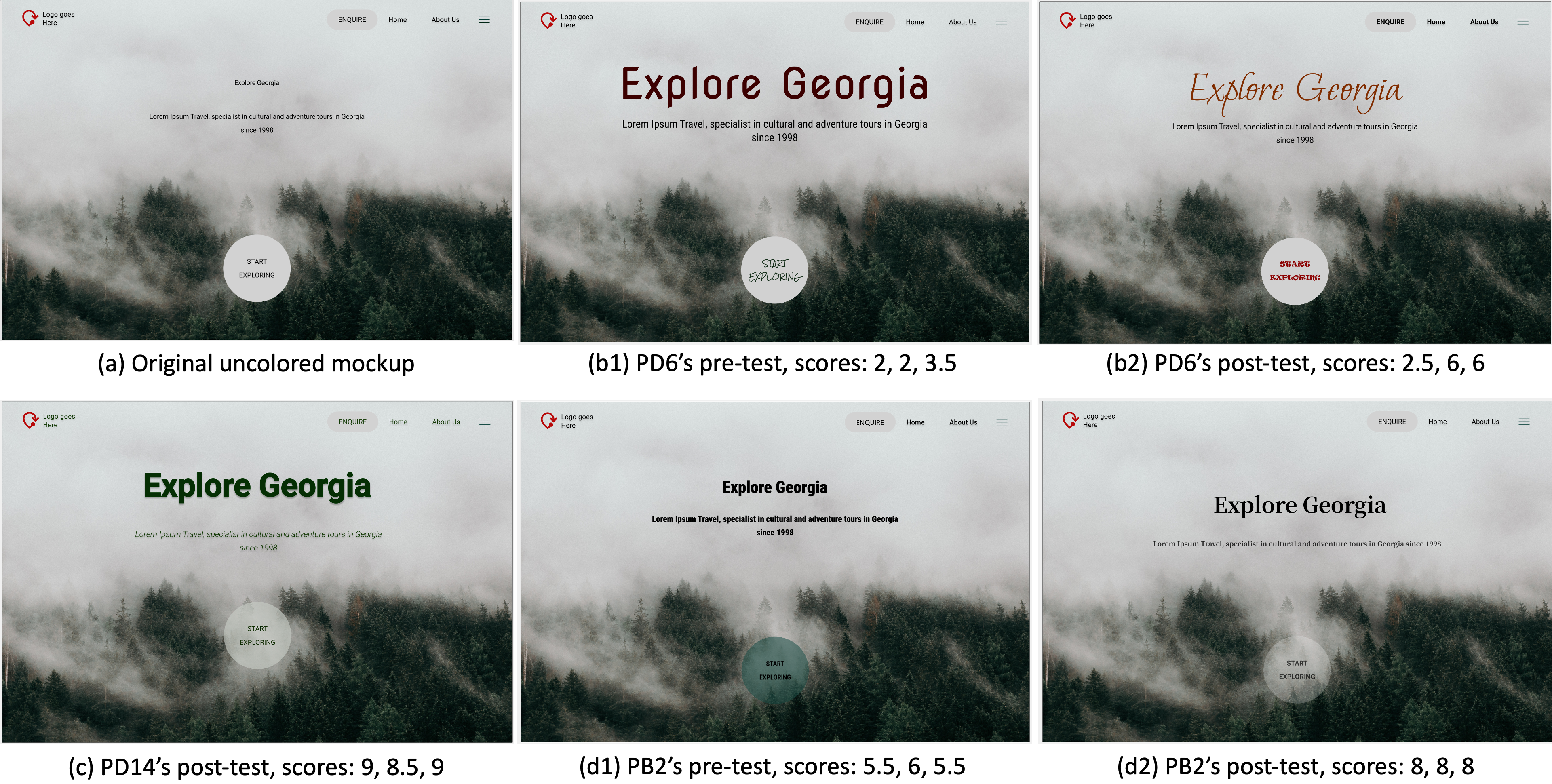}
  \caption{\peng{The original uncolored web page of a travel agency and the sampled outcomes from Participants in DesignQuizzer (note as PD) and Baseline (PB) groups. The average scores given by the two visual design experts are from aspects of consistency, distinction, and intention. This figure is better viewed in color.}}
  \label{fig:web_page_examples}
\end{figure}

\begin{figure}
  \centering
  \includegraphics[width=\linewidth]{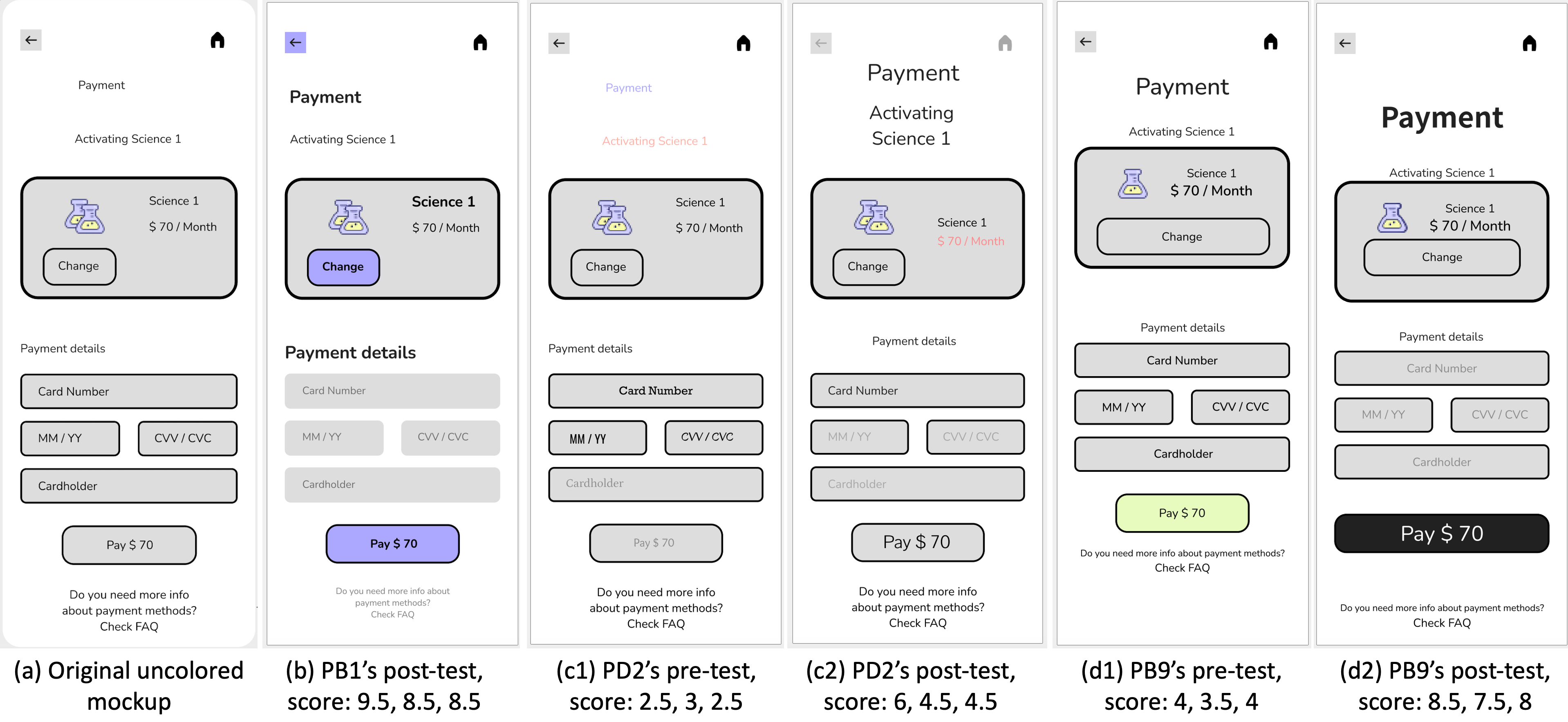}
  \caption{\peng{The original uncolored mobile payment page of an educational app and the sampled outcomes from Participants in DesignQuizzer (note as PD) and Baseline (PB) groups. The average scores given by the two visual design experts are from aspects of consistency, distinction, and intention. This figure is better viewed in color.}}
  \label{fig:mobile_page_examples}
\end{figure}

\autoref{fig:web_page_examples} and \autoref{fig:mobile_page_examples} show the sampled outcomes of the visual design activity. 
For example, PD6 (M, 23) did not perform well in the pre-test of the web page design (\autoref{fig:web_page_examples} b1), as commented by our design expert E1: 
\textit{``The designer tried to create distinction and hierarchy with purpose. Yet, each component has a unique color and font type, making the page very inconsistent to read. The dark red, the black, and the dark green of the button text are in conflict with each other''.} 
In the post-test (\autoref{fig:web_page_examples} b2), PD6 improved his performance in distinction and intention, with the red in the ``START EXPLORATION'' and the bold style in the menu items. 
PB2 (M, 21) in the baseline group also improved his performance in the post-test of mobile page design (\autoref{fig:web_page_examples} d1 and d2). 
Our expert E2 left a comment \cqy{on} PB2's post-test design: 
\textit{``This UI follows the consistency principle well. For example, the fonts of the text in the same module are consistent. The color system of the clickable color blocks is also united. Besides, the distinction between modules (e.g., the hero title and the description text below) is clear''.} 

For the visual design of a mobile payment page, PD2 (M, 21) improved his performance after the learning session (\autoref{fig:mobile_page_examples} c1 and c2). 
E2 judged his pre-test design: 
\textit{``It uses more than four font colors. The font styles are not uniform. What's worse, there is no distinction between titles and subtitles, the meaning of the spacing is unclear, and the meaning of the operation area is confusing''.} 
PB9 (M, 20) in the baseline group also applied what he learned in the community to improve his design performance (\autoref{fig:mobile_page_examples} d1 and d2). 
E2 commented \cqy{on} his post-test design: 
\textit{``This UI design clearly \cqy{distinguishes} each module. Its font and color styles are uniform''.}
}

\peng{
\subsubsection{Visual Design Experts' Feedback on DesignQuizzer}
We interviewed our two visual design experts to obtain their feedback on DesignQuizzer apart from inviting them to score the design outcome. 
We first walked them through \cqy{DesignQuizzer's features}. 
Then, we asked for their comments on the Quizzer's pros/cons for learning design knowledge and the ways to improve it for novices' long-term usages. 
Both experts viewed the DesignQuizzer as a personal design coach for novices and favored its low interaction load. 
\textit{``It looks like a coach of visual design. The conversation-style design with clickable buttons reduces our interaction effort''} (E2). 
However, E1 felt that its quizzes could be too easy for experienced designers. 
E1 suggested that it should involve professionally-generated content, e.g., the open-sourced design framework and the companies' design library, apart from the current user-generated content in the online design community. 
Both experts agreed that our classified visual elements and UI components could help users master design knowledge of interests. 
Nevertheless, to support long-term learning visual design, both experts recommended that DesignQuizzer should have more adaptive learning features. 
\textit{``It should allow users to archive the UI design examples and knowledge points, such that users can review them in the future. I would also expect that it could recommend related design examples and comments when I click a keyword in the current comment''} (E1). 
}

\peng{
\subsubsection{Summary of the Findings in Experiment \uppercase\expandafter{\romannumeral2}} 
Our experiment \uppercase\expandafter{\romannumeral2} provides empirical results on its primary RQ5. 
Both user groups of DesignQuizzer and the community-like baseline learned knowledge points about color and typography via exploring UI design examples and comments. 
In general, participants in both groups perform better in the visual design activity after the learning session. 
These results validate the motivation of this paper that the user-generated content in the online design community is promising to support the learning of visual design. 
We also support the experiment \uppercase\expandafter{\romannumeral1}'s findings that the DesignQuizzer is more engaging and perceived \cqy{as} more useful than the baseline interface for learning design knowledge. 
We additionally collect positive feedback from visual design experts on DesignQuizzer and gain more insights to improve it for novices' long-term usage. 
In the next section, we discuss the findings from our two experiments, \zhenhui{insights} for improving community-driven learning support tools, limitations, and future work. 
}

\section{Discussion}
\label{section:discussion}


In this paper, we have presented a community-powered method for learning elementary design concepts interactively with a chatbot.
The core of our method is a workflow for extracting and organizing comments into interactive quizzes.
\peng{Our two evaluation studies \cqy{support this approach. }
We found that novices using DesignQuizzer can explore helpful UI examples and comments more efficiently than those using the community-like baseline.} 
Participants attributed these benefits to our quiz pool, which leverages the computational workflow to filter out many lower-quality comments in the community discussion \cite{high-quality-cscw2012, high-quality-cscw2014,content_quality_1,structured_feedback}. 

Our results call for more research to study conversational agents for learning in creative domains. Our approach, which increased the level of engagement \zh{and effectiveness} in learning, complements the findings of previous works on educational chatbots, like QuizBot for learning factual knowledge \cite{quizbot10.1145/3290605.3300587}, Sara for learning programming knowledge \cite{sara10.1145/3313831.3376781}, and ArgueTutor for learning augmentation writings \cite{arguetutor}. 
\zh{After the learning sessions, participants with DesignQuizzer can recognize significantly more points of design principles about the learned visual elements that have or have not been used in unseen design examples. 

}
Our computational approach and the design in DesignQuizzer can serve as starting points for similar approaches in other domains, such as for learning about posters and logos \cite{reddit_graphic_design,reddit_design_critique} or programming scripts \cite{informal_learning_chi2022}. 
Researchers can follow our workflow detailed in \autoref{section:3} in order to generate learning materials. 
It first requires a labeled dataset regarding the useful content in online communities, e.g., sentences and keywords about a knowledge concept. 
Researchers can then explore appropriate pre-trained language models and fine-tune them on targeted downstream tasks, e.g., summarization, document/sentence/token classification and clustering, etc. 
%
However, we should be aware of the limitations of these computational models. 
First, they sometimes miss some useful content for learning (e.g., No.6 in \autoref{table:summarization_examples}) and make classification mistakes (e.g., No.5 in \autoref{table:sentence_classification}). 
Second, these models usually lack explanations for their results, e.g., what linguistic features would positively contribute to a meaningful rationale sentence. 
We also offer a promising way to generate a quiz pool to run a conversational agent based on the materials in an online community. 
As detailed in \autoref{section:quiz_pool}, researchers can first curate the publicly available data online and identify the useful content using the developed computational models above. 
Next, they should determine the conversation agent's interaction design and further structure the materials. 
We adopt a quiz-like design similar to QuizBot \cite{quizbot10.1145/3290605.3300587} and generate the single-choice questions by masking the right answers in the critique sentences. 
Other possible interaction strategies include 1) arranging a task and providing performance feedback after task completion \cite{arguetutor} and 2) prompting open-ended questions a critical thinking tasks \cite{peng2022crebot}. 
The later open-ended quizzes, as a survey by Wang et al. suggests, would be more valuable compared to the single-choice question \cite{wang2021seeing}. 



\peng{It is also promising to apply our computational approach to powering design support tools, as anticipated by our participants in experiment \uppercase\expandafter{\romannumeral2}. 
For example, when users want to create a new and unique UI design, they can seek inspiration from the design examples in online communities. 
Our workflow can help them distill the design examples with constructive feedback and insights on specific visual elements or UI components. 
Moreover, the workflow can enable a design support tool to help users reflect on their designs. 
For instance, it can detect the UI components of user designs with computer-vision techniques and retrieve related critiques and suggestions on these components when they appear in similar designs for reference. 
In a word, to promote reflection and active thinking on users' designs, a design support tool could adopt a question-driven, conversational design like DesignQuizzer. 
}

Our findings offer two design considerations for improving the effectiveness and user experience of future community-powered learning support tools.
First, they should incorporate professional knowledge from external resources into the user-generated learning materials in the communities. 
\peng{In the experiment \uppercase\expandafter{\romannumeral2}, both the DesignQuizzer and baseline user groups were able to apply what they learned to improve their performance in the visual design activity. 
This finding highlights the value of user-generated content in design communities as the learning materials for novice designers. 
Nevertheless, the participants in our two experiments pointed out that many comments under UI feedback-request posts often do not provide sufficient explanations for the critiques.}
As a result, learners can observe what is good or bad in the community-generated learning materials but often do not know why. 
Therefore, these tools can consider incorporating knowledge graphs/nets (e.g., ConceptNet \cite{liu2004conceptnet}) and structured documents mined from \cqy{textbooks} or course materials (e.g., \cite{best_practice}) to their knowledge base. 
Second, the community-powered learning support tools should provide more community-level features for users. 
For example, there are comments expressing diverse opinions on the design examples or telling jokes, which help to form a social, relaxing, and collaborative learning environment \cite{gray2019co}. 
The learning support tools should present users with \cqy{this beneficial information}, e.g., 
how many members have similar critiques on this UI example, how many upvotes/likes \cqy{does the meaningful feedback receive}, and how \cqy{do others agree or disagree} with the checked comment. 
The tools can further offer 1) a ``broad'' learning mode that encourages users to think from different aspects by showing competing or unrelated comments and 2) a ``deep'' mode that raises a set of questions on a specific topic (e.g., space) at each learning session.


\peng{Our DesignQuizzer can also incorporate the recent chat-tuned large language models (LLMs, e.g., ChatGPT) to power the dialog-based interaction flow. 
Our work has shown that the rule-based quiz-like dialog flow powered by our computational approaches can achieve good performance in helping users learn visual design. 
To step forward, we should find ways to fine-tune and control chat-tuned LLMs for DesignQuizzer that could converse with and guide the user more open-endedly.
For example, we could design prompts filled by the UI component and visual element keywords to fine-tune LLMs for generating related questions and providing feedback. 
Moreover, the advanced large multimodal  model (e.g., GPT4 \cite{openai2023gpt4}) can help to align the comments to the UI image from online communities (\autoref{table:experiment_2_pros_cons}). 
For instance, we could select a detected UI component in the comment and query GPT4 to highlight the related area of the component in the UI image, which can help users easily locate their interested area.}

Our work has several limitations that call for future work. 
\zh{First, while our computational models have acceptable performance in terms of automatic evaluation metrics, they sometimes make mistakes (e.g., \autoref{table:sentence_classification}), and they can not explicitly identify the relevancy among the critique, rationale, and suggestion sentences. 
Although our participants did not report these issues in the experiment, future work should improve our computational models with more training data.}
We can iteratively enlarge the training dataset by manually refining the machine-generated labels and re-train the models with the updated dataset. 
Second, we did not have a study separating the impact of organized comments and the quiz-like interaction design on our measured items. 
The qualitative feedback in the interviews \peng{in experiment \uppercase\expandafter{\romannumeral1}} suggests that the organized information contributed most to improving efficiency on exploring helpful examples and comments, and that the active question answering helped most with engagement. 
However, understanding the precise contributions of each system feature remains an important issue for future work.
\zh{Third, we evaluated the tools' impact on learners' user experience and outcome in \peng{two lab studies} but not a longer-term study. 
Thus, our empirical results can not expel the novelty effects of people using our DesignQuizzer for the first time. 
\peng{Fourth, our participants are novice designers who have little or no experience in learning UI visual design. It would be interesting to examine how regular designers can learn with DesignQuizzer and the online design community.  
For future work, we suggest measuring the long-term learning experience and effectiveness with users of diverse backgrounds. }} 
\section{Conclusion}
In this paper, we facilitate novices to learn visual elements in online communities by proposing a computational approach for organizing comments and a conversational agent DesignQuizzer. 
We presented methods to extract meaningful feedback from comments on UI designs, classify the feedback into critique, suggestion, and rationale sentences, recognize the keywords about visual elements and UI components in the sentences, and cluster the keywords into groups of higher-level concepts. 
This approach allows DesignQuizzer's quiz-like interaction with users. 
\peng{We compared DesignQuizzer with a community-like baseline interface in a within-subjects experiment \uppercase\expandafter{\romannumeral1} with 24 novices and a between-subjects experiment \uppercase\expandafter{\romannumeral2} with 28 novices. 
The results show that DesignQuizzer significantly improves \cqy{participants' efficiency and engagement} in learning visual design from examples and comments in the communities. 
Participants rated the Quizzer significantly more useful and easier to use for their learning tasks, and they favored its questions for promoting active thinking. 
Participants with DesignQuizzer can apply what they have learned to criticize others' UI designs and enhance their performance in the visual design activity. 
}
We also offered insights for generalizing our approach to other online communities and provided design considerations for improving the effectiveness and user experience of community-powered learning support tools. 
We hope our work will attract more researchers to leverage the resources in online communities to build intelligent learning support systems.

\begin{acks}
This work is supported by the Young Scientists Fund of the National Natural Science Foundation of China with Grant No. 62202509 and partially supported by the Research Grants Council of the Hong Kong Special Administrative Region under General Research Fund (GRF) with Grant No. 16203421. 
Antti Oulasvirta was supported by the Research Council of Finland (flagship program: Finnish Center for Artificial Intelligence, FCAI, grants 328400, 345604, 341763; Human Automata, grant 328813).
\end{acks}

\bibliographystyle{ACM-Reference-Format}
\bibliography{main}


\begin{thebibliography}{109}


\ifx \showCODEN    \undefined \def \showCODEN     #1{\unskip}     \fi
\ifx \showDOI      \undefined \def \showDOI       #1{#1}\fi
\ifx \showISBNx    \undefined \def \showISBNx     #1{\unskip}     \fi
\ifx \showISBNxiii \undefined \def \showISBNxiii  #1{\unskip}     \fi
\ifx \showISSN     \undefined \def \showISSN      #1{\unskip}     \fi
\ifx \showLCCN     \undefined \def \showLCCN      #1{\unskip}     \fi
\ifx \shownote     \undefined \def \shownote      #1{#1}          \fi
\ifx \showarticletitle \undefined \def \showarticletitle #1{#1}   \fi
\ifx \showURL      \undefined \def \showURL       {\relax}        \fi
\providecommand\bibfield[2]{#2}
\providecommand\bibinfo[2]{#2}
\providecommand\natexlab[1]{#1}
\providecommand\showeprint[2][]{arXiv:#2}

\bibitem[\protect\citeauthoryear{??}{red}{2008}]%
        {reddit_graphic_design}
 \bibinfo{year}{2008}\natexlab{}.
\newblock \bibinfo{title}{Reddit r/graphic\_design}.
\newblock
\newblock
\newblock
\shownote{Accessed in September, 2022 from
  \url{https://www.reddit.com/r/graphic_design/}.}


\bibitem[\protect\citeauthoryear{??}{red}{2010}]%
        {reddit_design_critique}
 \bibinfo{year}{2010}\natexlab{}.
\newblock \bibinfo{title}{Reddit r/design\_critiques}.
\newblock
\newblock
\newblock
\shownote{Accessed in September, 2022 from
  \url{https://www.reddit.com/r/design_critiques/}.}


\bibitem[\protect\citeauthoryear{Agichtein, Castillo, Donato, Gionis, and
  Mishne}{Agichtein et~al\mbox{.}}{2008}]%
        {content_quality_1}
\bibfield{author}{\bibinfo{person}{Eugene Agichtein}, \bibinfo{person}{Carlos
  Castillo}, \bibinfo{person}{Debora Donato}, \bibinfo{person}{Aristides
  Gionis}, {and} \bibinfo{person}{Gilad Mishne}.}
  \bibinfo{year}{2008}\natexlab{}.
\newblock \showarticletitle{Finding high-quality content in social media}. In
  \bibinfo{booktitle}{\emph{Proceedings of the 2008 international conference on
  web search and data mining}}. \bibinfo{pages}{183--194}.
\newblock


\bibitem[\protect\citeauthoryear{Alfieri, Nokes-Malach, and Schunn}{Alfieri
  et~al\mbox{.}}{2013}]%
        {case_comparison}
\bibfield{author}{\bibinfo{person}{Louis Alfieri}, \bibinfo{person}{Timothy~J.
  Nokes-Malach}, {and} \bibinfo{person}{Christian~D. Schunn}.}
  \bibinfo{year}{2013}\natexlab{}.
\newblock \showarticletitle{Learning Through Case Comparisons: A Meta-Analytic
  Review}.
\newblock \bibinfo{journal}{\emph{Educational Psychologist}}
  \bibinfo{volume}{48} (\bibinfo{year}{2013}), \bibinfo{pages}{113 -- 87}.
\newblock


\bibitem[\protect\citeauthoryear{Atkinson, Derry, Renkl, and Wortham}{Atkinson
  et~al\mbox{.}}{2000}]%
        {learnfromexamples}
\bibfield{author}{\bibinfo{person}{Robert~K. Atkinson},
  \bibinfo{person}{Sharon~J. Derry}, \bibinfo{person}{Alexander Renkl}, {and}
  \bibinfo{person}{Donald~W. Wortham}.} \bibinfo{year}{2000}\natexlab{}.
\newblock \showarticletitle{Learning from Examples: Instructional Principles
  from the Worked Examples Research}.
\newblock \bibinfo{journal}{\emph{Review of Educational Research}}
  \bibinfo{volume}{70} (\bibinfo{year}{2000}), \bibinfo{pages}{181 -- 214}.
\newblock


\bibitem[\protect\citeauthoryear{Barab and Duffy}{Barab and Duffy}{2012}]%
        {barab2012practice}
\bibfield{author}{\bibinfo{person}{Sasha~A Barab} {and} \bibinfo{person}{Thomas
  Duffy}.} \bibinfo{year}{2012}\natexlab{}.
\newblock \showarticletitle{From practice fields to communities of practice}.
\newblock In \bibinfo{booktitle}{\emph{Theoretical foundations of learning
  environments}}. \bibinfo{publisher}{Routledge}, \bibinfo{pages}{29--65}.
\newblock


\bibitem[\protect\citeauthoryear{Barrett}{Barrett}{1988}]%
        {Barrett1988ACO}
\bibfield{author}{\bibinfo{person}{Terry Barrett}.}
  \bibinfo{year}{1988}\natexlab{}.
\newblock \showarticletitle{A Comparison of The Goals of Studio Professors
  Conducting Critiques and Art Education Goals for Teaching Criticism}.
\newblock


\bibitem[\protect\citeauthoryear{Campbell, Aragon, Davis, Evans, Evans, and
  Randall}{Campbell et~al\mbox{.}}{2016a}]%
        {positive_reviews_cscw2016}
\bibfield{author}{\bibinfo{person}{Julie Campbell}, \bibinfo{person}{Cecilia
  Aragon}, \bibinfo{person}{Katie Davis}, \bibinfo{person}{Sarah Evans},
  \bibinfo{person}{Abigail Evans}, {and} \bibinfo{person}{David Randall}.}
  \bibinfo{year}{2016}\natexlab{a}.
\newblock \showarticletitle{Thousands of Positive Reviews: Distributed
  Mentoring in Online Fan Communities}. In
  \bibinfo{booktitle}{\emph{Proceedings of the 19th ACM Conference on
  Computer-Supported Cooperative Work \& Social Computing}} (San Francisco,
  California, USA) \emph{(\bibinfo{series}{CSCW '16})}.
  \bibinfo{publisher}{Association for Computing Machinery},
  \bibinfo{address}{New York, NY, USA}, \bibinfo{pages}{691–704}.
\newblock
\showISBNx{9781450335928}
\urldef\tempurl%
\url{https://doi.org/10.1145/2818048.2819934}
\showDOI{\tempurl}


\bibitem[\protect\citeauthoryear{Campbell, Aragon, Davis, Evans, Evans, and
  Randall}{Campbell et~al\mbox{.}}{2016b}]%
        {campbell2016thousands}
\bibfield{author}{\bibinfo{person}{Julie Campbell}, \bibinfo{person}{Cecilia
  Aragon}, \bibinfo{person}{Katie Davis}, \bibinfo{person}{Sarah Evans},
  \bibinfo{person}{Abigail Evans}, {and} \bibinfo{person}{David Randall}.}
  \bibinfo{year}{2016}\natexlab{b}.
\newblock \showarticletitle{Thousands of positive reviews: Distributed
  mentoring in online fan communities}. In
  \bibinfo{booktitle}{\emph{Proceedings of the 19th ACM Conference on
  Computer-Supported Cooperative Work \& Social Computing}}.
  \bibinfo{pages}{691--704}.
\newblock


\bibitem[\protect\citeauthoryear{Chandrasekharan, Pavalanathan, Srinivasan,
  Glynn, Eisenstein, and Gilbert}{Chandrasekharan et~al\mbox{.}}{2017}]%
        {chandrasekharan2017you}
\bibfield{author}{\bibinfo{person}{Eshwar Chandrasekharan},
  \bibinfo{person}{Umashanthi Pavalanathan}, \bibinfo{person}{Anirudh
  Srinivasan}, \bibinfo{person}{Adam Glynn}, \bibinfo{person}{Jacob
  Eisenstein}, {and} \bibinfo{person}{Eric Gilbert}.}
  \bibinfo{year}{2017}\natexlab{}.
\newblock \showarticletitle{You can't stay here: The efficacy of reddit's 2015
  ban examined through hate speech}.
\newblock \bibinfo{journal}{\emph{Proceedings of the ACM on Human-Computer
  Interaction}} \bibinfo{volume}{1}, \bibinfo{number}{CSCW}
  (\bibinfo{year}{2017}), \bibinfo{pages}{1--22}.
\newblock


\bibitem[\protect\citeauthoryear{Chaves and Gerosa}{Chaves and Gerosa}{2021}]%
        {chatbot_social_intelligence}
\bibfield{author}{\bibinfo{person}{Ana~Paula Chaves} {and}
  \bibinfo{person}{Marco~Aurelio Gerosa}.} \bibinfo{year}{2021}\natexlab{}.
\newblock \showarticletitle{How Should My Chatbot Interact? A Survey on Social
  Characteristics in Human–Chatbot Interaction Design}.
\newblock \bibinfo{journal}{\emph{International Journal of Human–Computer
  Interaction}} \bibinfo{volume}{37}, \bibinfo{number}{8}
  (\bibinfo{year}{2021}), \bibinfo{pages}{729--758}.
\newblock
\urldef\tempurl%
\url{https://doi.org/10.1080/10447318.2020.1841438}
\showDOI{\tempurl}
\showeprint{https://doi.org/10.1080/10447318.2020.1841438}


\bibitem[\protect\citeauthoryear{Cheng, Dasgupta, and Hill}{Cheng
  et~al\mbox{.}}{2022}]%
        {informal_learning_chi2022}
\bibfield{author}{\bibinfo{person}{Ruijia Cheng}, \bibinfo{person}{Sayamindu
  Dasgupta}, {and} \bibinfo{person}{Benjamin~Mako Hill}.}
  \bibinfo{year}{2022}\natexlab{}.
\newblock \showarticletitle{How Interest-Driven Content Creation Shapes
  Opportunities for Informal Learning in Scratch: A Case Study on Novices’
  Use of Data Structures}. In \bibinfo{booktitle}{\emph{Proceedings of the 2022
  CHI Conference on Human Factors in Computing Systems}} (New Orleans, LA, USA)
  \emph{(\bibinfo{series}{CHI '22})}. \bibinfo{publisher}{Association for
  Computing Machinery}, \bibinfo{address}{New York, NY, USA}, Article
  \bibinfo{articleno}{228}, \bibinfo{numpages}{16}~pages.
\newblock
\showISBNx{9781450391573}
\urldef\tempurl%
\url{https://doi.org/10.1145/3491102.3502124}
\showDOI{\tempurl}


\bibitem[\protect\citeauthoryear{Cheng and Hill}{Cheng and Hill}{2022a}]%
        {ruijia_cscw2022}
\bibfield{author}{\bibinfo{person}{Ruijia Cheng} {and}
  \bibinfo{person}{Benjamin~Mako Hill}.} \bibinfo{year}{2022}\natexlab{a}.
\newblock \showarticletitle{Many Destinations, Many Pathways: A Quantitative
  Analysis of Legitimate Peripheral Participation in Scratch}.
\newblock \bibinfo{journal}{\emph{Proc. ACM Hum.-Comput. Interact.}}
  \bibinfo{volume}{6}, \bibinfo{number}{CSCW2}, Article
  \bibinfo{articleno}{381} (\bibinfo{date}{nov} \bibinfo{year}{2022}),
  \bibinfo{numpages}{26}~pages.
\newblock
\urldef\tempurl%
\url{https://doi.org/10.1145/3555106}
\showDOI{\tempurl}


\bibitem[\protect\citeauthoryear{Cheng and Hill}{Cheng and Hill}{2022b}]%
        {ruijiacscw22}
\bibfield{author}{\bibinfo{person}{Ruijia Cheng} {and}
  \bibinfo{person}{Benjamin~Mako Hill}.} \bibinfo{year}{2022}\natexlab{b}.
\newblock \showarticletitle{Many Destinations, Many Pathways: A Quantitative
  Analysis of Legitimate Peripheral Participation in Scratch}.
\newblock \bibinfo{journal}{\emph{Proc. ACM Hum.-Comput. Interact.}}
  \bibinfo{volume}{6}, \bibinfo{number}{CSCW2}, Article
  \bibinfo{articleno}{381} (\bibinfo{date}{nov} \bibinfo{year}{2022}),
  \bibinfo{numpages}{26}~pages.
\newblock
\urldef\tempurl%
\url{https://doi.org/10.1145/3555106}
\showDOI{\tempurl}


\bibitem[\protect\citeauthoryear{Cheng and Zachry}{Cheng and Zachry}{2020}]%
        {ruijia_cscw2020}
\bibfield{author}{\bibinfo{person}{Ruijia Cheng} {and} \bibinfo{person}{Mark
  Zachry}.} \bibinfo{year}{2020}\natexlab{}.
\newblock \showarticletitle{Building Community Knowledge In Online
  Competitions: Motivation, Practices and Challenges}.
\newblock \bibinfo{journal}{\emph{Proc. ACM Hum.-Comput. Interact.}}
  \bibinfo{volume}{4}, \bibinfo{number}{CSCW2}, Article
  \bibinfo{articleno}{179} (\bibinfo{date}{oct} \bibinfo{year}{2020}),
  \bibinfo{numpages}{22}~pages.
\newblock
\urldef\tempurl%
\url{https://doi.org/10.1145/3415250}
\showDOI{\tempurl}


\bibitem[\protect\citeauthoryear{Cheng, Zeng, Liu, and Dow}{Cheng
  et~al\mbox{.}}{2020a}]%
        {critique_me}
\bibfield{author}{\bibinfo{person}{Ruijia Cheng}, \bibinfo{person}{Ziwen Zeng},
  \bibinfo{person}{Maysnow Liu}, {and} \bibinfo{person}{Steven Dow}.}
  \bibinfo{year}{2020}\natexlab{a}.
\newblock \showarticletitle{Critique Me: Exploring How Creators Publicly
  Request Feedback in an Online Critique Community}.
\newblock \bibinfo{journal}{\emph{Proc. ACM Hum.-Comput. Interact.}}
  \bibinfo{volume}{4}, \bibinfo{number}{CSCW2}, Article
  \bibinfo{articleno}{161} (\bibinfo{date}{oct} \bibinfo{year}{2020}),
  \bibinfo{numpages}{24}~pages.
\newblock
\urldef\tempurl%
\url{https://doi.org/10.1145/3415232}
\showDOI{\tempurl}


\bibitem[\protect\citeauthoryear{Cheng, Zeng, Liu, and Dow}{Cheng
  et~al\mbox{.}}{2020b}]%
        {cheng2020critique}
\bibfield{author}{\bibinfo{person}{Ruijia Cheng}, \bibinfo{person}{Ziwen Zeng},
  \bibinfo{person}{Maysnow Liu}, {and} \bibinfo{person}{Steven Dow}.}
  \bibinfo{year}{2020}\natexlab{b}.
\newblock \showarticletitle{Critique Me: Exploring How Creators Publicly
  Request Feedback in an Online Critique Community}.
\newblock \bibinfo{journal}{\emph{Proceedings of the ACM on Human-Computer
  Interaction}}  \bibinfo{volume}{4}, Article \bibinfo{articleno}{161}
  (\bibinfo{year}{2020}), \bibinfo{numpages}{24}~pages.
\newblock
Issue CSCW2.


\bibitem[\protect\citeauthoryear{Chi and Wylie}{Chi and Wylie}{2014}]%
        {ICAP_framework}
\bibfield{author}{\bibinfo{person}{Michelene T.~H. Chi} {and}
  \bibinfo{person}{Ruth Wylie}.} \bibinfo{year}{2014}\natexlab{}.
\newblock \showarticletitle{The ICAP Framework: Linking Cognitive Engagement to
  Active Learning Outcomes}.
\newblock \bibinfo{journal}{\emph{Educational Psychologist}}
  \bibinfo{volume}{49}, \bibinfo{number}{4} (\bibinfo{year}{2014}),
  \bibinfo{pages}{219--243}.
\newblock
\urldef\tempurl%
\url{https://doi.org/10.1080/00461520.2014.965823}
\showDOI{\tempurl}
\showeprint{https://doi.org/10.1080/00461520.2014.965823}


\bibitem[\protect\citeauthoryear{Chris~Bank}{Chris~Bank}{[n.d.]}]%
        {best_practice}
\bibfield{author}{\bibinfo{person}{Jerry~Cao Chris~Bank}.}
  \bibinfo{year}{[n.d.]}\natexlab{}.
\newblock \bibinfo{title}{Web UI Design Best Practices: UI Design From The
  Experts}.
\newblock
\newblock
\newblock
\shownote{Accessed on September 2022,
  \url{https://www.uxpin.com/studio/ebooks/web-ui-design-best-practices/}.}


\bibitem[\protect\citeauthoryear{Crain and Bailey}{Crain and Bailey}{2017}]%
        {crain2017share}
\bibfield{author}{\bibinfo{person}{Patrick~A Crain} {and}
  \bibinfo{person}{Brian~P Bailey}.} \bibinfo{year}{2017}\natexlab{}.
\newblock \showarticletitle{Share Once or Share Often? Exploring How Designers
  Approach Iteration in a Large Online Community}. \bibinfo{publisher}{ACM},
  \bibinfo{address}{New York, NY, USA}, \bibinfo{pages}{80--92}.
\newblock


\bibitem[\protect\citeauthoryear{Csikzentmihaly}{Csikzentmihaly}{1990}]%
        {csikszentmihalyi1990flow}
\bibfield{author}{\bibinfo{person}{Mihaly Csikzentmihaly}.}
  \bibinfo{year}{1990}\natexlab{}.
\newblock \bibinfo{booktitle}{\emph{Flow: The psychology of optimal
  experience}}. Vol.~\bibinfo{volume}{1990}.
\newblock \bibinfo{publisher}{Harper \& Row, New York, USA}.
\newblock


\bibitem[\protect\citeauthoryear{Dasgupta, Hale, Monroy-Hern{\'a}ndez, and
  Hill}{Dasgupta et~al\mbox{.}}{2016}]%
        {dasgupta2016remixing}
\bibfield{author}{\bibinfo{person}{Sayamindu Dasgupta},
  \bibinfo{person}{William Hale}, \bibinfo{person}{Andr{\'e}s
  Monroy-Hern{\'a}ndez}, {and} \bibinfo{person}{Benjamin~Mako Hill}.}
  \bibinfo{year}{2016}\natexlab{}.
\newblock \showarticletitle{Remixing as a pathway to computational thinking}.
  In \bibinfo{booktitle}{\emph{Proceedings of the 19th ACM Conference on
  Computer-Supported Cooperative Work \& Social Computing}}.
  \bibinfo{pages}{1438--1449}.
\newblock


\bibitem[\protect\citeauthoryear{Dasgupta and Hill}{Dasgupta and Hill}{2018}]%
        {dasgupta2018wide}
\bibfield{author}{\bibinfo{person}{Sayamindu Dasgupta} {and}
  \bibinfo{person}{Benjamin~Mako Hill}.} \bibinfo{year}{2018}\natexlab{}.
\newblock \showarticletitle{How “wide walls” can increase engagement:
  evidence from a natural experiment in Scratch}. In
  \bibinfo{booktitle}{\emph{Proceedings of the 2018 CHI Conference on Human
  Factors in Computing Systems}}. \bibinfo{pages}{1--11}.
\newblock


\bibitem[\protect\citeauthoryear{Devlin, Chang, Lee, and Toutanova}{Devlin
  et~al\mbox{.}}{2019}]%
        {bert_devlin-etal-2019-bert}
\bibfield{author}{\bibinfo{person}{Jacob Devlin}, \bibinfo{person}{Ming-Wei
  Chang}, \bibinfo{person}{Kenton Lee}, {and} \bibinfo{person}{Kristina
  Toutanova}.} \bibinfo{year}{2019}\natexlab{}.
\newblock \showarticletitle{{BERT}: Pre-training of Deep Bidirectional
  Transformers for Language Understanding}. In
  \bibinfo{booktitle}{\emph{Proceedings of the 2019 Conference of the North
  {A}merican Chapter of the Association for Computational Linguistics: Human
  Language Technologies, Volume 1 (Long and Short Papers)}}.
  \bibinfo{publisher}{Association for Computational Linguistics},
  \bibinfo{address}{Minneapolis, Minnesota}, \bibinfo{pages}{4171--4186}.
\newblock
\urldef\tempurl%
\url{https://doi.org/10.18653/v1/N19-1423}
\showDOI{\tempurl}


\bibitem[\protect\citeauthoryear{Divekar, Lepp, Chopade, Albin, Brenner, and
  Ramanarayanan}{Divekar et~al\mbox{.}}{2021}]%
        {divekar2021conversational}
\bibfield{author}{\bibinfo{person}{Rahul~R Divekar}, \bibinfo{person}{Haley
  Lepp}, \bibinfo{person}{Pravin Chopade}, \bibinfo{person}{Aaron Albin},
  \bibinfo{person}{Daniel Brenner}, {and} \bibinfo{person}{Vikram
  Ramanarayanan}.} \bibinfo{year}{2021}\natexlab{}.
\newblock \showarticletitle{Conversational Agents in Language Education: Where
  They Fit and Their Research Challenges}. In
  \bibinfo{booktitle}{\emph{International Conference on Human-Computer
  Interaction}}. Springer, \bibinfo{pages}{272--279}.
\newblock


\bibitem[\protect\citeauthoryear{doccano}{doccano}{2022}]%
        {annotation_tool}
\bibfield{author}{\bibinfo{person}{doccano}.} \bibinfo{year}{2022}\natexlab{}.
\newblock \bibinfo{title}{Text Annotation for Humans}.
\newblock
\newblock
\newblock
\shownote{Accessed in September, 2022 from
  \url{https://doccano.herokuapp.com/}.}


\bibitem[\protect\citeauthoryear{Dow, Kulkarni, Klemmer, and Hartmann}{Dow
  et~al\mbox{.}}{2012}]%
        {dow2012shepherding}
\bibfield{author}{\bibinfo{person}{Steven Dow}, \bibinfo{person}{Anand
  Kulkarni}, \bibinfo{person}{Scott Klemmer}, {and} \bibinfo{person}{Bj{\"o}rn
  Hartmann}.} \bibinfo{year}{2012}\natexlab{}.
\newblock \showarticletitle{Shepherding the crowd yields better work}. In
  \bibinfo{booktitle}{\emph{Proceedings of the ACM 2012 conference on computer
  supported cooperative work}}. \bibinfo{pages}{1013--1022}.
\newblock


\bibitem[\protect\citeauthoryear{Face}{Face}{2022a}]%
        {huggingface_sentence_transformers}
\bibfield{author}{\bibinfo{person}{Hugging Face}.}
  \bibinfo{year}{2022}\natexlab{a}.
\newblock \bibinfo{title}{Sentence Transformers}.
\newblock
\newblock
\newblock
\shownote{Accessed in September, 2022 from
  \url{https://huggingface.co/sentence-transformers}.}


\bibitem[\protect\citeauthoryear{Face}{Face}{2022b}]%
        {huggingface_summarization}
\bibfield{author}{\bibinfo{person}{Hugging Face}.}
  \bibinfo{year}{2022}\natexlab{b}.
\newblock \bibinfo{title}{Summarization}.
\newblock
\newblock
\newblock
\shownote{Accessed in September, 2022 from
  \url{https://huggingface.co/docs/transformers/tasks/summarization}.}


\bibitem[\protect\citeauthoryear{Face}{Face}{2022c}]%
        {huggingface_token_classification}
\bibfield{author}{\bibinfo{person}{Hugging Face}.}
  \bibinfo{year}{2022}\natexlab{c}.
\newblock \bibinfo{title}{Token Classification}.
\newblock
\newblock
\newblock
\shownote{Accessed in September, 2022 from
  \url{https://huggingface.co/course/chapter7/2?fw=pt}.}


\bibitem[\protect\citeauthoryear{Feldman}{Feldman}{1994}]%
        {feldman1994practical}
\bibfield{author}{\bibinfo{person}{Edmund~Burke Feldman}.}
  \bibinfo{year}{1994}\natexlab{}.
\newblock \bibinfo{booktitle}{\emph{Practical art criticism}}.
\newblock \bibinfo{publisher}{Pearson}.
\newblock


\bibitem[\protect\citeauthoryear{Foong, Dow, Bailey, and Gerber}{Foong
  et~al\mbox{.}}{2017}]%
        {online_feedback_exchange}
\bibfield{author}{\bibinfo{person}{Eureka Foong}, \bibinfo{person}{Steven~P.
  Dow}, \bibinfo{person}{Brian~P. Bailey}, {and} \bibinfo{person}{Elizabeth~M.
  Gerber}.} \bibinfo{year}{2017}\natexlab{}.
\newblock \showarticletitle{Online Feedback Exchange: A Framework for
  Understanding the Socio-Psychological Factors}. In
  \bibinfo{booktitle}{\emph{Proceedings of the 2017 CHI Conference on Human
  Factors in Computing Systems}} (Denver, Colorado, USA)
  \emph{(\bibinfo{series}{CHI '17})}. \bibinfo{publisher}{Association for
  Computing Machinery}, \bibinfo{address}{New York, NY, USA},
  \bibinfo{pages}{4454–4467}.
\newblock
\showISBNx{9781450346559}
\urldef\tempurl%
\url{https://doi.org/10.1145/3025453.3025791}
\showDOI{\tempurl}


\bibitem[\protect\citeauthoryear{Ford, Lustig, Banks, and Parnin}{Ford
  et~al\mbox{.}}{2018}]%
        {QA_chi2018}
\bibfield{author}{\bibinfo{person}{Denae Ford}, \bibinfo{person}{Kristina
  Lustig}, \bibinfo{person}{Jeremy Banks}, {and} \bibinfo{person}{Chris
  Parnin}.} \bibinfo{year}{2018}\natexlab{}.
\newblock \showarticletitle{"We Don't Do That Here": How Collaborative Editing
  with Mentors Improves Engagement in Social Q\&A Communities}. In
  \bibinfo{booktitle}{\emph{Proceedings of the 2018 CHI Conference on Human
  Factors in Computing Systems}} (Montreal QC, Canada)
  \emph{(\bibinfo{series}{CHI '18})}. \bibinfo{publisher}{Association for
  Computing Machinery}, \bibinfo{address}{New York, NY, USA},
  \bibinfo{pages}{1–12}.
\newblock
\showISBNx{9781450356206}
\urldef\tempurl%
\url{https://doi.org/10.1145/3173574.3174182}
\showDOI{\tempurl}


\bibitem[\protect\citeauthoryear{Frens, Walker, and Hsieh}{Frens
  et~al\mbox{.}}{2018}]%
        {frens2018supporting}
\bibfield{author}{\bibinfo{person}{John Frens}, \bibinfo{person}{Erin Walker},
  {and} \bibinfo{person}{Gary Hsieh}.} \bibinfo{year}{2018}\natexlab{}.
\newblock \showarticletitle{Supporting answerers with feedback in social Q\&A}.
  In \bibinfo{booktitle}{\emph{Proceedings of the Fifth Annual ACM Conference
  on Learning at Scale}}. \bibinfo{pages}{1--10}.
\newblock


\bibitem[\protect\citeauthoryear{Gan, Hill, and Dasgupta}{Gan
  et~al\mbox{.}}{2018}]%
        {gender_learner_community}
\bibfield{author}{\bibinfo{person}{Emilia~F. Gan},
  \bibinfo{person}{Benjamin~Mako Hill}, {and} \bibinfo{person}{Sayamindu
  Dasgupta}.} \bibinfo{year}{2018}\natexlab{}.
\newblock \showarticletitle{Gender, Feedback, and Learners' Decisions to Share
  Their Creative Computing Projects}.
\newblock \bibinfo{journal}{\emph{Proc. ACM Hum.-Comput. Interact.}}
  \bibinfo{volume}{2}, \bibinfo{number}{CSCW}, Article \bibinfo{articleno}{54}
  (\bibinfo{date}{nov} \bibinfo{year}{2018}), \bibinfo{numpages}{23}~pages.
\newblock
\urldef\tempurl%
\url{https://doi.org/10.1145/3274323}
\showDOI{\tempurl}


\bibitem[\protect\citeauthoryear{Gentner, Loewenstein, and Thompson}{Gentner
  et~al\mbox{.}}{2003}]%
        {learning_and_transfer}
\bibfield{author}{\bibinfo{person}{Dedre Gentner}, \bibinfo{person}{Jeffrey
  Loewenstein}, {and} \bibinfo{person}{Leigh Thompson}.}
  \bibinfo{year}{2003}\natexlab{}.
\newblock \showarticletitle{Learning and Transfer: A General Role for
  Analogical Encoding}.
\newblock \bibinfo{journal}{\emph{Journal of Educational Psychology}}
  \bibinfo{volume}{95} (\bibinfo{year}{2003}), \bibinfo{pages}{393--408}.
\newblock


\bibitem[\protect\citeauthoryear{Google}{Google}{2022}]%
        {materialdesign}
\bibfield{author}{\bibinfo{person}{Google}.} \bibinfo{year}{2022}\natexlab{}.
\newblock \bibinfo{title}{Material Design}.
\newblock
\newblock
\newblock
\shownote{Accessed in September, 2022 from
  \url{https://material.io/components}.}


\bibitem[\protect\citeauthoryear{Gorvine, Rosengren, Stein, and Biolsi}{Gorvine
  et~al\mbox{.}}{2017}]%
        {statistic_test_guide}
\bibfield{author}{\bibinfo{person}{Ben Gorvine}, \bibinfo{person}{Karl
  Rosengren}, \bibinfo{person}{Lisa Stein}, {and} \bibinfo{person}{Kevin
  Biolsi}.} \bibinfo{year}{2017}\natexlab{}.
\newblock \bibinfo{booktitle}{\emph{Research Methods: From Theory to Practice -
  Chapter 14: Analyzing Your Data II: Specific Approaches Inside Research}}.
\newblock \bibinfo{publisher}{Oxford University Press}.
\newblock
\showISBNx{978-0190201821}
\urldef\tempurl%
\url{https://global.oup.com/us/companion.websites/9780190201821/sr/outline/ch14/}
\showURL{%
\tempurl}


\bibitem[\protect\citeauthoryear{Gray and Kou}{Gray and Kou}{2019}]%
        {gray2019co}
\bibfield{author}{\bibinfo{person}{Colin~M Gray} {and} \bibinfo{person}{Yubo
  Kou}.} \bibinfo{year}{2019}\natexlab{}.
\newblock \showarticletitle{Co-producing, curating, and defining design
  knowledge in an online practitioner community}.
\newblock \bibinfo{journal}{\emph{CoDesign}} \bibinfo{volume}{15},
  \bibinfo{number}{1} (\bibinfo{year}{2019}), \bibinfo{pages}{41--58}.
\newblock


\bibitem[\protect\citeauthoryear{Halfaker, Keyes, and Taraborelli}{Halfaker
  et~al\mbox{.}}{2013}]%
        {Peripheral_cscw2013}
\bibfield{author}{\bibinfo{person}{Aaron Halfaker}, \bibinfo{person}{Os Keyes},
  {and} \bibinfo{person}{Dario Taraborelli}.} \bibinfo{year}{2013}\natexlab{}.
\newblock \showarticletitle{Making Peripheral Participation Legitimate: Reader
  Engagement Experiments in Wikipedia}. In
  \bibinfo{booktitle}{\emph{Proceedings of the 2013 Conference on Computer
  Supported Cooperative Work}} (San Antonio, Texas, USA)
  \emph{(\bibinfo{series}{CSCW '13})}. \bibinfo{publisher}{Association for
  Computing Machinery}, \bibinfo{address}{New York, NY, USA},
  \bibinfo{pages}{849–860}.
\newblock
\showISBNx{9781450313315}
\urldef\tempurl%
\url{https://doi.org/10.1145/2441776.2441872}
\showDOI{\tempurl}


\bibitem[\protect\citeauthoryear{Hart and Staveland}{Hart and
  Staveland}{1988}]%
        {nasatlx}
\bibfield{author}{\bibinfo{person}{Sandra~G. Hart} {and}
  \bibinfo{person}{Lowell~E. Staveland}.} \bibinfo{year}{1988}\natexlab{}.
\newblock \showarticletitle{Development of NASA-TLX (Task Load Index): Results
  of Empirical and Theoretical Research}.
\newblock In \bibinfo{booktitle}{\emph{Human Mental Workload}},
  \bibfield{editor}{\bibinfo{person}{Peter~A. Hancock} {and}
  \bibinfo{person}{Najmedin Meshkati}} (Eds.). \bibinfo{series}{Advances in
  Psychology}, Vol.~\bibinfo{volume}{52}. \bibinfo{publisher}{North-Holland},
  \bibinfo{pages}{139--183}.
\newblock
\showISSN{0166-4115}
\urldef\tempurl%
\url{https://doi.org/10.1016/S0166-4115(08)62386-9}
\showDOI{\tempurl}


\bibitem[\protect\citeauthoryear{Hegemann, Dayama, Iyer, Farhadi, Marchenko,
  and Oulasvirta}{Hegemann et~al\mbox{.}}{2023}]%
        {cocolor_iui2023}
\bibfield{author}{\bibinfo{person}{Lena Hegemann},
  \bibinfo{person}{Niraj~Ramesh Dayama}, \bibinfo{person}{Abhishek Iyer},
  \bibinfo{person}{Erfan Farhadi}, \bibinfo{person}{Ekaterina Marchenko}, {and}
  \bibinfo{person}{Antti Oulasvirta}.} \bibinfo{year}{2023}\natexlab{}.
\newblock \showarticletitle{CoColor: Interactive Exploration of Color Designs}.
  In \bibinfo{booktitle}{\emph{Proceedings of the 28th International Conference
  on Intelligent User Interfaces}} (Sydney, NSW, Australia)
  \emph{(\bibinfo{series}{IUI '23})}. \bibinfo{publisher}{Association for
  Computing Machinery}, \bibinfo{address}{New York, NY, USA},
  \bibinfo{pages}{106–127}.
\newblock
\showISBNx{9798400701061}
\urldef\tempurl%
\url{https://doi.org/10.1145/3581641.3584089}
\showDOI{\tempurl}


\bibitem[\protect\citeauthoryear{Herring, Chang, Krantzler, and Bailey}{Herring
  et~al\mbox{.}}{2009}]%
        {getting_inspire}
\bibfield{author}{\bibinfo{person}{Scarlett~R. Herring},
  \bibinfo{person}{Chia-Chen Chang}, \bibinfo{person}{Jesse Krantzler}, {and}
  \bibinfo{person}{Brian~P. Bailey}.} \bibinfo{year}{2009}\natexlab{}.
\newblock \showarticletitle{Getting Inspired! Understanding How and Why
  Examples Are Used in Creative Design Practice}. In
  \bibinfo{booktitle}{\emph{Proceedings of the SIGCHI Conference on Human
  Factors in Computing Systems}} (Boston, MA, USA) \emph{(\bibinfo{series}{CHI
  '09})}. \bibinfo{publisher}{Association for Computing Machinery},
  \bibinfo{address}{New York, NY, USA}, \bibinfo{pages}{87–96}.
\newblock
\showISBNx{9781605582467}
\urldef\tempurl%
\url{https://doi.org/10.1145/1518701.1518717}
\showDOI{\tempurl}


\bibitem[\protect\citeauthoryear{Hui, Gerber, and Dow}{Hui
  et~al\mbox{.}}{2014}]%
        {hui2014crowd}
\bibfield{author}{\bibinfo{person}{Julie~S Hui}, \bibinfo{person}{Elizabeth~M
  Gerber}, {and} \bibinfo{person}{Steven~P Dow}.}
  \bibinfo{year}{2014}\natexlab{}.
\newblock \showarticletitle{Crowd-based design activities: helping students
  connect with users online}. In \bibinfo{booktitle}{\emph{Proceedings of the
  2014 conference on Designing Interactive Systems}}.
  \bibinfo{pages}{875--884}.
\newblock


\bibitem[\protect\citeauthoryear{Kang, Amoako, Sengupta, and Dow}{Kang
  et~al\mbox{.}}{2018}]%
        {kang2018paragon}
\bibfield{author}{\bibinfo{person}{Hyeonsu~B Kang}, \bibinfo{person}{Gabriel
  Amoako}, \bibinfo{person}{Neil Sengupta}, {and} \bibinfo{person}{Steven~P
  Dow}.} \bibinfo{year}{2018}\natexlab{}.
\newblock \showarticletitle{Paragon: An online gallery for enhancing design
  feedback with visual examples}. In \bibinfo{booktitle}{\emph{Proceedings of
  the 2018 CHI Conference on Human Factors in Computing Systems}}.
  \bibinfo{pages}{1--13}.
\newblock


\bibitem[\protect\citeauthoryear{Kang, Sun, Wang, Huang, Wu, and Ma}{Kang
  et~al\mbox{.}}{2021}]%
        {MetaMap}
\bibfield{author}{\bibinfo{person}{Youwen Kang}, \bibinfo{person}{Zhida Sun},
  \bibinfo{person}{Sitong Wang}, \bibinfo{person}{Zeyu Huang},
  \bibinfo{person}{Ziming Wu}, {and} \bibinfo{person}{Xiaojuan Ma}.}
  \bibinfo{year}{2021}\natexlab{}.
\newblock \showarticletitle{MetaMap: Supporting Visual Metaphor Ideation
  through Multi-Dimensional Example-Based Exploration}. In
  \bibinfo{booktitle}{\emph{Proceedings of the 2021 CHI Conference on Human
  Factors in Computing Systems}} (Yokohama, Japan) \emph{(\bibinfo{series}{CHI
  '21})}. \bibinfo{publisher}{Association for Computing Machinery},
  \bibinfo{address}{New York, NY, USA}, Article \bibinfo{articleno}{427},
  \bibinfo{numpages}{15}~pages.
\newblock
\showISBNx{9781450380966}
\urldef\tempurl%
\url{https://doi.org/10.1145/3411764.3445325}
\showDOI{\tempurl}


\bibitem[\protect\citeauthoryear{Keikha, Park, and Croft}{Keikha
  et~al\mbox{.}}{2014}]%
        {keikha2014evaluating}
\bibfield{author}{\bibinfo{person}{Mostafa Keikha}, \bibinfo{person}{Jae~Hyun
  Park}, {and} \bibinfo{person}{W~Bruce Croft}.}
  \bibinfo{year}{2014}\natexlab{}.
\newblock \showarticletitle{Evaluating answer passages using summarization
  measures}. In \bibinfo{booktitle}{\emph{Proceedings of the 37th international
  ACM SIGIR conference on Research \& development in information retrieval}}.
  \bibinfo{pages}{963--966}.
\newblock


\bibitem[\protect\citeauthoryear{Kim, Choi, Choi, and Kim}{Kim
  et~al\mbox{.}}{2022}]%
        {kin2022stylette}
\bibfield{author}{\bibinfo{person}{Tae~Soo Kim}, \bibinfo{person}{DaEun Choi},
  \bibinfo{person}{Yoonseo Choi}, {and} \bibinfo{person}{Juho Kim}.}
  \bibinfo{year}{2022}\natexlab{}.
\newblock \showarticletitle{Stylette: Styling the Web with Natural Language}.
  In \bibinfo{booktitle}{\emph{Proceedings of the 2022 CHI Conference on Human
  Factors in Computing Systems}} \emph{(\bibinfo{series}{CHI '22})}.
  \bibinfo{publisher}{Association for Computing Machinery},
  \bibinfo{address}{New York, NY, USA}, Article \bibinfo{articleno}{5},
  \bibinfo{numpages}{17}~pages.
\newblock
\showISBNx{9781450391573}
\urldef\tempurl%
\url{https://doi.org/10.1145/3491102.3501931}
\showDOI{\tempurl}


\bibitem[\protect\citeauthoryear{Koch, Laszlo, Lucero, and Oulasvirta}{Koch
  et~al\mbox{.}}{2018}]%
        {koch2018}
\bibfield{author}{\bibinfo{person}{Janin Koch}, \bibinfo{person}{Magda Laszlo},
  \bibinfo{person}{Andres Lucero}, {and} \bibinfo{person}{Antti Oulasvirta}.}
  \bibinfo{year}{2018}\natexlab{}.
\newblock \showarticletitle{Surfing for Inspiration: digital inspirational
  material in design practice}. In \bibinfo{booktitle}{\emph{Design Research
  Society International Conference}}. Design Research Society,
  \bibinfo{pages}{1247--1260}.
\newblock


\bibitem[\protect\citeauthoryear{Kou and Gray}{Kou and Gray}{2017}]%
        {kou2017supporting}
\bibfield{author}{\bibinfo{person}{Yubo Kou} {and} \bibinfo{person}{Colin~M
  Gray}.} \bibinfo{year}{2017}\natexlab{}.
\newblock \showarticletitle{Supporting distributed critique through
  interpretation and sense-making in an online creative community}.
\newblock \bibinfo{journal}{\emph{Proceedings of the ACM on Human-Computer
  Interaction}} \bibinfo{volume}{1}, \bibinfo{number}{CSCW}
  (\bibinfo{year}{2017}), \bibinfo{pages}{1--18}.
\newblock


\bibitem[\protect\citeauthoryear{Krause, Garncarz, Song, Gerber, Bailey, and
  Dow}{Krause et~al\mbox{.}}{2017}]%
        {critique_style_guide}
\bibfield{author}{\bibinfo{person}{Markus Krause}, \bibinfo{person}{Tom
  Garncarz}, \bibinfo{person}{JiaoJiao Song}, \bibinfo{person}{Elizabeth~M.
  Gerber}, \bibinfo{person}{Brian~P. Bailey}, {and} \bibinfo{person}{Steven~P.
  Dow}.} \bibinfo{year}{2017}\natexlab{}.
\newblock \showarticletitle{Critique Style Guide: Improving Crowdsourced Design
  Feedback with a Natural Language Model}. In
  \bibinfo{booktitle}{\emph{Proceedings of the 2017 CHI Conference on Human
  Factors in Computing Systems}} (Denver, Colorado, USA)
  \emph{(\bibinfo{series}{CHI '17})}. \bibinfo{publisher}{Association for
  Computing Machinery}, \bibinfo{address}{New York, NY, USA},
  \bibinfo{pages}{4627–4639}.
\newblock
\showISBNx{9781450346559}
\urldef\tempurl%
\url{https://doi.org/10.1145/3025453.3025883}
\showDOI{\tempurl}


\bibitem[\protect\citeauthoryear{Krishna~Kumaran, McDonagh, and
  Bailey}{Krishna~Kumaran et~al\mbox{.}}{2017}]%
        {increase_quality_cscw2017}
\bibfield{author}{\bibinfo{person}{Sneha~R. Krishna~Kumaran},
  \bibinfo{person}{Deana~C. McDonagh}, {and} \bibinfo{person}{Brian~P.
  Bailey}.} \bibinfo{year}{2017}\natexlab{}.
\newblock \showarticletitle{Increasing Quality and Involvement in Online Peer
  Feedback Exchange}.
\newblock \bibinfo{journal}{\emph{Proc. ACM Hum.-Comput. Interact.}}
  \bibinfo{volume}{1}, \bibinfo{number}{CSCW}, Article \bibinfo{articleno}{63}
  (\bibinfo{date}{dec} \bibinfo{year}{2017}), \bibinfo{numpages}{18}~pages.
\newblock
\urldef\tempurl%
\url{https://doi.org/10.1145/3134698}
\showDOI{\tempurl}


\bibitem[\protect\citeauthoryear{Kulkarni, Wei, Le, Chia, Papadopoulos, Cheng,
  Koller, and Klemmer}{Kulkarni et~al\mbox{.}}{2013}]%
        {peer_assessment_classes}
\bibfield{author}{\bibinfo{person}{Chinmay Kulkarni}, \bibinfo{person}{Koh~Pang
  Wei}, \bibinfo{person}{Huy Le}, \bibinfo{person}{Daniel Chia},
  \bibinfo{person}{Kathryn Papadopoulos}, \bibinfo{person}{Justin Cheng},
  \bibinfo{person}{Daphne Koller}, {and} \bibinfo{person}{Scott~R. Klemmer}.}
  \bibinfo{year}{2013}\natexlab{}.
\newblock \showarticletitle{Peer and Self Assessment in Massive Online
  Classes}.
\newblock \bibinfo{journal}{\emph{ACM Trans. Comput.-Hum. Interact.}}
  \bibinfo{volume}{20}, \bibinfo{number}{6}, Article \bibinfo{articleno}{33}
  (\bibinfo{date}{dec} \bibinfo{year}{2013}), \bibinfo{numpages}{31}~pages.
\newblock
\showISSN{1073-0516}
\urldef\tempurl%
\url{https://doi.org/10.1145/2505057}
\showDOI{\tempurl}


\bibitem[\protect\citeauthoryear{Kulkarni, Bernstein, and Klemmer}{Kulkarni
  et~al\mbox{.}}{2015}]%
        {kulkarni2015peerstudio}
\bibfield{author}{\bibinfo{person}{Chinmay~E Kulkarni},
  \bibinfo{person}{Michael~S Bernstein}, {and} \bibinfo{person}{Scott~R
  Klemmer}.} \bibinfo{year}{2015}\natexlab{}.
\newblock \showarticletitle{PeerStudio: rapid peer feedback emphasizes revision
  and improves performance}. In \bibinfo{booktitle}{\emph{Proceedings of the
  second (2015) ACM conference on learning@ scale}}. \bibinfo{pages}{75--84}.
\newblock


\bibitem[\protect\citeauthoryear{Lan, Chen, Goodman, Gimpel, Sharma, and
  Soricut}{Lan et~al\mbox{.}}{2019}]%
        {lan2019albert}
\bibfield{author}{\bibinfo{person}{Zhenzhong Lan}, \bibinfo{person}{Mingda
  Chen}, \bibinfo{person}{Sebastian Goodman}, \bibinfo{person}{Kevin Gimpel},
  \bibinfo{person}{Piyush Sharma}, {and} \bibinfo{person}{Radu Soricut}.}
  \bibinfo{year}{2019}\natexlab{}.
\newblock \showarticletitle{Albert: A lite bert for self-supervised learning of
  language representations}.
\newblock \bibinfo{journal}{\emph{arXiv preprint arXiv:1909.11942}}
  (\bibinfo{year}{2019}).
\newblock


\bibitem[\protect\citeauthoryear{Lee, Srivastava, Kumar, Brafman, and
  Klemmer}{Lee et~al\mbox{.}}{2010}]%
        {designing_with_examples}
\bibfield{author}{\bibinfo{person}{Brian Lee}, \bibinfo{person}{Savil
  Srivastava}, \bibinfo{person}{Ranjitha Kumar}, \bibinfo{person}{Ronen
  Brafman}, {and} \bibinfo{person}{Scott~R. Klemmer}.}
  \bibinfo{year}{2010}\natexlab{}.
\newblock \showarticletitle{Designing with Interactive Example Galleries}. In
  \bibinfo{booktitle}{\emph{Proceedings of the SIGCHI Conference on Human
  Factors in Computing Systems}} (Atlanta, Georgia, USA)
  \emph{(\bibinfo{series}{CHI '10})}. \bibinfo{publisher}{Association for
  Computing Machinery}, \bibinfo{address}{New York, NY, USA},
  \bibinfo{pages}{2257–2266}.
\newblock
\showISBNx{9781605589299}
\urldef\tempurl%
\url{https://doi.org/10.1145/1753326.1753667}
\showDOI{\tempurl}


\bibitem[\protect\citeauthoryear{Liao, Gruen, and Miller}{Liao
  et~al\mbox{.}}{2020}]%
        {questioningAI}
\bibfield{author}{\bibinfo{person}{Q.~Vera Liao}, \bibinfo{person}{Daniel
  Gruen}, {and} \bibinfo{person}{Sarah Miller}.}
  \bibinfo{year}{2020}\natexlab{}.
\newblock \showarticletitle{Questioning the AI: Informing Design Practices for
  Explainable AI User Experiences}. In \bibinfo{booktitle}{\emph{Proceedings of
  the 2020 CHI Conference on Human Factors in Computing Systems}}.
  \bibinfo{publisher}{Association for Computing Machinery},
  \bibinfo{address}{New York, NY, USA}, \bibinfo{pages}{1–15}.
\newblock
\showISBNx{9781450367080}
\urldef\tempurl%
\url{https://doi.org/10.1145/3313831.3376590}
\showURL{%
\tempurl}


\bibitem[\protect\citeauthoryear{Lin}{Lin}{2004}]%
        {lin2004rouge}
\bibfield{author}{\bibinfo{person}{Chin-Yew Lin}.}
  \bibinfo{year}{2004}\natexlab{}.
\newblock \showarticletitle{Rouge: A package for automatic evaluation of
  summaries}. In \bibinfo{booktitle}{\emph{Text summarization branches out}}.
  \bibinfo{pages}{74--81}.
\newblock


\bibitem[\protect\citeauthoryear{Liu, Huang, Liu, Zhou, Peng, and Ma}{Liu
  et~al\mbox{.}}{2022}]%
        {planhelper}
\bibfield{author}{\bibinfo{person}{Chengzhong Liu}, \bibinfo{person}{Zeyu
  Huang}, \bibinfo{person}{Dingdong Liu}, \bibinfo{person}{Shixu Zhou},
  \bibinfo{person}{Zhenhui Peng}, {and} \bibinfo{person}{Xiaojuan Ma}.}
  \bibinfo{year}{2022}\natexlab{}.
\newblock \showarticletitle{PlanHelper: Supporting Activity Plan Construction
  with Answer Posts in Community-Based QA Platforms}.
\newblock \bibinfo{journal}{\emph{Proc. ACM Hum.-Comput. Interact.}}
  \bibinfo{volume}{6}, \bibinfo{number}{CSCW2}, Article
  \bibinfo{articleno}{454} (\bibinfo{date}{nov} \bibinfo{year}{2022}),
  \bibinfo{numpages}{26}~pages.
\newblock
\urldef\tempurl%
\url{https://doi.org/10.1145/3555555}
\showDOI{\tempurl}


\bibitem[\protect\citeauthoryear{Liu and Singh}{Liu and Singh}{2004}]%
        {liu2004conceptnet}
\bibfield{author}{\bibinfo{person}{Hugo Liu} {and} \bibinfo{person}{Push
  Singh}.} \bibinfo{year}{2004}\natexlab{}.
\newblock \showarticletitle{ConceptNet—a practical commonsense reasoning
  tool-kit}.
\newblock \bibinfo{journal}{\emph{BT technology journal}} \bibinfo{volume}{22},
  \bibinfo{number}{4} (\bibinfo{year}{2004}), \bibinfo{pages}{211--226}.
\newblock


\bibitem[\protect\citeauthoryear{Liu, Ott, Goyal, Du, Joshi, Chen, Levy, Lewis,
  Zettlemoyer, and Stoyanov}{Liu et~al\mbox{.}}{2019}]%
        {reberta}
\bibfield{author}{\bibinfo{person}{Yinhan Liu}, \bibinfo{person}{Myle Ott},
  \bibinfo{person}{Naman Goyal}, \bibinfo{person}{Jingfei Du},
  \bibinfo{person}{Mandar Joshi}, \bibinfo{person}{Danqi Chen},
  \bibinfo{person}{Omer Levy}, \bibinfo{person}{Mike Lewis},
  \bibinfo{person}{Luke Zettlemoyer}, {and} \bibinfo{person}{Veselin
  Stoyanov}.} \bibinfo{year}{2019}\natexlab{}.
\newblock \showarticletitle{RoBERTa: {A} Robustly Optimized {BERT} Pretraining
  Approach}.
\newblock \bibinfo{journal}{\emph{CoRR}}  \bibinfo{volume}{abs/1907.11692}
  (\bibinfo{year}{2019}).
\newblock
\showeprint[arxiv]{1907.11692}
\urldef\tempurl%
\url{http://arxiv.org/abs/1907.11692}
\showURL{%
\tempurl}


\bibitem[\protect\citeauthoryear{Lubold, Walker, Pon-Barry, and Ogan}{Lubold
  et~al\mbox{.}}{2018}]%
        {lubold2018automated}
\bibfield{author}{\bibinfo{person}{Nichola Lubold}, \bibinfo{person}{Erin
  Walker}, \bibinfo{person}{Heather Pon-Barry}, {and} \bibinfo{person}{Amy
  Ogan}.} \bibinfo{year}{2018}\natexlab{}.
\newblock \showarticletitle{Automated pitch convergence improves learning in a
  social, teachable robot for middle school mathematics}. In
  \bibinfo{booktitle}{\emph{International conference on artificial intelligence
  in education}}. Springer, \bibinfo{pages}{282--296}.
\newblock


\bibitem[\protect\citeauthoryear{Lundstrom and Baker}{Lundstrom and
  Baker}{2009}]%
        {Lundstrom2009ToGI}
\bibfield{author}{\bibinfo{person}{Kristi Lundstrom} {and}
  \bibinfo{person}{Wendy Baker}.} \bibinfo{year}{2009}\natexlab{}.
\newblock \showarticletitle{To give is better than to receive: The benefits of
  peer review to the reviewer's own writing}.
\newblock \bibinfo{journal}{\emph{Journal of Second Language Writing}}
  \bibinfo{volume}{18} (\bibinfo{year}{2009}), \bibinfo{pages}{30--43}.
\newblock


\bibitem[\protect\citeauthoryear{Luther, Tolentino, Wu, Pavel, Bailey,
  Agrawala, Hartmann, and Dow}{Luther et~al\mbox{.}}{2015}]%
        {luther2015structuring}
\bibfield{author}{\bibinfo{person}{Kurt Luther}, \bibinfo{person}{Jari-Lee
  Tolentino}, \bibinfo{person}{Wei Wu}, \bibinfo{person}{Amy Pavel},
  \bibinfo{person}{Brian~P Bailey}, \bibinfo{person}{Maneesh Agrawala},
  \bibinfo{person}{Bj{\"o}rn Hartmann}, {and} \bibinfo{person}{Steven~P Dow}.}
  \bibinfo{year}{2015}\natexlab{}.
\newblock \showarticletitle{Structuring, aggregating, and evaluating
  crowdsourced design critique}. In \bibinfo{booktitle}{\emph{Proceedings of
  the 18th ACM conference on computer supported cooperative work \& social
  computing}}. \bibinfo{pages}{473--485}.
\newblock


\bibitem[\protect\citeauthoryear{LW, DR, PW, KA, Mayer, PR, Raths, and MC}{LW
  et~al\mbox{.}}{2001}]%
        {bloom_taxonomy}
\bibfield{author}{\bibinfo{person}{Anderson LW}, \bibinfo{person}{Krathwohl
  DR}, \bibinfo{person}{Airasian PW}, \bibinfo{person}{Cruikshank KA},
  \bibinfo{person}{Richard Mayer}, \bibinfo{person}{Pintrich PR},
  \bibinfo{person}{J. Raths}, {and} \bibinfo{person}{Wittrock MC}.}
  \bibinfo{year}{2001}\natexlab{}.
\newblock \bibinfo{booktitle}{\emph{A Taxonomy for Learning, Teaching, and
  Assessing: A Revision of Bloom's Taxonomy of Educational Objectives}}.
\newblock \bibinfo{publisher}{Pearson; 1st edition}.
\newblock
\showISBNx{ISBN: 080131903X}


\bibitem[\protect\citeauthoryear{Mann and Whitney}{Mann and Whitney}{1947}]%
        {mann1947test}
\bibfield{author}{\bibinfo{person}{Henry~B Mann} {and}
  \bibinfo{person}{Donald~R Whitney}.} \bibinfo{year}{1947}\natexlab{}.
\newblock \showarticletitle{On a test of whether one of two random variables is
  stochastically larger than the other}.
\newblock \bibinfo{journal}{\emph{The annals of mathematical statistics}}
  (\bibinfo{year}{1947}), \bibinfo{pages}{50--60}.
\newblock


\bibitem[\protect\citeauthoryear{Marlow and Dabbish}{Marlow and
  Dabbish}{2014}]%
        {high-quality-cscw2014}
\bibfield{author}{\bibinfo{person}{Jennifer Marlow} {and}
  \bibinfo{person}{Laura Dabbish}.} \bibinfo{year}{2014}\natexlab{}.
\newblock \showarticletitle{From rookie to all-star: professional development
  in a graphic design social networking site}. In
  \bibinfo{booktitle}{\emph{Proceedings of the 17th ACM conference on Computer
  supported cooperative work \& social computing}}. \bibinfo{pages}{922--933}.
\newblock


\bibitem[\protect\citeauthoryear{MasterClass}{MasterClass}{2021}]%
        {visual_elements}
\bibfield{author}{\bibinfo{person}{MasterClass}.}
  \bibinfo{year}{2021}\natexlab{}.
\newblock \bibinfo{title}{Elements of Design: Understanding the 7 Elements of
  Design}.
\newblock
\newblock
\newblock
\shownote{Accessed in September, 2022 from
  \url{https://www.masterclass.com/articles/elements-of-design-explained}.}


\bibitem[\protect\citeauthoryear{Newman and Landay}{Newman and Landay}{2000}]%
        {graphic_design_common}
\bibfield{author}{\bibinfo{person}{Mark~W. Newman} {and}
  \bibinfo{person}{James~A. Landay}.} \bibinfo{year}{2000}\natexlab{}.
\newblock \showarticletitle{Sitemaps, Storyboards, and Specifications: A Sketch
  of Web Site Design Practice}. In \bibinfo{booktitle}{\emph{Proceedings of the
  3rd Conference on Designing Interactive Systems: Processes, Practices,
  Methods, and Techniques}} (New York City, New York, USA)
  \emph{(\bibinfo{series}{DIS '00})}. \bibinfo{publisher}{Association for
  Computing Machinery}, \bibinfo{address}{New York, NY, USA},
  \bibinfo{pages}{263–274}.
\newblock
\showISBNx{1581132190}
\urldef\tempurl%
\url{https://doi.org/10.1145/347642.347758}
\showDOI{\tempurl}


\bibitem[\protect\citeauthoryear{Ngoon, Fraser, Weingarten, Dontcheva, and
  Klemmer}{Ngoon et~al\mbox{.}}{2018}]%
        {InteractiveGuidance}
\bibfield{author}{\bibinfo{person}{Tricia~J. Ngoon}, \bibinfo{person}{C.~Ailie
  Fraser}, \bibinfo{person}{Ariel~S. Weingarten}, \bibinfo{person}{Mira
  Dontcheva}, {and} \bibinfo{person}{Scott Klemmer}.}
  \bibinfo{year}{2018}\natexlab{}.
\newblock \bibinfo{booktitle}{\emph{Interactive Guidance Techniques for
  Improving Creative Feedback}}.
\newblock \bibinfo{publisher}{Association for Computing Machinery},
  \bibinfo{address}{New York, NY, USA}, \bibinfo{pages}{1–11}.
\newblock
\showISBNx{9781450356206}
\urldef\tempurl%
\url{https://doi.org/10.1145/3173574.3173629}
\showURL{%
\tempurl}


\bibitem[\protect\citeauthoryear{Nye, Graesser, and Hu}{Nye
  et~al\mbox{.}}{2014}]%
        {auto_tutor}
\bibfield{author}{\bibinfo{person}{Benjamin Nye}, \bibinfo{person}{Arthur
  Graesser}, {and} \bibinfo{person}{Xiangen Hu}.}
  \bibinfo{year}{2014}\natexlab{}.
\newblock \showarticletitle{AutoTutor and Family: A Review of 17 Years of
  Natural Language Tutoring}.
\newblock \bibinfo{journal}{\emph{International Journal of Artificial
  Intelligence in Education}}  \bibinfo{volume}{24} (\bibinfo{date}{12}
  \bibinfo{year}{2014}).
\newblock
\urldef\tempurl%
\url{https://doi.org/10.1007/s40593-014-0029-5}
\showDOI{\tempurl}


\bibitem[\protect\citeauthoryear{O'Brien}{O'Brien}{2016}]%
        {OBrien2016}
\bibfield{author}{\bibinfo{person}{Heather O'Brien}.}
  \bibinfo{year}{2016}\natexlab{}.
\newblock \bibinfo{booktitle}{\emph{Theoretical Perspectives on User
  Engagement}}.
\newblock \bibinfo{publisher}{Springer International Publishing},
  \bibinfo{address}{Cham}, \bibinfo{pages}{1--26}.
\newblock
\showISBNx{978-3-319-27446-1}
\urldef\tempurl%
\url{https://doi.org/10.1007/978-3-319-27446-1_1}
\showDOI{\tempurl}


\bibitem[\protect\citeauthoryear{OpenAI}{OpenAI}{2023}]%
        {openai2023gpt4}
\bibfield{author}{\bibinfo{person}{OpenAI}.} \bibinfo{year}{2023}\natexlab{}.
\newblock \bibinfo{title}{GPT-4 Technical Report}.
\newblock
\newblock
\showeprint[arxiv]{2303.08774}~[cs.CL]


\bibitem[\protect\citeauthoryear{Oppenlaender, Kuosmanen, Lucero, and
  Hosio}{Oppenlaender et~al\mbox{.}}{2021}]%
        {peerfeedback_chi2021}
\bibfield{author}{\bibinfo{person}{Jonas Oppenlaender}, \bibinfo{person}{Elina
  Kuosmanen}, \bibinfo{person}{Andr\'{e}s Lucero}, {and} \bibinfo{person}{Simo
  Hosio}.} \bibinfo{year}{2021}\natexlab{}.
\newblock \showarticletitle{Hardhats and Bungaloos: Comparing Crowdsourced
  Design Feedback with Peer Design Feedback in the Classroom}. In
  \bibinfo{booktitle}{\emph{Proceedings of the 2021 CHI Conference on Human
  Factors in Computing Systems}} (Yokohama, Japan) \emph{(\bibinfo{series}{CHI
  '21})}. \bibinfo{publisher}{Association for Computing Machinery},
  \bibinfo{address}{New York, NY, USA}, Article \bibinfo{articleno}{570},
  \bibinfo{numpages}{14}~pages.
\newblock
\showISBNx{9781450380966}
\urldef\tempurl%
\url{https://doi.org/10.1145/3411764.3445380}
\showDOI{\tempurl}


\bibitem[\protect\citeauthoryear{Peng, Guo, Tsang, and Ma}{Peng
  et~al\mbox{.}}{2020}]%
        {zhenhui2020}
\bibfield{author}{\bibinfo{person}{Zhenhui Peng}, \bibinfo{person}{Qingyu Guo},
  \bibinfo{person}{Ka~Wing Tsang}, {and} \bibinfo{person}{Xiaojuan Ma}.}
  \bibinfo{year}{2020}\natexlab{}.
\newblock \showarticletitle{Exploring the Effects of Technological Writing
  Assistance for Support Providers in Online Mental Health Community}. In
  \bibinfo{booktitle}{\emph{Proceedings of the 2020 CHI Conference on Human
  Factors in Computing Systems}} (Honolulu, HI, USA)
  \emph{(\bibinfo{series}{CHI '20})}. \bibinfo{publisher}{ACM},
  \bibinfo{address}{New York, NY, USA}, \bibinfo{pages}{556--567}.
\newblock
\showISBNx{978-1-4503-6708-0}
\urldef\tempurl%
\url{https://doi.org/10.1145/3313831.3376695}
\showDOI{\tempurl}


\bibitem[\protect\citeauthoryear{Peng, Kwon, Lu, Wu, and Ma}{Peng
  et~al\mbox{.}}{2019}]%
        {Zhenhui2019}
\bibfield{author}{\bibinfo{person}{Zhenhui Peng}, \bibinfo{person}{Yunhwan
  Kwon}, \bibinfo{person}{Jiaan Lu}, \bibinfo{person}{Ziming Wu}, {and}
  \bibinfo{person}{Xiaojuan Ma}.} \bibinfo{year}{2019}\natexlab{}.
\newblock \showarticletitle{Design and Evaluation of Service Robot's
  Proactivity in Decision-Making Support Process}. In
  \bibinfo{booktitle}{\emph{Proceedings of the 2019 CHI Conference on Human
  Factors in Computing Systems}} (Glasgow, Scotland Uk)
  \emph{(\bibinfo{series}{CHI '19})}. \bibinfo{publisher}{Association for
  Computing Machinery}, \bibinfo{address}{New York, NY, USA},
  \bibinfo{pages}{1–13}.
\newblock
\showISBNx{9781450359702}
\urldef\tempurl%
\url{https://doi.org/10.1145/3290605.3300328}
\showDOI{\tempurl}


\bibitem[\protect\citeauthoryear{Peng, Liu, Zhou, Xu, and Ma}{Peng
  et~al\mbox{.}}{2022}]%
        {peng2022crebot}
\bibfield{author}{\bibinfo{person}{Zhenhui Peng}, \bibinfo{person}{Yuzhi Liu},
  \bibinfo{person}{Hanqi Zhou}, \bibinfo{person}{Zuyu Xu}, {and}
  \bibinfo{person}{Xiaojuan Ma}.} \bibinfo{year}{2022}\natexlab{}.
\newblock \showarticletitle{CReBot: Exploring interactive question prompts for
  critical paper reading}.
\newblock \bibinfo{journal}{\emph{International Journal of Human-Computer
  Studies}}  \bibinfo{volume}{167} (\bibinfo{year}{2022}),
  \bibinfo{pages}{102898}.
\newblock


\bibitem[\protect\citeauthoryear{Peng, Ma, Yang, Tsang, and Guo}{Peng
  et~al\mbox{.}}{2021}]%
        {zhenhui2021chi}
\bibfield{author}{\bibinfo{person}{Zhenhui Peng}, \bibinfo{person}{Xiaojuan
  Ma}, \bibinfo{person}{Diyi Yang}, \bibinfo{person}{Ka~Wing Tsang}, {and}
  \bibinfo{person}{Qingyu Guo}.} \bibinfo{year}{2021}\natexlab{}.
\newblock \showarticletitle{Effects of Support-Seekers’ Community Knowledge
  on Their Expressed Satisfaction with the Received Comments in Mental Health
  Communities}. In \bibinfo{booktitle}{\emph{Proceedings of the 2021 CHI
  Conference on Human Factors in Computing Systems}} (Yokohama, Japan)
  \emph{(\bibinfo{series}{CHI '21})}. \bibinfo{publisher}{Association for
  Computing Machinery}, \bibinfo{address}{New York, NY, USA}, Article
  \bibinfo{articleno}{536}, \bibinfo{numpages}{12}~pages.
\newblock
\showISBNx{9781450380966}
\urldef\tempurl%
\url{https://doi.org/10.1145/3411764.3445446}
\showDOI{\tempurl}


\bibitem[\protect\citeauthoryear{Pushshift}{Pushshift}{2022}]%
        {pushshift}
\bibfield{author}{\bibinfo{person}{Pushshift}.}
  \bibinfo{year}{2022}\natexlab{}.
\newblock \bibinfo{title}{Reddit Statistics - pushshift.io}.
\newblock
\newblock
\newblock
\shownote{Accessed in September, 2022 from \url{https://pushshift.io/}.}


\bibitem[\protect\citeauthoryear{Raffel, Shazeer, Roberts, Lee, Narang, Matena,
  Zhou, Li, and Liu}{Raffel et~al\mbox{.}}{2020}]%
        {2020t5}
\bibfield{author}{\bibinfo{person}{Colin Raffel}, \bibinfo{person}{Noam
  Shazeer}, \bibinfo{person}{Adam Roberts}, \bibinfo{person}{Katherine Lee},
  \bibinfo{person}{Sharan Narang}, \bibinfo{person}{Michael Matena},
  \bibinfo{person}{Yanqi Zhou}, \bibinfo{person}{Wei Li}, {and}
  \bibinfo{person}{Peter~J. Liu}.} \bibinfo{year}{2020}\natexlab{}.
\newblock \showarticletitle{Exploring the Limits of Transfer Learning with a
  Unified Text-to-Text Transformer}.
\newblock \bibinfo{journal}{\emph{Journal of Machine Learning Research}}
  \bibinfo{volume}{21}, \bibinfo{number}{140} (\bibinfo{year}{2020}),
  \bibinfo{pages}{1--67}.
\newblock
\urldef\tempurl%
\url{http://jmlr.org/papers/v21/20-074.html}
\showURL{%
\tempurl}


\bibitem[\protect\citeauthoryear{Resnick, Maloney, Monroy-Hern\'{a}ndez, Rusk,
  Eastmond, Brennan, Millner, Rosenbaum, Silver, Silverman, and Kafai}{Resnick
  et~al\mbox{.}}{2009}]%
        {scratch}
\bibfield{author}{\bibinfo{person}{Mitchel Resnick}, \bibinfo{person}{John
  Maloney}, \bibinfo{person}{Andr\'{e}s Monroy-Hern\'{a}ndez},
  \bibinfo{person}{Natalie Rusk}, \bibinfo{person}{Evelyn Eastmond},
  \bibinfo{person}{Karen Brennan}, \bibinfo{person}{Amon Millner},
  \bibinfo{person}{Eric Rosenbaum}, \bibinfo{person}{Jay Silver},
  \bibinfo{person}{Brian Silverman}, {and} \bibinfo{person}{Yasmin Kafai}.}
  \bibinfo{year}{2009}\natexlab{}.
\newblock \showarticletitle{Scratch: Programming for All}.
\newblock \bibinfo{journal}{\emph{Commun. ACM}} \bibinfo{volume}{52},
  \bibinfo{number}{11} (\bibinfo{date}{nov} \bibinfo{year}{2009}),
  \bibinfo{pages}{60–67}.
\newblock
\showISSN{0001-0782}
\urldef\tempurl%
\url{https://doi.org/10.1145/1592761.1592779}
\showDOI{\tempurl}


\bibitem[\protect\citeauthoryear{Ruan, Jiang, Xu, Tham, Qiu, Zhu, Murnane,
  Brunskill, and Landay}{Ruan et~al\mbox{.}}{2019}]%
        {quizbot10.1145/3290605.3300587}
\bibfield{author}{\bibinfo{person}{Sherry Ruan}, \bibinfo{person}{Liwei Jiang},
  \bibinfo{person}{Justin Xu}, \bibinfo{person}{Bryce Joe-Kun Tham},
  \bibinfo{person}{Zhengneng Qiu}, \bibinfo{person}{Yeshuang Zhu},
  \bibinfo{person}{Elizabeth~L. Murnane}, \bibinfo{person}{Emma Brunskill},
  {and} \bibinfo{person}{James~A. Landay}.} \bibinfo{year}{2019}\natexlab{}.
\newblock \showarticletitle{QuizBot: A Dialogue-Based Adaptive Learning System
  for Factual Knowledge}. In \bibinfo{booktitle}{\emph{Proceedings of the 2019
  CHI Conference on Human Factors in Computing Systems}} (Glasgow, Scotland Uk)
  \emph{(\bibinfo{series}{CHI ’19})}. \bibinfo{publisher}{Association for
  Computing Machinery}, \bibinfo{address}{New York, NY, USA},
  \bibinfo{pages}{1–13}.
\newblock
\showISBNx{9781450359702}
\urldef\tempurl%
\url{https://doi.org/10.1145/3290605.3300587}
\showDOI{\tempurl}


\bibitem[\protect\citeauthoryear{Sanh, Debut, Chaumond, and Wolf}{Sanh
  et~al\mbox{.}}{2019}]%
        {sanh2019distilbert}
\bibfield{author}{\bibinfo{person}{Victor Sanh}, \bibinfo{person}{Lysandre
  Debut}, \bibinfo{person}{Julien Chaumond}, {and} \bibinfo{person}{Thomas
  Wolf}.} \bibinfo{year}{2019}\natexlab{}.
\newblock \showarticletitle{DistilBERT, a distilled version of BERT: smaller,
  faster, cheaper and lighter}.
\newblock \bibinfo{journal}{\emph{arXiv preprint arXiv:1910.01108}}
  (\bibinfo{year}{2019}).
\newblock


\bibitem[\protect\citeauthoryear{Scaffidi and Chambers}{Scaffidi and
  Chambers}{2012}]%
        {scaffidi2012skill}
\bibfield{author}{\bibinfo{person}{Christopher Scaffidi} {and}
  \bibinfo{person}{Christopher Chambers}.} \bibinfo{year}{2012}\natexlab{}.
\newblock \showarticletitle{Skill progression demonstrated by users in the
  Scratch animation environment}.
\newblock \bibinfo{journal}{\emph{International Journal of Human-Computer
  Interaction}} \bibinfo{volume}{28}, \bibinfo{number}{6}
  (\bibinfo{year}{2012}), \bibinfo{pages}{383--398}.
\newblock


\bibitem[\protect\citeauthoryear{Sharma and De~Choudhury}{Sharma and
  De~Choudhury}{2018}]%
        {sharma2018mental}
\bibfield{author}{\bibinfo{person}{Eva Sharma} {and} \bibinfo{person}{Munmun
  De~Choudhury}.} \bibinfo{year}{2018}\natexlab{}.
\newblock \showarticletitle{Mental health support and its relationship to
  linguistic accommodation in online communities}. In
  \bibinfo{booktitle}{\emph{Proceedings of the 2018 CHI conference on human
  factors in computing systems}}. \bibinfo{pages}{1--13}.
\newblock


\bibitem[\protect\citeauthoryear{Shorey, Hill, and Woolley}{Shorey
  et~al\mbox{.}}{2021}]%
        {shorey2021hanging}
\bibfield{author}{\bibinfo{person}{Samantha Shorey},
  \bibinfo{person}{Benjamin~Mako Hill}, {and} \bibinfo{person}{Samuel
  Woolley}.} \bibinfo{year}{2021}\natexlab{}.
\newblock \showarticletitle{From hanging out to figuring it out: Socializing
  online as a pathway to computational thinking}.
\newblock \bibinfo{journal}{\emph{New Media \& Society}} \bibinfo{volume}{23},
  \bibinfo{number}{8} (\bibinfo{year}{2021}), \bibinfo{pages}{2327--2344}.
\newblock


\bibitem[\protect\citeauthoryear{Song, Ren, Liang, Li, Ma, and de~Rijke}{Song
  et~al\mbox{.}}{2017}]%
        {song2017summarizing}
\bibfield{author}{\bibinfo{person}{Hongya Song}, \bibinfo{person}{Zhaochun
  Ren}, \bibinfo{person}{Shangsong Liang}, \bibinfo{person}{Piji Li},
  \bibinfo{person}{Jun Ma}, {and} \bibinfo{person}{Maarten de Rijke}.}
  \bibinfo{year}{2017}\natexlab{}.
\newblock \showarticletitle{Summarizing answers in non-factoid community
  question-answering}. In \bibinfo{booktitle}{\emph{Proceedings of the Tenth
  ACM International Conference on Web Search and Data Mining}}.
  \bibinfo{pages}{405--414}.
\newblock


\bibitem[\protect\citeauthoryear{Syed, Collins-Thompson, Bennett, Tang,
  Williams, Iqbal, and Tay}{Syed et~al\mbox{.}}{2020}]%
        {question_generation}
\bibfield{author}{\bibinfo{person}{Rohail Syed}, \bibinfo{person}{Kevyn
  Collins-Thompson}, \bibinfo{person}{Paul Bennett}, \bibinfo{person}{Mengqiu
  Tang}, \bibinfo{person}{Shane Williams}, \bibinfo{person}{Shamsi Iqbal},
  {and} \bibinfo{person}{Wendy Tay}.} \bibinfo{year}{2020}\natexlab{}.
\newblock \showarticletitle{Improving Learning Outcomes with Gaze Tracking and
  Automatic Question Generation}. In \bibinfo{booktitle}{\emph{The Web
  Conference 2020 (formerly WWW conference)}}.
\newblock
\urldef\tempurl%
\url{https://www.microsoft.com/en-us/research/publication/improving-learning-outcomes-with-gaze-tracking-and-automatic-question-generation/}
\showURL{%
\tempurl}


\bibitem[\protect\citeauthoryear{Tausczik and Wang}{Tausczik and Wang}{2017}]%
        {toshare_cscw2017}
\bibfield{author}{\bibinfo{person}{Yla Tausczik} {and} \bibinfo{person}{Ping
  Wang}.} \bibinfo{year}{2017}\natexlab{}.
\newblock \showarticletitle{To Share, or Not to Share? Community-Level
  Collaboration in Open Innovation Contests}.
\newblock \bibinfo{journal}{\emph{Proc. ACM Hum.-Comput. Interact.}}
  \bibinfo{volume}{1}, \bibinfo{number}{CSCW}, Article \bibinfo{articleno}{100}
  (\bibinfo{date}{dec} \bibinfo{year}{2017}), \bibinfo{numpages}{23}~pages.
\newblock
\urldef\tempurl%
\url{https://doi.org/10.1145/3134735}
\showDOI{\tempurl}


\bibitem[\protect\citeauthoryear{Tausczik, Kittur, and Kraut}{Tausczik
  et~al\mbox{.}}{2014}]%
        {tausczik2014collaborative}
\bibfield{author}{\bibinfo{person}{Yla~R Tausczik}, \bibinfo{person}{Aniket
  Kittur}, {and} \bibinfo{person}{Robert~E Kraut}.}
  \bibinfo{year}{2014}\natexlab{}.
\newblock \showarticletitle{Collaborative problem solving: A study of
  mathoverflow}. In \bibinfo{booktitle}{\emph{Proceedings of the 17th ACM
  conference on Computer supported cooperative work \& social computing}}.
  \bibinfo{pages}{355--367}.
\newblock


\bibitem[\protect\citeauthoryear{Umar, Squicciarini, and Rajtmajer}{Umar
  et~al\mbox{.}}{2019}]%
        {umar2019detection}
\bibfield{author}{\bibinfo{person}{Prasanna Umar}, \bibinfo{person}{Anna
  Squicciarini}, {and} \bibinfo{person}{Sarah Rajtmajer}.}
  \bibinfo{year}{2019}\natexlab{}.
\newblock \showarticletitle{Detection and analysis of self-disclosure in online
  news commentaries}. In \bibinfo{booktitle}{\emph{The World Wide Web
  Conference}}. \bibinfo{pages}{3272--3278}.
\newblock


\bibitem[\protect\citeauthoryear{Venkatesh and Bala}{Venkatesh and
  Bala}{2008}]%
        {Technology_acceptance_model:doi:10.1111/j.1540-5915.2008.00192.x}
\bibfield{author}{\bibinfo{person}{Viswanath Venkatesh} {and}
  \bibinfo{person}{Hillol Bala}.} \bibinfo{year}{2008}\natexlab{}.
\newblock \showarticletitle{Technology Acceptance Model 3 and a Research Agenda
  on Interventions}.
\newblock \bibinfo{journal}{\emph{Decision Sciences}} \bibinfo{volume}{39},
  \bibinfo{number}{2} (\bibinfo{year}{2008}), \bibinfo{pages}{273--315}.
\newblock
\urldef\tempurl%
\url{https://doi.org/10.1111/j.1540-5915.2008.00192.x}
\showDOI{\tempurl}
\showeprint{https://onlinelibrary.wiley.com/doi/pdf/10.1111/j.1540-5915.2008.00192.x}


\bibitem[\protect\citeauthoryear{Wambsganss, Kueng, Soellner, and
  Leimeister}{Wambsganss et~al\mbox{.}}{2021}]%
        {arguetutor}
\bibfield{author}{\bibinfo{person}{Thiemo Wambsganss}, \bibinfo{person}{Tobias
  Kueng}, \bibinfo{person}{Matthias Soellner}, {and} \bibinfo{person}{Jan~Marco
  Leimeister}.} \bibinfo{year}{2021}\natexlab{}.
\newblock \bibinfo{booktitle}{\emph{ArgueTutor: An Adaptive Dialog-Based
  Learning System for Argumentation Skills}}.
\newblock \bibinfo{publisher}{Association for Computing Machinery},
  \bibinfo{address}{New York, NY, USA}.
\newblock
\showISBNx{9781450380966}
\urldef\tempurl%
\url{https://doi.org/10.1145/3411764.3445781}
\showURL{%
\tempurl}


\bibitem[\protect\citeauthoryear{Wambsganss, Winkler, S{\"o}llner, and
  Leimeister}{Wambsganss et~al\mbox{.}}{2020}]%
        {wambsganss2020conversational}
\bibfield{author}{\bibinfo{person}{Thiemo Wambsganss}, \bibinfo{person}{Rainer
  Winkler}, \bibinfo{person}{Matthias S{\"o}llner}, {and}
  \bibinfo{person}{Jan~Marco Leimeister}.} \bibinfo{year}{2020}\natexlab{}.
\newblock \showarticletitle{A conversational agent to improve response quality
  in course evaluations}. In \bibinfo{booktitle}{\emph{Extended abstracts of
  the 2020 CHI conference on human factors in computing systems}}.
  \bibinfo{pages}{1--9}.
\newblock


\bibitem[\protect\citeauthoryear{Wang, Rose, and Koedinger}{Wang
  et~al\mbox{.}}{2021}]%
        {wang2021seeing}
\bibfield{author}{\bibinfo{person}{Xu Wang}, \bibinfo{person}{Carolyn Rose},
  {and} \bibinfo{person}{Ken Koedinger}.} \bibinfo{year}{2021}\natexlab{}.
\newblock \showarticletitle{Seeing beyond expert blind spots: Online learning
  design for scale and quality}. In \bibinfo{booktitle}{\emph{Proceedings of
  the 2021 CHI Conference on Human Factors in Computing Systems}}.
  \bibinfo{pages}{1--14}.
\newblock


\bibitem[\protect\citeauthoryear{Weber, Wambsganß, Rüttimann, and
  Söllner}{Weber et~al\mbox{.}}{2021}]%
        {weber_pca}
\bibfield{author}{\bibinfo{person}{Florian Weber}, \bibinfo{person}{Thiemo
  Wambsganß}, \bibinfo{person}{Dominic Rüttimann}, {and}
  \bibinfo{person}{Matthias Söllner}.} \bibinfo{year}{2021}\natexlab{}.
\newblock \showarticletitle{Pedagogical Agents for Interactive Learning: A
  Taxonomy of Conversational Agents in Education Completed Research Paper}.
\newblock


\bibitem[\protect\citeauthoryear{Weinman, Drucker, Barik, and DeLine}{Weinman
  et~al\mbox{.}}{2021}]%
        {forkit}
\bibfield{author}{\bibinfo{person}{Nathaniel Weinman},
  \bibinfo{person}{Steven~M. Drucker}, \bibinfo{person}{Titus Barik}, {and}
  \bibinfo{person}{Robert DeLine}.} \bibinfo{year}{2021}\natexlab{}.
\newblock \showarticletitle{Fork It: Supporting Stateful Alternatives in
  Computational Notebooks}. In \bibinfo{booktitle}{\emph{Proceedings of the
  2021 CHI Conference on Human Factors in Computing Systems}} (Yokohama, Japan)
  \emph{(\bibinfo{series}{CHI '21})}. \bibinfo{publisher}{Association for
  Computing Machinery}, \bibinfo{address}{New York, NY, USA}, Article
  \bibinfo{articleno}{307}, \bibinfo{numpages}{12}~pages.
\newblock
\showISBNx{9781450380966}
\urldef\tempurl%
\url{https://doi.org/10.1145/3411764.3445527}
\showDOI{\tempurl}


\bibitem[\protect\citeauthoryear{Winkler, Hobert, Salovaara, S\"{o}llner, and
  Leimeister}{Winkler et~al\mbox{.}}{2020}]%
        {sara10.1145/3313831.3376781}
\bibfield{author}{\bibinfo{person}{Rainer Winkler}, \bibinfo{person}{Sebastian
  Hobert}, \bibinfo{person}{Antti Salovaara}, \bibinfo{person}{Matthias
  S\"{o}llner}, {and} \bibinfo{person}{Jan~Marco Leimeister}.}
  \bibinfo{year}{2020}\natexlab{}.
\newblock \showarticletitle{Sara, the Lecturer: Improving Learning in Online
  Education with a Scaffolding-Based Conversational Agent}. In
  \bibinfo{booktitle}{\emph{Proceedings of the 2020 CHI Conference on Human
  Factors in Computing Systems}} (Honolulu, HI, USA)
  \emph{(\bibinfo{series}{CHI ’20})}. \bibinfo{publisher}{Association for
  Computing Machinery}, \bibinfo{address}{New York, NY, USA},
  \bibinfo{pages}{1–14}.
\newblock
\showISBNx{9781450367080}
\urldef\tempurl%
\url{https://doi.org/10.1145/3313831.3376781}
\showDOI{\tempurl}


\bibitem[\protect\citeauthoryear{Wolf, Debut, Sanh, Chaumond, Delangue, Moi,
  Cistac, Rault, Louf, Funtowicz, et~al\mbox{.}}{Wolf et~al\mbox{.}}{2020}]%
        {wolf2020transformers}
\bibfield{author}{\bibinfo{person}{Thomas Wolf}, \bibinfo{person}{Lysandre
  Debut}, \bibinfo{person}{Victor Sanh}, \bibinfo{person}{Julien Chaumond},
  \bibinfo{person}{Clement Delangue}, \bibinfo{person}{Anthony Moi},
  \bibinfo{person}{Pierric Cistac}, \bibinfo{person}{Tim Rault},
  \bibinfo{person}{R{\'e}mi Louf}, \bibinfo{person}{Morgan Funtowicz},
  {et~al\mbox{.}}} \bibinfo{year}{2020}\natexlab{}.
\newblock \showarticletitle{Transformers: State-of-the-art natural language
  processing}. In \bibinfo{booktitle}{\emph{Proceedings of the 2020 conference
  on empirical methods in natural language processing: system demonstrations}}.
  \bibinfo{pages}{38--45}.
\newblock


\bibitem[\protect\citeauthoryear{Woolson}{Woolson}{2008}]%
        {wolcoxon_test}
\bibfield{author}{\bibinfo{person}{Robert Woolson}.}
  \bibinfo{year}{2008}\natexlab{}.
\newblock \bibinfo{booktitle}{\emph{Wilcoxon Signed-Rank Test}}.
\newblock
\showISBNx{047146242X}
\urldef\tempurl%
\url{https://doi.org/10.1002/9780471462422.eoct979}
\showDOI{\tempurl}


\bibitem[\protect\citeauthoryear{Xu and Bailey}{Xu and Bailey}{2012}]%
        {high-quality-cscw2012}
\bibfield{author}{\bibinfo{person}{Anbang Xu} {and} \bibinfo{person}{Brian
  Bailey}.} \bibinfo{year}{2012}\natexlab{}.
\newblock \showarticletitle{What do you think? A case study of benefit,
  expectation, and interaction in a large online critique community}. In
  \bibinfo{booktitle}{\emph{Proceedings of the acm 2012 conference on computer
  supported cooperative work}}. \bibinfo{pages}{295--304}.
\newblock


\bibitem[\protect\citeauthoryear{Xu, Huang, and Bailey}{Xu
  et~al\mbox{.}}{2014}]%
        {structured_feedback}
\bibfield{author}{\bibinfo{person}{Anbang Xu}, \bibinfo{person}{Shih-Wen
  Huang}, {and} \bibinfo{person}{Brian Bailey}.}
  \bibinfo{year}{2014}\natexlab{}.
\newblock \showarticletitle{Voyant: Generating Structured Feedback on Visual
  Designs Using a Crowd of Non-Experts}. In
  \bibinfo{booktitle}{\emph{Proceedings of the 17th ACM Conference on Computer
  Supported Cooperative Work and Social Computing}} (Baltimore, Maryland, USA)
  \emph{(\bibinfo{series}{CSCW '14})}. \bibinfo{publisher}{Association for
  Computing Machinery}, \bibinfo{address}{New York, NY, USA},
  \bibinfo{pages}{1433–1444}.
\newblock
\showISBNx{9781450325400}
\urldef\tempurl%
\url{https://doi.org/10.1145/2531602.2531604}
\showDOI{\tempurl}


\bibitem[\protect\citeauthoryear{Yan, Glassman, and Zhang}{Yan
  et~al\mbox{.}}{2021}]%
        {visualize_example}
\bibfield{author}{\bibinfo{person}{Litao Yan}, \bibinfo{person}{Elena~L.
  Glassman}, {and} \bibinfo{person}{Tianyi Zhang}.}
  \bibinfo{year}{2021}\natexlab{}.
\newblock \showarticletitle{Visualizing Examples of Deep Neural Networks at
  Scale}. In \bibinfo{booktitle}{\emph{Proceedings of the 2021 CHI Conference
  on Human Factors in Computing Systems}} (Yokohama, Japan)
  \emph{(\bibinfo{series}{CHI '21})}. \bibinfo{publisher}{Association for
  Computing Machinery}, \bibinfo{address}{New York, NY, USA}, Article
  \bibinfo{articleno}{313}, \bibinfo{numpages}{14}~pages.
\newblock
\showISBNx{9781450380966}
\urldef\tempurl%
\url{https://doi.org/10.1145/3411764.3445654}
\showDOI{\tempurl}


\bibitem[\protect\citeauthoryear{Yang, Kraut, Smith, Mayfield, and
  Jurafsky}{Yang et~al\mbox{.}}{2019a}]%
        {yangseeker}
\bibfield{author}{\bibinfo{person}{Diyi Yang}, \bibinfo{person}{Robert~E.
  Kraut}, \bibinfo{person}{Tenbroeck Smith}, \bibinfo{person}{Elijah Mayfield},
  {and} \bibinfo{person}{Dan Jurafsky}.} \bibinfo{year}{2019}\natexlab{a}.
\newblock \showarticletitle{Seekers, Providers, Welcomers, and Storytellers:
  Modeling Social Roles in Online Health Communities}. In
  \bibinfo{booktitle}{\emph{Proceedings of the 2019 CHI Conference on Human
  Factors in Computing Systems}} (Glasgow, Scotland Uk)
  \emph{(\bibinfo{series}{CHI '19})}. \bibinfo{publisher}{Association for
  Computing Machinery}, \bibinfo{address}{New York, NY, USA},
  \bibinfo{pages}{1–14}.
\newblock
\showISBNx{9781450359702}
\urldef\tempurl%
\url{https://doi.org/10.1145/3290605.3300574}
\showDOI{\tempurl}


\bibitem[\protect\citeauthoryear{Yang, Yao, Seering, and Kraut}{Yang
  et~al\mbox{.}}{2019b}]%
        {yang2019channel}
\bibfield{author}{\bibinfo{person}{Diyi Yang}, \bibinfo{person}{Zheng Yao},
  \bibinfo{person}{Joseph Seering}, {and} \bibinfo{person}{Robert Kraut}.}
  \bibinfo{year}{2019}\natexlab{b}.
\newblock \showarticletitle{The channel matters: Self-disclosure, reciprocity
  and social support in online cancer support groups}. In
  \bibinfo{booktitle}{\emph{Proceedings of the 2019 chi conference on human
  factors in computing systems}}. \bibinfo{pages}{1--15}.
\newblock


\bibitem[\protect\citeauthoryear{Yang, Domeniconi, Revelle, Sweeney, Gelman,
  Beckley, and Johri}{Yang et~al\mbox{.}}{2015}]%
        {yang2015uncovering}
\bibfield{author}{\bibinfo{person}{Seungwon Yang}, \bibinfo{person}{Carlotta
  Domeniconi}, \bibinfo{person}{Matt Revelle}, \bibinfo{person}{Mack Sweeney},
  \bibinfo{person}{Ben~U Gelman}, \bibinfo{person}{Chris Beckley}, {and}
  \bibinfo{person}{Aditya Johri}.} \bibinfo{year}{2015}\natexlab{}.
\newblock \showarticletitle{Uncovering trajectories of informal learning in
  large online communities of creators}. In
  \bibinfo{booktitle}{\emph{Proceedings of the Second (2015) ACM Conference on
  Learning@ Scale}}. \bibinfo{pages}{131--140}.
\newblock


\bibitem[\protect\citeauthoryear{Yen and Dow}{Yen and Dow}{2022}]%
        {example_for_learning}
\bibfield{author}{\bibinfo{person}{Yu-Chun~Grace Yen} {and}
  \bibinfo{person}{Steven~P. Dow}.} \bibinfo{year}{2022}\natexlab{}.
\newblock \showarticletitle{Seeking Exemplars in the Wild: Exploring How
  Students Find Design Examples to Support Personalized Learning}. In
  \bibinfo{booktitle}{\emph{Proceedings of the Ninth ACM Conference on Learning
  @ Scale}} (New York City, NY, USA) \emph{(\bibinfo{series}{L@S '22})}.
  \bibinfo{publisher}{Association for Computing Machinery},
  \bibinfo{address}{New York, NY, USA}, \bibinfo{pages}{418–421}.
\newblock
\showISBNx{9781450391580}
\urldef\tempurl%
\url{https://doi.org/10.1145/3491140.3528303}
\showDOI{\tempurl}


\bibitem[\protect\citeauthoryear{Yen, Kim, and Bailey}{Yen
  et~al\mbox{.}}{2020}]%
        {yen2020decipher}
\bibfield{author}{\bibinfo{person}{Yu-Chun~Grace Yen}, \bibinfo{person}{Joy~O
  Kim}, {and} \bibinfo{person}{Brian~P Bailey}.}
  \bibinfo{year}{2020}\natexlab{}.
\newblock \showarticletitle{Decipher: an interactive visualization tool for
  interpreting unstructured design feedback from multiple providers}. In
  \bibinfo{booktitle}{\emph{Proceedings of the 2020 CHI Conference on Human
  Factors in Computing Systems}}. \bibinfo{pages}{1--13}.
\newblock


\bibitem[\protect\citeauthoryear{Yuan, Luther, Krause, Vennix, Dow, and
  Hartmann}{Yuan et~al\mbox{.}}{2016}]%
        {yuan2016almost}
\bibfield{author}{\bibinfo{person}{Alvin Yuan}, \bibinfo{person}{Kurt Luther},
  \bibinfo{person}{Markus Krause}, \bibinfo{person}{Sophie~Isabel Vennix},
  \bibinfo{person}{Steven~P Dow}, {and} \bibinfo{person}{Bjorn Hartmann}.}
  \bibinfo{year}{2016}\natexlab{}.
\newblock \showarticletitle{Almost an expert: The effects of rubrics and
  expertise on perceived value of crowdsourced design critiques}. In
  \bibinfo{booktitle}{\emph{Proceedings of the 19th ACM Conference on
  Computer-Supported Cooperative Work \& Social Computing}}.
  \bibinfo{pages}{1005--1017}.
\newblock


\end{thebibliography}


\end{document}